\documentclass[11pt]{article}
\usepackage[margin=1in]{geometry}
\usepackage{bm}

\usepackage{multirow}
\usepackage{amssymb,amsmath,amsthm,amsfonts}
\usepackage{hyperref}
\usepackage[capitalise,nameinlink]{cleveref}
\usepackage{aliascnt}
\usepackage{letltxmacro}
\usepackage{cancel}
\usepackage{bm}
\usepackage{mathtools}
\usepackage{multicol}
\usepackage{enumitem}
\usepackage{authblk}
\usepackage{graphicx}
\usepackage{subcaption}
\usepackage{tablefootnote} 
\usepackage{float}
\usepackage{physics}
\usepackage{footnote}
\usepackage{xcolor}
\usepackage{mathrsfs}
\usepackage{bbm}
\usepackage{makecell}
\usepackage{pbox}
\usepackage{multirow}%
\usepackage[title]{appendix}%
\usepackage{textcomp}%
\usepackage{manyfoot}%
\usepackage{booktabs}%
\usepackage{algpseudocode}%
\usepackage{listings}%

\usepackage{xcolor}
\usepackage{amsmath}
\usepackage[most]{tcolorbox}
\usepackage{lmodern}
\usepackage{dsfont}
\usepackage{tikz}
\usepackage{standalone}

\allowdisplaybreaks[4] 

\renewcommand{\d}{\mathrm{d}}
\newcommand{\im}{\mathrm{i}}

\renewcommand{\Tr}{\mathrm{Tr}}  
\def\lr#1{\left\langle {#1} \right\rangle}
\newcommand{\lrb}[1]{\left(#1\right)}
\newcommand{\lrc}[1]{\left\{#1 \right \}}
\newcommand{\lrs}[1]{\left[#1 \right]}
\newcommand{\lrnorm}[1]{\left\|#1 \right \|}

\newcommand{\CX}{{\mathcal{X}}}
\newcommand{\BR}{{\mathbb{R}}}
\newcommand{\CB}{{\mathcal{B}}}
\newcommand{\BC}{{\mathbb{C}}}
\newcommand{\CS}{{\mathcal{S}}}
\newcommand{\CU}{{\mathcal{U}}}
\newcommand{\CP}{{\mathcal{P}}}
\newcommand{\BI}{{\mathbb{I}}}

\numberwithin{equation}{section}

\newtheorem{theorem_}{Theorem}[section]

\newaliascnt{assumption_}{theorem_}
\newtheorem{assumption_}[assumption_]{Assumption}
\aliascntresetthe{assumption_}
\crefname{assumption_}{Assumption}{Assumptions}
\Crefname{assumption_}{Assumption}{Assumptions}

\newaliascnt{lemma_}{theorem_}
\newtheorem{lemma_}[lemma_]{Lemma}
\aliascntresetthe{lemma_}
\crefname{lemma_}{Lemma}{Lemmas}
\Crefname{lemma_}{Lemma}{Lemmas}

\newaliascnt{corollary_}{theorem_}
\newtheorem{corollary_}[corollary_]{Corollary}
\aliascntresetthe{corollary_}
\crefname{corollary_}{Corollary}{Corollaries}
\Crefname{corollary_}{Corollary}{Corollaries}

\newaliascnt{proposition_}{theorem_}
\newtheorem{proposition_}[proposition_]{Proposition}
\aliascntresetthe{proposition_}
\crefname{proposition_}{Proposition}{Propositions}
\Crefname{proposition_}{Proposition}{Propositions}

\newaliascnt{definition_}{theorem_}

\aliascntresetthe{definition_}
\crefname{definition_}{Definition}{Definitions}
\Crefname{definition_}{Definition}{Definitions}

\newaliascnt{remark_}{theorem_}
\newtheorem{remark_}[remark_]{Remark}
\aliascntresetthe{remark_}
\crefname{remark_}{Remark}{Remarks}
\Crefname{remark_}{Remark}{Remarks}

\newaliascnt{question_}{theorem_}

\aliascntresetthe{question_}
\crefname{question_}{Question}{Questions}
\Crefname{question_}{Question}{Questions}

\newaliascnt{fact}{theorem_}

\aliascntresetthe{fact}
\crefname{fact}{Fact}{Facts}
\Crefname{fact}{Fact}{Facts}

\makeatletter
\g@addto@macro\appendix{%
  \crefalias{section}{appendix}%
  \crefalias{subsection}{appendix}%
  \crefalias{subsubsection}{appendix}%
}
\makeatother

\AtBeginDocument{%
}
\renewcommand{\eqref}[1]{\cref{#1}}
\AtBeginDocument{\let\ref\cref}

\newcommand{\rev}[1]{{\color{black}{#1}}}
\graphicspath{{figs/}}

\newcommand{\lind}{{\mathcal{L}}}

\DeclareMathOperator{\SWAP}{{SWAP}}

\DeclareMathOperator{\Gap}{{Gap}}
\DeclareMathOperator{\Spec}{{Spec}}

\newcommand{\mc}[1]{\mathcal{#1}}
\newcommand{\supp}{\mathrm{supp}}
\newcommand{\Span}{\mathrm{span}}
\newcommand{\diag}{\mathrm{diag}}
\newcommand{\offdiag}{\mathrm{off}}

\newcommand{\erfc}{\mathrm{erfc}}

\DeclareMathOperator{\poly}{{poly}}

\newcommand{\rest}{\mathrm{rest}}

\newcommand{\DIAG}{{\mathsf{Diag}}}
\newcommand{\OFFDIAG}{{\mathsf{Off}}}

\begin{document}

\title{Quantum Replica Exchange}
\author[1] {Zherui Chen}
\author[1] {Joao Basso}
\author[2] {Zhiyan Ding}
\author[1,3,*] {Lin Lin}
\affil[1]{Department of Mathematics, University of California, Berkeley, CA USA}
\affil[2]{Department of Mathematics, University of Michigan, Ann Arbor, MI USA}
\affil[3]{Applied Mathematics and Computational Research Division, Lawrence Berkeley National Laboratory, Berkeley, CA USA}

\date{}
\maketitle

\renewcommand{\thefootnote}{*}
\footnotetext{\href{mailto:linlin@math.berkeley.edu}{linlin@math.berkeley.edu}}
\renewcommand{\thefootnote}{\arabic{footnote}}

\begin{abstract}
The presence of energy barriers in the state space of a physical system can lead to exponentially slow convergence for sampling algorithms like Markov chain Monte Carlo (MCMC). In the classical setting, replica exchange (or parallel tempering) is a powerful heuristic to accelerate mixing in these scenarios. In the quantum realm, preparing Gibbs states of Hamiltonians faces a similar challenge, where bottlenecks can dramatically increase the mixing time of quantum dynamical semigroups. In this work, we introduce a quantum analogue of the replica exchange method. We define a Lindbladian on a joint system of two replicas and prove that it can accelerate mixing for a class of Hamiltonians with local energy barriers. We provide a rigorous lower bound on the spectral gap of the combined system's Lindbladian, which leads to an exponential improvement in spectral gap with respect to the barrier height. We showcase the applicability of our method with several proof-of-concept examples that satisfy a commuting-cut assumption, including the defected 1D Ising model at arbitrary constant temperature, and defected non-commuting local Hamiltonians at high temperature. 
\end{abstract}

\vspace{0in}
\setlength{\columnsep}{25pt}
\begin{multicols}{2}
{\small \tableofcontents}
\end{multicols}
\thispagestyle{empty}

\section{Introduction} \label{sec:intro}

Markov chain Monte Carlo (MCMC) is a widely used sampling method with applications ranging from statistical physics and chemistry to probability and machine learning \cite{hastings1970monte,robert1999monte}. Although standard theory guarantees that a Markov chain with appropriately chosen transitions eventually converges to a prescribed target distribution, in practice one must control the rate of convergence. The \emph{mixing time} is defined as the time required for the distribution of the chain, started from the worst initial state, to approach the target distribution within a specified distance \cite{levin2017markov,montenegro2006mathematical}. Bounds on the mixing time are therefore crucial for assessing the efficiency and reliability of MCMC algorithms.

A typical cause of large mixing times is the presence of an \emph{energy barrier} in the target distribution \cite{schlichting2019poincare,montenegro2006mathematical}. Pictorially, if the distribution places significant mass in several regions separated by low-probability ``valleys'', then the Markov chain must cross these valleys (bottlenecks) in order to move between modes. When these bottlenecks are high, the time required to cross them is often exponential in the barrier height, leading to exponentially slow convergence. A prototypical example is the Ising model with a local energy barrier,
\begin{equation}\label{eqn:local_bottleneck}
H(\vec{z}) = -J z_1 z_2 + H_{\text{rest}}(\vec{z}),
\end{equation}
where $J \gg 1$ and the operator norm of $H_{\text{rest}}(\vec{z})$ does not scale with $J$. The large local coupling $J$ creates a strong bond between $z_1$ and $z_2$, so that transitions between configurations with $z_1 = z_2$ and those with $z_1 \neq z_2$ are exponentially suppressed in $J$. For standard local dynamics this produces an exponentially large mixing time in $J$ \cite{sokal1997monte,bovier2002metastability,bovier2016metastability}.

Although this Ising model is very simple, it already captures the mechanism of \emph{local} energy barriers that appears in many areas of physics, chemistry, and biology. In lattice spin systems describing magnetic materials, a strong local coupling models an impurity or pinning center that traps a domain wall and produces long-lived metastable magnetization patterns. In molecular systems, a large local interaction term is the discrete analogue of a stiff bond or a hydrogen-bond network; rearranging it corresponds to crossing an activation barrier for a chemical reaction or for a conformational change of a complex molecule. In biological systems, rare transitions among metastable states between macroscopically distinct configurations are often governed by local rearrangements, such as opening a loop in a polymer or breaking a key contact in a protein fold.  In each case, a small region of the system functions as a gate through which global reconfiguration must pass, and the resulting local barrier dictates the mixing time. The spin model with a local defect studied in this work should be viewed as a minimal example of this widely occurring phenomenon.

To overcome such bottlenecks, one often augments local dynamics with additional global moves that can bypass the barrier. Replica exchange, also known as parallel tempering, implements this idea by running MCMC on several replicas of the system at different temperatures and occasionally proposing swaps of their configurations. The high-temperature replica is less affected by the barrier and explores the state space more freely; by coupling the replicas through exchanges, the high-temperature chain provides a path ``around'' the energy barrier for the low-temperature chain (see \cref{fig:barrier_and_swap}).

Replica exchange was introduced for classical spin-glass simulations \cite{swendsen1986replica,hukushima1996exchange} and later extended to molecular dynamics \cite{sugita1999replica}. It has since become a widely used tool in statistical physics, chemistry, and biology. Its efficiency has been investigated rigorously for classical spin systems, where one can establish rapid mixing and spectral-gap bounds for representative models \cite{madras2003swapping,woodard2009conditions}. For continuous-variable targets such as mixtures of Gaussian components, spectral analyses of replica exchange Langevin dynamics also provide convergence guarantees \cite{woodard2009conditions,dong2022spectral}.
Other recent progress includes analyses of the infinite-swapping limit of the replica exchange process \cite{lu2013infinite,lu2019methodological}, and the application of replica exchange techniques to other areas such as nonconvex learning \cite{chen2020accelerating}.


In quantum algorithms, the task analogous to sampling from a classical distribution is to prepare a Gibbs state
\[
\rho = \frac{e^{-\beta H}}{\Tr[e^{-\beta H}]},
\]
given an inverse temperature $\beta$ and a Hamiltonian $H$. There has recently been a surge of work designing algorithms akin to MCMC to prepare such states \cite{davies1974markovian,terhal2000problem,temme2011quantum,yung2012quantum,moussa2019low,ChenKastoryanoBrandaoEtAl2023,shtanko2021preparing,chen2021fast,wocjan2023szegedy,rall2023thermal,chen2023efficient,zhang2023dissipative,jiang2024quantum,ding2025efficient,gilyen2024quantum,ding2024polynomial,bakshi2024high,ding2025end,chen2025efficient}. Several works have established rigorous bounds on quantum mixing times for dissipative dynamics and quantum Markov chains \cite{TemmeKastoryanoRuskaiEtAl2010,kastoryano2013quantum,BardetCapelGaoEtAl2023,rouze2025efficient,ding2024polynomial,kochanowski2024rapid,rouze2024optimal,tong2024fast,vsmid2025polynomial}. As in the classical case, however, energy barriers in the Hamiltonian generically lead to very large quantum mixing times, and rigorous lower bounds show that local bottlenecks can cause slow convergence even in small quantum systems \cite{gamarnik2024slow,rakovszky2024bottlenecks}. This raises a natural question: can a quantum version of replica exchange accelerate convergence in the presence of such barriers?

In this work, we introduce a natural quantum analogue of replica exchange that is designed to accelerate convergence when the underlying dynamics is governed by a slowly mixing generator. Concretely, we consider two replicas: one governed by a hard-to-mix generator $\lind_{\rm hard}$ and the other by an easy-to-mix generator $\lind_{\rm easy}$. We then couple them by a swap-type generator $\lind_{\rm SWAP}$ that exchanges the two replicas, and obtain a joint generator acting on the tensor product system,
\begin{equation}
\begin{split}
\lind
=
\lind_{\rm hard} \otimes \mathbb{I}
+
\mathbb{I} \otimes \lind_{\rm easy}
+
\lind_{\rm SWAP}.
\end{split}
\end{equation}
Under suitable assumptions, we show in \cref{thm:Gap_lower_bound} that this joint generator is rapidly mixing even when $\lind_{\rm hard}$ alone mixes slowly. The construction is the quantum analogue of coupling a low-temperature chain to a high-temperature one in classical parallel tempering, with $\lind_{\rm easy}$ playing the role of the fast, barrier-free dynamics.

\rev{We illustrate our theoretical guarantees with proof-of-concept models: the defected Ising model, and a noncommuting spin system with a commuting cut and a local strong-coupling bottleneck. Each satisfies the structural assumptions used in our analysis.} These examples serve both as paradigmatic instances of local energy barriers and as simplified models of the much more complex energy landscapes encountered in realistic physical, chemical, and biological systems.

\subsection{Scope and assumptions}

Our proofs address slow thermalization caused by a \emph{local} bottleneck, similar to that in \cref{eqn:local_bottleneck}. We focus on situations in which the slow behavior is localized to a small region $A$ of the system, while the complement $B$ can already be mixed efficiently by standard local updates. Our goal is to accelerate convergence by coupling the target system to an auxiliary system that mixes rapidly, and then occasionally swapping the degrees of freedom in $A$ between the two systems. In other words, the swap we use is \emph{local}: it exchanges only the registers in $A$ between the replicas and leaves the rest of the system $B$ untouched. The motivation for this choice is efficiency. Global swaps are always possible, but when the obstruction is localized they can be unnecessarily costly.

\rev{For the rigorous analysis in this work,} we analyze models that admit what we informally call a \emph{commuting cut}: operators that act across the boundary of $A$ and $B$ commute with all operators supported strictly inside $A$ or $B$. This structure allows for a transparent proof that the combined dynamics has a large spectral gap whenever $A$ is small and $B$ already mixes rapidly under local updates. The commuting cut should be viewed as a simplifying assumption, which we introduce to formalize the argument cleanly. The underlying mechanism relies only on the idea that the auxiliary system can carry information ``around'' the barrier in $A$ and return it to the target system after mixing.

\paragraph{Relation to structural results of quantum Gibbs states}

Recent structural results on the local Markov properties of quantum Gibbs states and on quantum metastability provide a complementary viewpoint on the dynamical behavior studied here \cite{chen2025quantumLocalMarkov,bergamaschi2025structural}. For local Hamiltonians with bounded interaction strength, the authors show that metastable states arise as configurations trapped in free-energy wells whose barriers are global in nature. A defining characteristic of these states is their robustness to arbitrary localized perturbations, a property formalized through an approximate local Markov condition.

This work, in contrast, investigates a complementary regime where the effective energy barriers are local, generated by defects such as terms with a large operator norm. As a result, the corresponding metastable state does not satisfy the approximate local Markov properties established in~\cite{bergamaschi2025structural}. Despite this distinction in the barrier's origin, the qualitative behavior is similar: the system becomes trapped in a local minimum of the energy landscape, causing slow mixing. To overcome these specific local barriers, we introduce a quantum replica exchange mechanism, which provides a nonlocal means to escape these metastable basins.

\subsection{Implications and generalizations} \label{sec:Implicationsgeneralizations}

The most direct implication of our results is a rigorous acceleration scheme for models with local energy barriers. In the examples we consider, the method substantially improves the dependence of the mixing time on the barrier height, typically converting an exponential dependence into a polynomial one. The required coupling is simple: a swap on $A$ decomposed into $\mathcal{O}(|A|)$ two-qubit gates, together with standard local dissipative updates on each replica. From the perspective of implementation in near-term or modular architectures, this is favorable because it involves only local control on $A$ and no long-range coherent operations across the full system.

Replica exchange is not the only viable strategy for incorporating nonlocal moves into the dynamics. One could attempt to design problem-specific global moves that jump directly across the barrier, or, when $|A|$ is small, augment the generator with \emph{all} possible nonlocal updates on $A$ (whose total number grows exponentially in $|A|$). From a purely Markov-chain perspective, such a brute-force scheme also removes the dependence of the mixing time on the barrier height, because every configuration of $A$ can be reached in a single step. The difference is quantitative and algorithmic: the number of distinct updates on $A$ is exponential in $|A|$, so implementing them requires an exponentially large family of jump operators, control pulses, or measurement-and-feedback rules. By contrast, quantum replica exchange achieves essentially the same barrier-insensitive scaling using a single family of local swap operations of size $\mathcal{O}(|A|)$, together with an easy-to-mix auxiliary dynamics. Thus the exponential improvement in the dependence on the barrier height that we prove is not merely a consequence of allowing larger-support updates, but of using the temperature-coupling mechanism to realize the effect of exponentially many collective moves through a small number of physically local operations.

From the standpoint of physical applications, this distinction mirrors long-standing practice in classical simulations. For instance, in practical molecular dynamics simulations, 
it is rarely feasible to implement an exhaustive set of nonlocal moves. Instead, practitioners rely on methods such as classical replica exchange to discover transition paths automatically: a high-temperature replica surmounts the local barrier by ordinary local moves, and occasional swaps transfer the resulting transitions to the low-temperature replica. Our quantum scheme plays the same role. It can be used as a drop-in acceleration primitive whenever a localized obstruction is suspected: one couples the slow region $A$ to an auxiliary replica that is easy to mix and performs local swaps, allowing the auxiliary to explore rare activated events and feed them back to the target system.  Moreover, the construction can be adapted to the case where the bottleneck location is not known in advance by running the same scheme on a family of candidate regions $A$ and letting the dynamics itself indicate which regions yield the largest acceleration.

\rev{The rigorous guarantee proved here applies to a structured local-barrier
regime, in which the cut is commuting and the complement mixes rapidly. It
should therefore not be read as a general acceleration result for systems
with global or hierarchically organized barriers. For example, the present
analysis does not cover the Sherrington--Kirkpatrick model or generic random
local Hamiltonians with frustrated interactions. Even in the classical
setting, rigorous results for replica exchange are available only for a
relatively limited class of structured models~\cite{madras2003swapping,woodard2009conditions,dong2022spectral}.
}

Although our proofs rely on the presence of a commuting cut for simplicity, the underlying acceleration mechanism is more general. In essence, it requires only a fast component that can efficiently refresh $B$ and an operation that swaps the state of $A$ between replicas. \rev{This mechanism may extend to some noncommuting Hamiltonians without an exact commuting cut, but establishing such an extension is beyond our present analysis.} Physically, the slowness of $\lind_{\rm hard}$ can arise from local energy barriers (as in our defect models), from global energy barriers (as in symmetry-breaking or topological transitions), or from low-temperature regimes $\beta \gg 1$ where thermal fluctuations are too weak to induce efficient transitions. \rev{For more global obstructions, variants based on global swaps between full replicas at different temperatures may also be worth investigating.} A systematic exploration of alternative quantum replica exchange protocols, including multiple replicas arranged along a temperature ladder or other control parameters, is an interesting direction for future work.

\subsection{Organization}
We start with background on bottlenecks (\cref{sec:bottlenecks}), the classical replica exchange technique (\cref{sec:classical_replica_exchange}) and the Lindbladian construction from~\cite{chen2023efficient} for quantum Gibbs sampling (\cref{sec:quantum_gibbs_recall}). \cref{sec:main_theorem} introduces the quantum replica exchange framework, while \cref{sec:tech_main_result} contains the main \rev{theorem} (\cref{thm:Gap_lower_bound}) with the acceleration guarantee. Finally, example applications of the main \rev{theorem} are worked out in \cref{sec:applications}.
\rev{For convenience, a consolidated \hyperref[app:notation]{summary of the notation used throughout the paper} is provided at the beginning of the appendices.}

\section{Preliminaries}

\subsection{Bottlenecks}\label{sec:bottlenecks}
\begin{figure}[h!]
\centering
 \centering
  \includegraphics[width=0.5\linewidth]{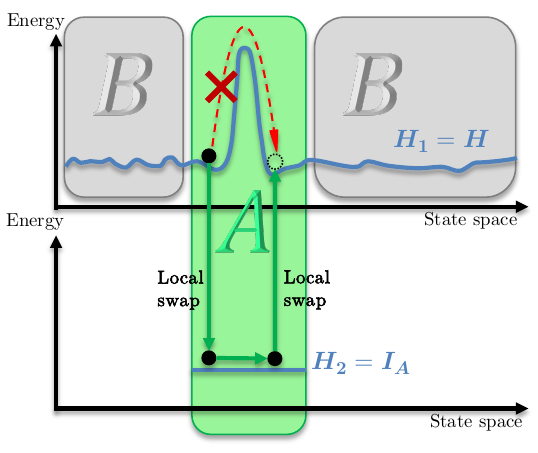}
  \caption{Illustration of a local energy barrier with replica exchange. Region $A$ of the state space has a local energy barrier which slows down mixing in $H$. However, by coupling to $I_A$, which has a flat energy landscape, the barrier can be traversed by going to the second copy, going across $A$ there, and then returning to the first copy (path in green). More generally, $H_2$ may be replaced by any fast-mixing Hamiltonian that assists in overcoming barriers in $H_1$, as discussed in \cref{sec:Implicationsgeneralizations}.} 
  \label{fig:barrier_and_swap}
\end{figure}

The slowness in the mixing of a Markov chain can usually be attributed to some form of a bottleneck. Intuitively, a Markov chain has a bottleneck if some regions of the state space are hard to reach from others, which leads to slow convergence depending on the starting point. See \cref{fig:barrier_and_swap} for an illustration.

In classical Markov chains, this is quantified by the {\it bottleneck ratio} $\Phi_*$ (for a pedagogical introduction, see \cite[Chapter 7]{levin2017markov}). Let $\pi$ be the stationary distribution of a Markov chain and $W$ be the transition matrix. The probability flow from a set of states $S$ to its complement $S^c$ under stationarity is given by $Q(S,S^c) = \sum_{x\in S, y \in S^c} \pi(x) W_{x,y}$. The Cheeger constant is then defined as the minimum flow across any bipartition of the state space, normalized by the size of the smaller part:
\begin{align}
\Phi_* := \min_{S \colon \pi(S) \le 1/2} \frac{Q(S,S^c)}{\pi(S)}.
\label{eq:cheeger_constant}
\end{align}
Intuitively, $\Phi_*$ quantifies the tightest bottleneck in the state space. A small value implies the existence of a partition with low probability flow between its components, which corresponds to slow mixing.

The work \cite{gamarnik2024slow} initiated the study of bottlenecks in quantum Markov chains. They roughly show that if $\Pi_A, \Pi_C$ are orthogonal projectors on a Hilbert space of dimension $2^n$ {that are not connected by a single jump}, $\sigma_H$ is the Gibbs state, and also
\begin{align}
\Tr[\Pi_A \sigma_H], \Tr[\Pi_C \sigma_H] = \Omega(1) \qquad \text{ while } \qquad \Tr[({I} - \Pi_A - \Pi_C) \sigma_H] = \exp[-\Omega(\poly(n))],\label{eqn:rare_region}
\end{align}
then the mixing time is exponential in $n$. That is, $\Pi_B = {I} - \Pi_A - \Pi_C$ constitutes a bottleneck for crossing between $\Pi_A$ and $\Pi_C$. For further details, see \cref{sec:slow_defect}.

Some models may have a {\it local} bottleneck, whereby certain local moves would cause a large energy increase, thus being rejected with large probability. If those moves are the only way to reach a certain region of the state space, they witness a bottleneck.  Returning to the example in \cref{sec:intro}, consider the 1D Ising model on a ring, but with a defect whereby an edge has a large weight $J \gg 1$:
\begin{align}\label{eq:classical_defected_ising}
    H(\vec{z}) = - J z_1 z_2 - \sum_{i=2}^N z_i z_{i+1},
\end{align}
where $\vec{z} \in \{\pm 1\}^N$ and $z_{N+1} \equiv z_1$. For a classical chain running Glauber dynamics on $H$ at inverse temperature $\beta$, if we are at the state $\vec{z}$ with $z_1 = z_2=1$, to reach $z_1 = z_2=-1$ we must flip $z_1$ or $z_2$, incurring a large energy penalty of  $2J$, which is only accepted with probability $\mathcal{O}(\exp(-2\beta J))$. In fact, one can check that $\Phi_* = \mathcal{O}(\exp(-2\beta J))$.  Thus, crossing between $\{\vec{z} \in \{\pm 1\}^N \colon z_1 = z_2 = 1\}$ and $\{\vec{z} \in \{\pm 1\}^N \colon z_1 = z_2 = -1\}$ constitutes a bottleneck.

In this work, we focus on overcoming {\it local} bottlenecks, such as those coming from defects much like in the example above. For acceleration results for this and other models, see \cref{sec:applications}.

\subsection{Classical replica exchange}\label{sec:classical_replica_exchange}
To address the sampling bottlenecks described in  \cref{sec:bottlenecks}, which are typically caused by high energy barriers separating different regions of the state space, the classical replica exchange method introduces a clever solution. 

Let $\beta_{1}$ be an inverse temperature and $\mathcal{X}$ be a state space. We consider a {continuous-time Markov chain} $(X_t)_{t \geq 0}$ on $\mathcal{X}$ whose dynamics we want to accelerate. The evolution of this chain is fully described by its {generator}, $L_{\beta_1}$.
For any function $f : \mathcal{X} \to \mathbb{R}$, its expected value evolves according to the {master equation}:
\begin{align}
\frac{\d}{\d t}\mathbb{E}[f(X_t)] = \mathbb{E}\big[(L_{\beta_1} f)(X_t)\big].
\end{align}
This chain is designed to relax towards a unique stationary (or equilibrium) distribution 
$
\pi_{\beta_1}(x) \propto \exp\big(-\beta_1 H(x)\big)
$
for some potential energy function $H : \mathcal{X} \to \mathbb{R}$.

When  the energy landscape defined by $H(x)$ is rugged, sampling from $\pi_{\beta_1}$ can be extremely slow due to high energy barriers trapping the system in local minima. The classical replica exchange method (also known as parallel tempering) accelerates sampling by running simulations at different temperatures in parallel and allowing them to exchange configurations. Specifically, we introduce an auxiliary replica at a higher temperature, 
i.e., $\beta_2 < \beta_1$, and define a new Markov chain $L_{\beta_1, \beta_2}$ on the larger space $\CX \times \CX$. Given $f \colon \CX \times \CX \to \BR$ and $x_1,x_2 \in \CX$:
\begin{align}\label{eq:classical_generator_with_swap}
(L_{\beta_1,\beta_2} f)(x_1, x_2)
\coloneqq (L_{\beta_1} f(\cdot, x_2))(x_1) + (L_{\beta_2} f(x_1, \cdot))(x_2) + L_{\mathrm{SWAP},\beta_1,\beta_2}(f)(x_1, x_2).
\end{align}
The first two terms represent independent runs of the Markov chain on the two copies. The third term is the swap term defined as
\begin{align}\label{eq:classical_swap_generator}
(L_{\SWAP, \beta_1, \beta_2} f)(x_1, x_2) 
\coloneqq  
\min\left\{ 1, \exp\left[ (\beta_1-\beta_2) (H(x_1) - H(x_2)) \right] \right\} \left( f(x_2,x_1) - f(x_1, x_2) \right)\,,
\end{align}
{where the prefactor is the Metropolis–Hastings acceptance probability, obtained from the ratio of the joint Boltzmann weights of the post-swap and pre-swap states}
\begin{align}
\frac{\pi_{\beta_1}(x_2) \pi_{\beta_2}(x_1)}{\pi_{\beta_1}(x_1) \pi_{\beta_2}(x_2)} &= \frac{e^{-\beta_1 H(x_2)} e^{-\beta_2 H(x_1)}}{e^{-\beta_1 H(x_1)} e^{-\beta_2 H(x_2)}}\,;
\end{align}
see for instance \cite{chen2020accelerating,dong2022spectral,leng2025operator}. 

When sampling from non-logconcave distributions, low-temperature dynamics can get trapped in metastable configurations for an exponentially long time due to weak thermal fluctuations. A swapping process resolves this by transferring well-explored configurations from a parallel, high-temperature system to the low-temperature one, since the former is better able to surpass energy barriers. This dramatically reduces the overall mixing time. Specifically, 
the ability of replica exchange to accelerate over-damped Langevin diffusion is extensively supported by numerical \rev{experiments}~\cite{sindhikara2010exchange,lu2013infinite,lu2019methodological,chen2020accelerating} and, to a lesser extent, by theoretical proofs, such as those on mixtures of Gaussians \cite{dong2022spectral}.

\subsection{Quantum Gibbs sampling}\label{sec:quantum_gibbs_recall}
While classical replica exchange speeds up sampling by running classical Gibbs samplers at different temperatures, its quantum analogue will aid in efficiently preparing quantum Gibbs states. In this part, we recall a quantum primitive for preparing the Gibbs state $ \sigma_H  = e^{-\beta H}/\Tr\left(e^{-\beta H}\right)$ for a given quantum Hamiltonian $H$ at inverse temperature $\beta$. Our approach is to simulate the dynamics of a quantum semigroup generated by a Lindbladian $\lind$. We adopt the construction of \cite{chen2023efficient}, which was the first to provide an efficiently simulatable Lindbladian satisfying exact detailed balance. More recently, \cite{ding2025efficient} introduced a quantum Gibbs sampler with greater design flexibility and a simpler implementation, which subsumes the framework of \cite{chen2023efficient}. While the method of \cite{ding2025efficient} could also be integrated into our replica exchange framework, we choose to rely on \cite{chen2023efficient}, since our later application of quantum replica exchange in \cref{sec:applications} builds upon \cite{rouze2025efficient}, which is based on \cite{chen2023efficient}.

Given an $n$-qubit Hamiltonian  $H$, inverse temperature $\beta$, a set of coupling operator $\CS = \{S^a\}$, and a weight function $\gamma \colon \mathbb{R} \to \mathbb{R}$, we define the Lindbladian in the Heisenberg picture as:
\begin{align}
    \lind_{(H, \beta, \CS, \gamma)}\lrb{X}  = \im \lrs{G_{\CS},X} +\sum_{S^a \in \CS} \int_{-\infty}^\infty \d\omega~ \gamma(\omega) \left( \hat{S}^{a\dagger}(\omega) X \hat{S}^a(\omega) -\frac{1}{2} \acomm{\hat{S}^{a\dagger}(\omega) \hat{S}^a(\omega)}{X} \right)\,.
    \label{eq:KMS_def}
\end{align}
When clear from context, we might drop some of the subscripts in $\lind_{(H, \beta, \CS, \gamma)}$. The evolution of an observable $X$ under the Heisenberg picture is given by the Lindblad equation $\frac{\d X}{\d t} =  \lind \lrb{X}$. Equivalently, the evolution of a density matrix $\rho$ under the Schr\"odinger picture is given by the dual Lindblad equation $\frac{\d \rho}{\d t} = \lind^\dagger \lrb{\rho}$, where $\lind^\dagger$ is the adjoint of $\lind$ with respect to the Hilbert-Schmidt inner product. 
The Lindblad operators $\hat{S}^a (\omega)$ are spectrally-filtered versions of the coupling operators $S^a$:
\begin{align}
    \hat{S}^a(\omega) = \frac{1}{\sqrt{2\pi}} \int_{-\infty}^\infty \d t~f(t) e^{\im Ht} S^a e^{-\im Ht} e^{-\im \omega t},
    \label{eq:S_hat_omega}
\end{align}
where $f$ is chosen so that its Fourier transform $\hat{f}$ is Gaussian:
\begin{align}
    \hat{f}(\omega) = \sqrt{\frac{\beta}{\sqrt{2\pi}}} \exp\left(-\frac{\beta^2 \omega^2}{4}\right).
    \label{eq:gaussian_filter_CKG}
    \end{align}
The coherent term takes the form
\begin{align}
    G_\CS = \sum_{S^a \in \CS} \, \sum_{\nu_1,\nu_2\in B(H)} \frac{\tanh(-\beta(\nu_1 - \nu_2)/4)}{2\im } \alpha _{\nu_1,\nu_2} \lrb{S^{a}_{\nu_2}}^{\dagger} S^a_{\nu_1},
    \label{eq:G_def_CKG}
\end{align}
where
\begin{align}\label{eq:alpha_omega1_omega2_CKG}
\alpha_{\nu_1,\nu_2} =\int_{-\infty}^\infty \d \omega \gamma(\omega) \hat{f}(\omega - \nu_1)\overline{\hat{f}(\omega - \nu_2)},
\end{align}
the set of Bohr frequencies is
\begin{align}
    B(H) \coloneqq \left\{ \nu =\lambda_i -\lambda_j \colon \lambda_i,\lambda_j \in \mathrm{Spec}(H) \right\}\,,
\end{align}
and, if we have the eigendecomposition $H = \sum_i \lambda_i \ketbra{i}{i}$, then
\begin{align}
    S^a_\nu \coloneqq \sum_{i,j \text{ s.t. } \lambda_i - \lambda_j = \nu} \ketbra{i}{i} S^a \ketbra{j}{j}.
    \label{eq:S_nu}
\end{align}

Possible choices of weight functions $\gamma(\omega)$ include the Gaussian- or the Metropolis-type
(\cite[Eqs. (1.5), (1.8)]{chen2023efficient}):
\begin{align}\label{eq:metropolis_weight_gamma}
\gamma_G(\omega) = \exp\left[ -\frac{\left( \beta\omega+1 \right)^2}{2} \right], \qquad \gamma_M(\omega) = \exp\left[ -\beta \max\left\{\omega +\frac{1}{2\beta}, 0\right\} \right].
\end{align}
When the range of energy transitions $|\omega|$ is small, the Gaussian weight $\gamma_G$ and the Metropolis weight $\gamma_M$ behave similarly. The distinction becomes pronounced for large $|\omega|$: the Gaussian filter $\gamma_G$ decays exponentially in both directions, suppressing positive and negative $\omega$ alike, while the Metropolis rule accepts all negative $\omega$ and only exponentially suppresses positive $\omega$. The asymmetric behavior of the Metropolis filter is crucial for enabling the Lindbladian to efficiently overcome {energy barriers}  (\cref{eq:gen_L_def} and \cref{sec:fast_mixing_local_39})~{by facilitating transitions between eigenvectors with large energy differences \cite[Page 4]{chen2023efficient}.}

Let the Kubo--Martin--Schwinger (KMS) inner product be defined as 
\[
\lr{ X, Y }_{ \sigma_H } = \Tr\lrs{ { \sigma_H }^{1/2} X^\dag { \sigma_H }^{1/2} Y }\,.
\]
We say that $\lind$ satisfies quantum detailed balance if it is self-adjoint in the KMS inner product, i.e., $\langle X, \lind( Y) \rangle_{\sigma_H} = \langle \lind (X), Y \rangle_{\sigma_H}$.
The quantum detailed balance condition furnishes a set of criteria that characterize the fixed points of Lindbladian. 
As illustrated by \cite[Page 11]{rouze2025efficient}, when the coupling operator set $\mc{S} = \mc{P}_{[n]}$, $ \sigma_H $ is the unique fixed point of the evolution generated by $\mathcal{L}^\dagger$. 
For a precision $\epsilon>0$, the mixing time of a Lindbladian $\mathcal{L}^\dagger $ with unique fixed point $ \sigma_H $  can be defined as
\begin{equation}\label{eq:tmix}
t_{\mathrm{mix}}(\epsilon)
:=\inf\Big\{t\ge 0 \ \Big|\ \lrnorm{e^{t\mathcal{L}^\dagger}(\rho)- \sigma_H  }_{\Tr} \le \epsilon,\ \forall\ \text{quantum state }\rho\Big\},
\end{equation}
where $\|O\|_\Tr=\Tr\lrs{\sqrt{O^{\dag} O}}$ denotes the trace norm.
Specifically, for a primitive Lindbladian that satisfies quantum detailed balance with respect to the full-rank fixed point $\sigma_H$, the following two-sided bound holds (see \cref{lem:gap-mixing} for a proof):
\begin{equation}
\frac{1}{g}\,\log\!\left(\frac{\lambda_{\min}(\sigma_H)}{2\,\epsilon}\right)
\; \leq \;
t_{\mathrm{mix}}(\epsilon) \;\le\; \frac{1}{2g}\,\log\!\left(\frac{1}{\lambda_{\min}(\sigma_H)\,\epsilon^2}\right),
\label{eq:mixingTime_pre}
\end{equation}
where $\lambda_{\min} \lrb{ \sigma_H }$ is the smallest eigenvalue of $ \sigma_H $ and $g$ is the spectral gap of $\mathcal{L}$, defined following 
\cite[Definition 10]{kastoryano2013quantum} as
\begin{align}
g = {\Gap}(\mathcal{L})  \coloneqq  \inf_{X \neq 0\,, \lr{I,X}_{ \sigma_H } = 0} \frac{\lr{ X, -\mathcal{L} (X) }_{ \sigma_H }}{\lr{ X, X}_{ \sigma_H }}\,.
\label{eq:gap_def}
\end{align}
The associated operator norm of $\mathcal{L}$ is given by 
\begin{align}
    \lrnorm{\mathcal{L}}_{ \sigma_H } = \sup_{
    X\neq 0}
    \frac{\lr{X, - \mathcal{L}(X)}_{ \sigma_H }}
    { \lr{X, X}_{ \sigma_H }} \,.
    \label{eqn:gap_up_bound_local_Adiag_swap}
\end{align}

\section{Quantum replica exchange} \label{sec:main_theorem}
We now introduce the direct quantum analogue of the replica exchange in \cref{eq:classical_generator_with_swap,eq:classical_swap_generator}. Let 
$$\lind_{\lrb{H_1, \beta_1, \CS_1, \gamma_1}} \colon \CB \lrb{\BC^{2^n}} \to \CB \lrb{\BC^{2^n}}$$ 
be the Lindbladian on the first system of $n$ qubits. To accelerate its mixing, we introduce the second system of $n$ qubits where we operate with another Lindbladian 
$$\lind_{\lrb{H_2, \beta_2, \CS_2, \gamma_2}} \colon \CB\lrb{\BC^{2^n}} \to \CB\lrb{\BC^{2^n}}.$$ 
{Building upon this setup, we propose the replica exchange Lindbladian on the joint system, $\lind \colon \CB\lrb{\BC^{2^{2n}}}\to \CB\lrb{\BC^{2^{2n}}}$,  in the following general form:}
\begin{align}
\lind = \lind_{(H_1, \beta_1, \CS_1, \gamma_1)}  \otimes \mathbb{I}_n  + \mathbb{I}_n \otimes \lind_{(H_2, \beta_2, \CS_2, \gamma_2)} + \lind_{\lrb{\lrc{\SWAP}}}\,.
\label{eq:gen_L_def}
\end{align}
The term 
 $\lind_{\lrb{\lrc{\SWAP}}}$ couples the first and second systems together and is defined using the compact notation:
\begin{align}
\lind_{\lrb{\lrc{\SWAP}}} = \lind_{\lrb{\beta_1 H_1 \otimes {I}_n + \beta_2 {I}_n \otimes H_2,\, \beta_{\SWAP} = 1,\,\{\SWAP\},\,\gamma_M}}\,\rev{.}
\end{align}
Here, \(\SWAP\) can be  a swap-like unitary between the two replicas, 
which may be taken to be either global or local. The global swap is defined by its action on states
\begin{equation}\label{eq:global_swap}
  \SWAP\!\left(\ket{\psi}_1\otimes\ket{\phi}_2\right)
  = \ket{\phi}_1\otimes\ket{\psi}_2,
  \qquad \forall\,\ket{\psi},\ket{\phi}\in\mathbb{C}^{2^n},
\end{equation}
while the local variant \(\mathcal{U}\) is specified later in \cref{eqn:swap_operator_local}.
Since $\lind_{\lrb{\lrc{\SWAP}}}^\dagger$  is constructed by \cref{eq:KMS_def}, it is also a detailed-balanced Lindbladian that preserves the full system Gibbs state $\sigma_1 \otimes \sigma_2$, where $\sigma_i =e^{-\beta_i H_i} \big/ \Tr\lrs{e^{-\beta_i H_i} }$ is the fixed-point of $\lind_{\lrb{H_i,\beta_i,\CS_i,\gamma_i}}^\dagger $, $i=1,2$. Therefore, the fixed point of $\mathcal{L}^\dagger$ is still $\sigma_1\otimes \sigma_2$. 

We note that although an additional swap operator is introduced into the Lindbladian dynamics, the overall simulation cost remains comparable to that of the original dynamics, which can be efficiently simulated~\cite{ChenKastoryanoBrandaoEtAl2023,chen2023efficient,ding2025efficient,PRXQuantum.5.020332,cleve2016efficient,LW22}. Specifically, let the set of coupling operators be {$\{D_a\}=\mathcal{S}_1 \cup \mathcal{S}_2 \cup \{\mathrm{SWAP}\}$}. Since $\|\mathrm{SWAP}\|=1$ and $\sum_{D\in \mathcal{S}_1\cup \mathcal{S}_2}\|D\|=\mathcal{O}(n)$, by~\cite[Theorem I.2]{chen2023efficient}, the dynamics generated by~\cref{eq:gen_L_def} can be simulated up to time $T$ with the following cost: $\widetilde{\mathcal{O}}(\beta nT)$ total Hamiltonian simulation $\exp(-i(H_1\otimes I_n+I_n\otimes H_2)t)$, $\widetilde{\mathcal{O}}(nT)$ queries to the controlled block encodings of the coupling operators $\sum_{a}\ket{a}\otimes D_a$, and $\widetilde{\mathcal{O}}(nT)$ additional two-qubit gates.

In the conventional setting of classical replica exchange, setting $H_1 = H_2 = H$ makes our construction in \cref{eq:gen_L_def} a direct quantum analogue of the method described in \cref{eq:classical_generator_with_swap,eq:classical_swap_generator}, where one accelerates the mixing of the first replica by coupling it to a second copy at a higher temperature (i.e., $\beta_2 < \beta_1$). {More generally, we may couple the system of interest $H_1$ to an auxiliary system with a different Hamiltonian $H_2\neq H_1$, and $H_1,H_2$ need not be defined on the same Hilbert space.} There is additional flexibility in choosing the SWAP operator. In \cref{sec:fast_mixing_local_39} and the applications in \cref{sec:applications}, for instance, we choose a swap that acts locally. 
{
Specifically,
fix a bipartition $A\sqcup B =[n]$.
We denote vectors on $A$ and $B$ by $\ket{\psi_A}\in \mathbb{C}^{d_A}$ and  $\ket{\xi_B}\in \mathbb{C}^{d_B}$, respectively.
Assume first that $H_1$ and $H_2$ act on isomorphic spaces $\mathbb{C}^{d_A}\otimes \mathbb{C}^{d_B}$
(i.e., the two replicas each have $A$ and $B$ registers of the same size).
We use the \emph{local} swap on $A$ only, defined as
$
\SWAP_{\mathrm{local}} \bigl(\ket{\psi_A}_1\ket{\xi_B}_1\otimes\ket{\phi_A}_2\ket{\eta_B}_2\bigr)
= \ket{\phi_A}_1\ket{\xi_B}_1\otimes\ket{\psi_A}_2\ket{\eta_B}_2.
$
With this choice, \cref{eq:gen_L_def} is a direct “partial-swap’’  coupling on $A$ between two full replicas.
The above $\SWAP_{\mathrm{local}}$ also motivates removing the redundant $B$ register of $H_2$.
In \cref{sec:fast_mixing_local_39} we specialize to a smaller auxiliary system that carries only the $A$ register:
Let $H_2$ act only on $\mathbb{C}^{d_A}$ (there is no $B$ register on $H_2$ side).
The coupling then swaps only the two $A$ registers and leaves $B$ of $H_1$ untouched:
\begin{align}
\mc{U}\bigl(\ket{\psi_A}_1\ket{\xi_B}_1\otimes\ket{\phi_A}_2\bigr)
= \ket{\phi_A}_1\ket{\xi_B}_1 \otimes\ket{\psi_A}_2,
\label{eqn:swap_operator_local_11}
\end{align}
which is exactly the operator $\CU$ defined in \cref{eqn:swap_operator_local}.
With $H_1=H$, $\beta_1=\beta$, $\CS_1=\CP_{[n]}$ and $H_2=I_A$, $\beta_2=\beta$, $\CS_2=\CP_A$,
substituting these choices into \cref{eq:gen_L_def} gives \cref{eq:total_lind_p2} \footnote{Note that we also need to generalize $\lind_{\lrb{H_2, \beta_2, \CS_2, \gamma_2}} \colon \CB\lrb{\BC^{d_A}} \to \CB\lrb{\BC^{d_A}}$ as well. }.
}

\section{From bottlenecks to acceleration}\label{sec:tech_main_result}

We first show how a localized obstruction, such as a strong ferromagnetic bond $-J\sigma_i^z\sigma_j^z$, creates an energy barrier that results in an exponentially small spectral gap for dynamics driven by single-site coupling operators (\cref{sec:slow_defect_main}). We emphasize that the focus on a single strong ferromagnetic bond is merely a convenient example to illustrate the concept. In reality, many local structures can create similar slow-mixing bottlenecks; we use this specific construction because it simplifies the proof of the core idea. We then introduce a quantum replica exchange protocol that uses a local swap acting only on the slow-mixing subsystem. Unlike a global swap, whose performance degrades with system size, this targeted approach effectively removes the local barrier and reduces the mixing time (\cref{sec:fast_mixing_local_39}).

\subsection{Slow mixing induced by a single strong ferromagnetic bond} \label{sec:slow_defect_main}
This section shows that a single strong ferromagnetic bond $-J \sigma_i^z \sigma_j^z$ with $J \gg 1$ can induce mixing that is exponentially slow in $J$. The bond creates a $2J$ energy barrier between aligned and anti-aligned spin configurations on $(i,j)$, producing a bottleneck in state space. To ensure this bottleneck is well defined, we require the remainder of the Hamiltonian to commute with $\sigma_i^z$ and $\sigma_j^z$, so that the dynamics preserve the decomposition into four eigenspaces of $-J\sigma_i^z\sigma_j^z$ with eigenvalues $-J, -J, +J, +J$. We also assume that the jump operator set $\mc{S}$ used in constructing $\lind_{(H,\beta,\mc{S},\gamma)}$ of \cref{eq:KMS_def}  consists only of single-site flips, making it impossible to surmount the $2J$ barrier in a single jump. 

The above considerations yield the following proposition, whose full proof is given in \cref{sec:slow_defect}.

\begin{proposition_}[Slow mixing Hamiltonians, informal] 
\label{prop:slow_mixing_prop_informal}
    Consider the Hamiltonian $H = - J\sigma^z_i\sigma^z_j + H_\rest$. Assume that $[H_\rest, \sigma^z_i] = [H_\rest, \sigma^z_j] = 0$, and $\lrnorm{H_\rest}\leq C_0$ for a constant $C_0$ independent of $J$. Then, for a sufficiently large coupling $J$ at a fixed inverse temperature $\beta$, the mixing time of $ \lind_{(H, \beta,\mathcal{S}, \gamma)}$ satisfies: 
    \begin{align}
    t_{\mathrm{mix}} \lrb{\lind_{\lrb{H, \beta, \mathcal{S}, \gamma}}}= \exp\!\left( \Omega(J) \right),
\end{align} 
    where the coupling operators in $\mathcal{S}$ are single-site.
\end{proposition_}
From the above proposition, we can see that a single strong ferromagnetic bond can create an exponentially difficult bottleneck for Hamiltonians with a commuting cut, which leads to an exponentially small spectral gap with respect to $J$ by \cref{eq:mixingTime_pre}. More generally, the same effect can arise from a variety of local features that impede single-site dynamics, not only the ferromagnetic bond.

\subsection{Accelerated mixing via quantum replica exchange}\label{sec:fast_mixing_local_39}
While \rev{the global swap operation (\cref{eq:global_swap}) may accelerate the mixing of the entire system $H_1$}, it suffers from the curse of dimensionality. Namely, as the dimension of the state space increases,  the acceptance probability for such full-state swaps decays with system size unless the temperatures are very close, which significantly increases the mixing time \cite{lee2023improved,woodard2009conditions}. 

This inefficiency is particularly pronounced in scenarios where slow mixing is localized. For instance, high energy barriers might be confined to a specific subsystem $A$ of $H_1$, while the rest of the system mixes rapidly. In such cases, performing a global swap is inefficient. A more targeted approach is a local swap between subsystem $A$ and an auxiliary system $H_2$ with the same dimension as subsystem $A$. This can be sufficient to overcome the local energy barriers in subsystem $A$ of $H_1$ if $H_2$ is inherently easier to mix, thus yielding a shorter mixing time than a generic global swap would.
See \cref{fig:barrier_and_swap} for an illustration of the above protocol.

To formalize the above intuition, we assume that $H_1$ can be decomposed according to the following assumption. Note that, in the remainder of this section, we will denote the Hamiltonian of the first system as $H$, i.e., $H_1=H$. Our goal is to accelerate the mixing to its Gibbs state $e^{-\beta H} / \Tr\lrs{e^{-\beta H}}$.

\begin{assumption_}[Hamiltonian decomposition]\label{assumption:H_decomposition}
Given a Hamiltonian $H$ on $n$ qubits, suppose there are two sets $A, B \subseteq [n]$ with  $B=[n]\setminus A$, and Hamiltonians $H_A, H_B, H_{AB}$ which satisfy:

\begin{enumerate}
    \item $\supp(H_A) \subseteq A$ and $\supp(H_B) \subseteq B$.

    \item \label{item:H_AB_bounded} The interaction term $H_{AB}$  is given by  
    $$H_{AB} = \sum_{k=1}^{K} V_A^{(k)} \otimes V_B^{(k)}\,,$$ 
    where, for all $k\in [K]$,  $V_A^{(k)}$ and $V_B^{(k)}$ are Hermitian operators with  $\supp\lrb{V_A^{(k)}} \subseteq A$ and $\supp\lrb{V_B^{(k)}} \subseteq B$. 
    {Moreover, 
    the operator norm of each individual interaction term is upper bounded by $V_{\max} \coloneqq  \max_{k\in [K]}\lrc{\lrnorm{V_A^{(k)} \otimes V_B^{(k)} }} $.}
    
    \item \label{item:mutual_commute} The operators within each partition $A$ and $B$ mutually commute, i.e.,
    \begin{itemize}
        \item The set of operators $\{H_A\} \cup \lrc{V_A^{(k)}}_k$ mutually commute.
        \item The set of operators $\{H_B\} \cup \lrc{V_B^{(k)}}_k$ mutually commute.
    \end{itemize}

    \item The total Hamiltonian $H$ can be decomposed as:
    \begin{align}
        H = H_A \otimes I_B + I_A \otimes H_B + H_{AB}.
        \label{eq:H_decomposition_2547}
    \end{align}
\end{enumerate}
\end{assumption_}
{
A direct consequence of \cref{item:H_AB_bounded} in \cref{assumption:H_decomposition} is the bound
\begin{align}
\lrnorm{H_{AB}} \leq K V_{\max}.
\label{eq:H_AB_bounded}
\end{align}
Heuristically, a large $V_{\max}$ could create a local energy barrier due to the interaction term $H_{AB}$, which may slow down mixing; by contrast, taking $V_{\max}=\mathcal{O}(1)$ guarantees no pronounced $H_{AB}$-induced bottleneck across the cut, consistent with the picture that the principal barrier is localized within $A$. All applications we discuss in \cref{sec:applications} fall into this latter category. We emphasize, however, that a large $\lrnorm{H_{AB}}$ by itself does {not} guarantee an energy barrier between $A$ and $B$. The bounded-$\lrnorm{H_{AB}}$ regime should therefore be viewed as a {sufficient regularity assumption} under which our later analysis of \cref{thm:Gap_lower_bound} yields clean, quantitative bounds. }

The mutual commutativity condition (\cref{item:mutual_commute}  in \cref{assumption:H_decomposition}) guarantees the existence of a common eigenbasis for the operators within each partition. We define these as follows:
\begin{align}
    \lrc{\ket{i_A}}_{i_A} \quad & \text{is an orthonormal eigenbasis that simultaneously diagonalizes } \{H_A\} \cup \lrc{V_A^{(k)}}_k.
    \label{eq:basis_A}
    \\
    \lrc{\ket{j_B}}_{j_B} \quad & \text{is an orthonormal eigenbasis that simultaneously diagonalizes } \{H_B\} \cup \lrc{V_B^{(k)}}_k.
    \label{eq:basis_B}
\end{align}
The product states $\lrc{\ket{i_A j_B}}_{i_A,j_B}$ thus form a common eigenbasis for the  Hamiltonians $H_{AB}$ and $H$. 

\paragraph{\rev{Role of the commuting cut.}}
\rev{As discussed in the above paragraph, \cref{item:mutual_commute} is a commuting cut between the two subsystems $A$ and $B$, which
provides a common product eigenbasis $\lrc{\ket{i_Aj_B}}$. In this basis, the
spectrally filtered $B$-supported jumps preserve the $i_A$-label and reduce
within each $A$-sector to the conditional dynamics generated by
$\bra{i_A}H\ket{i_A}$. At the level of the bare jump operator $\mathcal{U}$ as defined in \cref{eqn:swap_operator_local_11}, the local swap
acts only on subsystem $A$ and leaves subsystem $B$ unchanged, but the swap Lindbladian is constructed from operators containing}
\[
\rev{
e^{\mathrm{i}tH_{\mathrm{tot}}}\mathcal U
e^{-\mathrm{i}tH_{\mathrm{tot}}},
\qquad
H_{\mathrm{tot}}=H\otimes I_A+I_n\otimes I_A.
}
\]
\rev{The common product eigenbasis ensures that these filtered swap operators preserve the \(j_B\)-label. We can therefore analyze the two generators within the corresponding invariant sectors. 
 }

To capture the notion that the bottleneck is localized to $A$ while the partition $B$ mixes rapidly, we entail a gap lower bound $g_B$ on a \emph{partial Lindbladian} $\lind_{\lrb{\bra{i_A} H \ket{i_A}, \beta, \CP_B,\gamma_1}}$ (\cref{assumption:fast_mixing}) on the restricted system $\bra{i_A}H\ket{i_A}$. Here, $\bra{i_A}H\ket{i_A} = \lrb{\bra{i_A}\otimes I_B}H\lrb{\ket{i_A}\otimes I_B} $.  For more on partial Lindbladians, see \cref{app:partial_lindbladian}.

We now describe an instantiation of replica exchange that provably overcomes local energy barriers. We set the Hamiltonian of the second system to be $I_A$, which is inherently easy to mix. Furthermore, we select a local swap operator $\CU$ which only swaps the subsystem \(A\) of $H$ with $I_A$, while leaving the subsystem \(B\) of $H$ unchanged. 
The action of $\mathcal{U}$ on the basis states is defined as
\begin{align}
\mc{U} {\ket{i_A j_{B}}}\otimes \ket{m_A} = {\ket{m_{A} j_{B}}}\otimes \ket{i_A}\,,
\label{eqn:swap_operator_local}
\end{align}
where $\ket{i_A}$ and $\ket{j_{B}}$ are the eigenvectors of $H_A$ and $H_B$ as introduced in \cref{eq:basis_A,eq:basis_B}, and $\ket{m_{A}}$ denotes an eigenvector of the second system $I_A$. For later algebraic convenience we take $\lrc{\ket{m_A}}$ to be the same orthonormal basis as $\lrc{\ket{i_A}}$. 
{Here, we also emphasize that implementing $\mathcal{U}$ does not require diagonalizing $H_A$ (and certainly not $H_B$). The reason is that $\mathcal{U}$ is basis-independent: for any unitaries $U_A$ on $A$ and $U_B$ on $B$,
$
\bigl(U_A \otimes U_B \otimes U_A \bigr)\,\mathcal{U}\,\bigl(U_A^\dagger \otimes U_B^\dagger \otimes U_A^\dagger \bigr)
= \mathcal{U}.
$
Equivalently, $\mathcal{U}$ is simply the permutation that exchanges the two $A$ registers and acts trivially on $B$.
}

The acceleration of quantum replica exchange with local swap is guaranteed by the following Theorem.

{
\begin{theorem_}[Main result]\label{thm:Gap_lower_bound}
Let $H$ be a Hamiltonian satisfying \cref{assumption:H_decomposition} with constants $K$ and $V_{\max}$ as therein, and suppose that the {partial Lindbladians} $\lind_{\lrb{\bra{i_A} H \ket{i_A}, \beta, \CP_B,\gamma_1}}$ have a gap lower bound of $g_B$ for all $\ket{i_A}$ in \cref{eq:basis_A}:
\begin{align}
\Gap\lrb{\lind_{\lrb{\bra{i_A} H \ket{i_A}, \beta, \CP_B,\gamma_1}}} \geq g_B.
\label{assumption:fast_mixing}
\end{align}
Consider the quantum replica exchange Lindbladian $\mathcal{L}$ given by
\begin{align}
\lind = \lind_{\lrb{H, \beta, \mathcal{P}_{\lrs{n}},\gamma_1}} \otimes \BI_A + \BI_n \otimes \lind_{\lrb{I_A, \beta, \mathcal{P}_A,\gamma_2}} + \lind_{\lrb{H\otimes I_A+ I_n\otimes I_A, \beta, \{\CU\},\gamma_M}},
\label{eq:total_lind_p2}
\end{align}
for any $\gamma_1,\gamma_2 \in \lrc{\gamma_G,\gamma_M}$, where $\mathcal{U}$ is the local swap operator on subsystem $A$ defined in \cref{eqn:swap_operator_local}\rev{. Then} the following gap lower bound holds:
\begin{align}
    \Gap(\lind) =\Omega\lrb{\frac{\min\left\{ g_B, 1 \right\}}{2^{|A|}  \exp(4 \beta KV_{\max})}} .
    \label{eq:gap_RE_L}
\end{align}

\end{theorem_}
}
{
This result shows that when $|A|$ is constant and the interaction strength across the cut is bounded (i.e., $KV_{\max} = \mathcal{O}(1)$), the gap of the replica exchange Lindbladian is proportional to the gap of the fast-mixing subsystem $B$, up to a constant factor. This holds for any constant inverse temperature $\beta$.

\rev{The fixed point of the joint replica exchange Lindbladian is
$\sigma=\sigma_H\otimes I_A/2^{|A|}$, and hence
$\lambda_{\min}(\sigma)=\lambda_{\min}(\sigma_H)/2^{|A|}$.
Combining \cref{eq:gap_RE_L} with \cref{eq:mixingTime_pre}, we obtain}
\begin{equation}
\rev{
t_{\mathrm{mix}}(\varepsilon)
\;\le\;
\frac{C\,2^{|A|}\exp(4\beta K V_{\max})}{\min\{g_B,1\}}
\log\!\left(
\frac{2^{|A|}}{\lambda_{\min}(\sigma_H)\varepsilon^2}
\right),
}
\end{equation}
\rev{for a universal constant \(C>0\). Under the constant \(|A|\), bounded \(KV_{\max}\), and constant \(\beta\) assumptions stated above, \(2^{|A|}\exp(4\beta K V_{\max})=\mathcal{O}(1)\) and \(\log(2^{|A|})=\mathcal{O}(1)\).} Consequently, the mixing time is polynomial in the system size and depends on the barrier height \(J\) only through the logarithmic factor \(\log \bigl(1/\lambda_{\min}(\sigma_H)\bigr)\). Specifically, in the setting of \cref{sec:slow_defect_main}, where the local energy barrier has height \(\Theta(J)\) deriving that \(\lambda_{\min}(\sigma_H)\sim e^{-\Theta(J)}\), the resulting overhead scales at most linearly in \(J\), representing an exponential improvement over the \(\exp(\Omega(J))\) mixing time of \cref{prop:slow_mixing_prop_informal}.

\rev{The factor \(d_A=2^{|A|}\) is not merely a proof artifact, but reflects
an intrinsic worst case obstruction. Machta's classical golf-course
analysis exhibits this obstruction: with \(\mathcal{O}(1)\) ground states
among \(D\) microstates and a constant number of replicas, replica exchange
yields only a replica count parallelism gain, leaving an inverse
equilibration time scale of order \(D^{-1}\)~\cite{machta2009strengths}. The
diagonal sector of our generator contains an analogous 
exchange mechanism, indicating that the \(d_A^{-1}\) penalty can arise
without further structure on \(A\), although structured instances may mix
faster.}

\begin{figure}[H]
    \centering
    \includegraphics[width=0.95\linewidth]{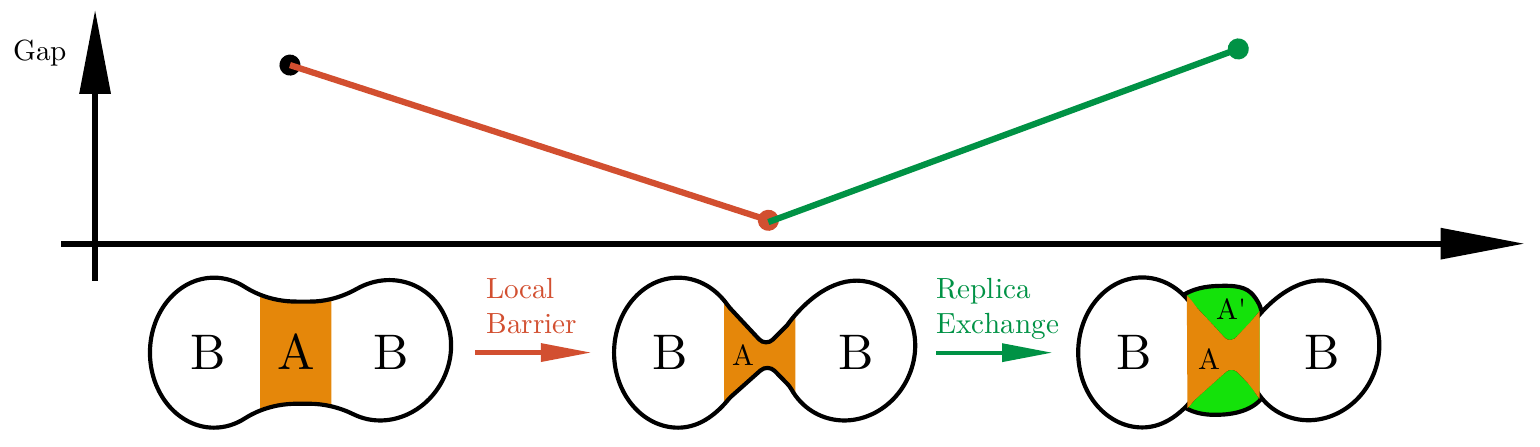}
    \caption{When a local barrier is introduced to the initial system, it creates a bottleneck in $A$ which decreases the gap. The application of replica exchange amplifies the gap by introducing a copy of region $A$ ($A'$, in green), which removes the bottleneck and restores  the gap of the fast-mixing subsystem $B$, up to a constant factor.  }
    \label{fig:2placeholder}
\end{figure}

We provide a proof sketch of \cref{thm:Gap_lower_bound} here, while a full proof can be found in \cref{app:Gap_lower_bound}.

\paragraph{\textbf{Proof sketch.}} 
Intuitively, \cref{assumption:fast_mixing} ensures that $ \lind_{\lrb{H, \beta, \mathcal{P}_{\lrs{n}}, \gamma_1}}$ mixes rapidly within subsystem $B$ of $H$, which means $ \lind_{\lrb{H, \beta, \mathcal{P}_B, \gamma_1}}$ has a good gap $g_B$, whereas energy barriers in $H_A$ make mixing on subsystem $A$ difficult. However, the second system with Hamiltonian $I_A$ always mixes quickly, since there are no energy barriers in $I_A$. Consequently, performing swaps on subsystem $A$ between the two systems allows the first chain to overcome the barriers in $H_A$ and achieve fast mixing. Our proof formalizes this intuition.
We are seeking a lower bound on the spectral gap of \cref{eq:total_lind_p2}. By \cref{it:gapA_B} of \cref{lem:gap_lemma}, it suffices to bound
\begin{align}
    \lind_{\lrb{H, \beta, \mathcal{P}_B, \gamma_1}} \otimes \BI_A \;+\; \BI_n \otimes \lind_{\lrb{I_A, \beta, \mathcal{P}_A, \gamma_2}} \;+\; \lind_{\lrb{H\otimes I_A+ I_n\otimes I_A, \beta, \{\CU\}, \gamma_M}},
    \label{eq:proof_sketch_5710}
\end{align}
(note that we dropped the jumps on $A$ from the first Lindbladian) which, by the above intuition, should mix quickly. In fact, \cref{eqn:partial_lindbladian} of \cref{lem:partial_lindbladian} implies that for any observable $ O_{A,B} $, considering the $A$-diagonal block is sufficient to prove fast mixing. We therefore decompose $ O_{A,B} $ in the eigenbasis of $H_A$ into its $A$-diagonal and $A$-off-diagonal parts, and define $\tilde{\mc{L}}_{\lrb{H, \beta, \mathcal{P}_B, \gamma_1}}$ to be $\lind_{\lrb{H, \beta, \mathcal{P}_B, \gamma_1}}$ restricted to the $A$-diagonal block as in \cref{eqn:tilde_L_H1_def}, whose gap is at least $g_B$ by \cref{lem:gap_lemma_tilde_L_H1}. Moreover, $\Gap\lrb{\lind_{\lrb{I_A, \beta, \mathcal{P}_A, \gamma_2}} }=\Theta(1)$ by \cref{lem:gap_lemma_H2}, because there is no energy barrier in $I_A$. So the first two terms of \cref{eq:proof_sketch_5710} have spectral gap $\min\lrc{g_B,1}$ by \cref{it:commut_AB} of \cref{lem:gap_lemma}.
Then using \cref{it:inner_B} of \cref{lem:gap_lemma}, it is therefore sufficient to lower bound
\begin{align}
\lr{X,- \lind_{\lrb{\{\CU\}}}(X)}_\sigma
\text{ ~for normalized } X\in \ker \lrb{\tilde{\mc{L}}_{\lrb{H, \beta, \mathcal{P}_B, \gamma_1}} \otimes \BI_A + \BI_n \otimes \lind_{\lrb{I_A, \beta, \mathcal{P}_A, \gamma_2}} } ,
\end{align}
where $\lr{X,I} _\sigma = 0$, and $\sigma $ is the Gibbs state of joint system, i.e., $\sigma  = \frac{e^{-\beta H}} {\Tr\lrs{e^{-\beta H}}}\otimes \frac{I_A}{2^\abs{A}}$. 
A key observation is that $\mathcal{L}_{\lrb{\lrc{\mc{U}}}}$ only swaps the subsystem $A$ between the two \rev{replicas}, while leaving $B$ invariant. Consequently, if $X$ is block diagonal in $A$, it remains block diagonal in $A$ after swapping; if $X$ is block off-diagonal in $A$, it remains block off-diagonal in $A$ after swapping. The same argument holds for $B$. Because the inner product $\lr{\cdot, \cdot}_{\sigma}$ between diagonal and off-diagonal matrices is always zero, we can decompose $X$ into its $A$-/$B$-block diagonal and off-diagonal components and estimate each case separately, as illustrated in \Cref{prop:gap_lower_bound_local_Adiag_swap,prop:gap_lower_bound_local_Aoffdiag_swap,prop:gap_lower_bound_local_AoffdiagBoff_swap}. Combining all cases, we obtain 
$
\lr{X,-\mathcal{L}_{\lrb{\lrc{\mc{U}}}}\lrb{X}}_\sigma = \Omega\lrb{1\big/\lrb{2^{|A|}  \exp(4 \beta KV_{\max})}}
$, as desired. Accounting for the $\min\lrc{g_B,1}$ factor coming from the first two terms of \cref{eq:proof_sketch_5710}, \cref{it:inner_B} of \cref{lem:gap_lemma} then implies the bound $\Omega\lrb{\min\lrc{g_B,1}\big/ \lrb{2^{|A|}  \exp(4 \beta KV_{\max})}}$.

\par\vspace{1em}

We emphasize the nontriviality of this result. In the classical setting, although numerous numerical experiments demonstrate the empirical success of replica exchange,  rigorous proofs of its acceleration are rare, and are limited to specific models and settings~\cite{madras2003swapping,woodard2009conditions,dong2022spectral}. In the quantum setting, the situation is even more subtle due to the non-commuting nature of quantum operations. \rev{To the best of our knowledge, \cref{thm:Gap_lower_bound} provides the first rigorous proof of principle for quantum replica exchange acceleration in this structured local-barrier setting.}

\section{Applications}\label{sec:applications}
In this section, 
we present two illustrative examples that fall within the scope of \cref{prop:slow_mixing_prop_informal} and \cref{thm:Gap_lower_bound}.
\subsection{Defected 1D Ising model}\label{sec:defected_1D_ising}
Recall the defected 1D Ising model we introduced in \cref{sec:bottlenecks}.
Consider $n$ spins on a ring structure, whose Hamiltonian is 
\begin{align}
    H =-J \sigma_1^z \sigma_2^z  - \sum_{i=2}^n \sigma_{i}^z \sigma_{i+1}^z\,,
    \label{eq:defected_1D_Ising}
\end{align}
where $\sigma_i^z$ is the Pauli-Z operator acting on the $i$-th spin, and periodic boundary conditions are imposed via $\sigma_{n+1}^z = \sigma_1^z$. See \cref{fig:1Dising} for an illustration. Note that $H$ satisfies \cref{assumption:H_decomposition}, if we define $A = \lrc{1,2}$ and $B=\lrc{3,\ldots, n}$ and $H = H_A+H_B+H_{AB}$, where $H_A = - J \sigma_1^z \sigma_2^z ,$
$
H_B =  - \sum_{i=3}^{n-1} \sigma_i^z \sigma_{i+1}^z\,,
$
and
$
H_{AB} =  - \sigma_2^z \sigma_3^z - \sigma_n^z \sigma_1^z\,
$. {Consequently, $\|H_{AB}\|\leq KV_{\max} = 2$}.
Additionally, \cite{ding2024polynomial,alicki2009thermalization} demonstrate that, for any inverse temperature $\beta = \mathcal{O}(1)$,
\begin{align}
\Gap\left(\lind_{\lrb{\bra{i_A}H \ket{i_A}, \beta, \CP_B, \gamma_1}}\right) = \Omega(1),
\end{align}
which derives that $g_B$ of \cref{assumption:fast_mixing} is $\Omega(1)$.
\begin{figure}[!htbp]
    \centering
     \centering
        \includegraphics[width=0.7\linewidth]{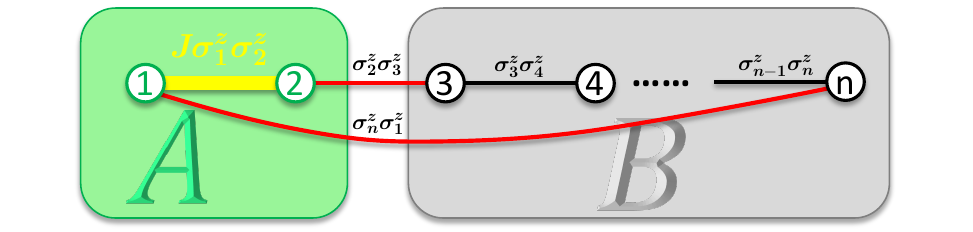}
        \caption{Defected 1D Ising model (schematic)} 
        \label{fig:1Dising}
\end{figure}
Now, we fix $\beta $ and focus on the regime $J \gg 1$.
This defect induces an energy barrier causing 
$\lind_{\lrb{H, \beta, \mathcal{P}_{[n]}, \gamma_1}}$ to have a mixing time that scales as $ \exp\!\left( \Omega(J) \right)$ by \cref{prop:slow_mixing_prop_informal}.
In contrast, by employing our replica exchange Lindbladian in  \cref{eq:total_lind_p2},
the gap is $\Gap(\mc{L})=\Omega(1)$ (by \cref{thm:Gap_lower_bound} since $|A|=2$), implying a mixing time that is polynomial in system size but {at most linear in $J$}, yielding an exponential improvement with respect to $J$.

Beyond the defected 1D Ising model, \rev{a similar analysis can be extended to} the defected 2D Toric code \cite{ding2024polynomial,hwang2024gibbs}, yielding a comparable speedup.

\subsection{Defected 2D Heisenberg model} 

\rev{As a proof-of-concept example within the scope of \cref{thm:rouze_fast_mixing}, we consider the Hamiltonian shown in \cref{fig:noncommuting_spin_example}. It combines Heisenberg interactions in the two-dimensional bulk with a commuting Ising interface and a strong defect of strength $J\gg1$:}
{
Let  $G=(V,E)$ be a graph with a partition $A \sqcup B= V$. Define the {boundary-vertex set of the partition}
\begin{align}
    V_{AB} \;:=\; \bigl\{\, u\in V \;:\; \exists\, v\in V \text{ with } (u,v)\in E
  \,, \text{ and }(u\in A,\ v\in B)\ \text{or}\ (u\in B,\ v\in A)  \,\bigr\},
\end{align}
which are the vertices of edges that cross $A$ and $B$.
Assign each edge a two-body Hamiltonian $h_{ij}$ by the rule
\begin{align}
    h_{ij} \;=\;
\begin{cases}
\sigma_i^z \sigma_j^z, & \text{if } \{i,j\}\cap V_{AB}\neq \varnothing,\\
\sigma_i^x \sigma_j^x \;+\; \sigma_i^y \sigma_j^y \;+\; \sigma_i^z \sigma_j^z, & \text{otherwise}.
\end{cases}
\end{align}
Then define
\begin{align}
    H_A = - \sum_{\substack{(i,j)\in E,\\ i,j\in A}} h_{ij} - (J-1) \sigma_{i^\star} ^z  \sigma_{j^\star} ^z ,
\qquad
H_B = - \sum_{\substack{(i,j)\in E,\\ i,j\in B}} h_{ij},
\qquad
H_{AB} = - \sum_{\substack{(i,j)\in E,\\ i\in A,\ j\in B}} \sigma_i^z \sigma_j^z,
\label{eq:2D_heisenberg_4710}
\end{align}
and
$
H = H_A + H_{AB} + H_B .
$ 
Here, $(i^\star,j^\star)\in E$ with $\,i^\star,j^\star \in V_{AB} \,\cap \,A$ introduces a single strong ferromagnetic bond as  \cref{sec:slow_defect_main}.
By construction, every term touching any boundary site $u\in V_{AB}$ involves $\sigma_u^z$. 
\rev{Here, the Ising interface serves only as a convenient example of a boundary condition realizing the commuting-cut requirement in \cref{assumption:H_decomposition}, while the bulk Hamiltonian remains noncommuting.}
}
Additionally, $KV_{\max}$ in \cref{item:H_AB_bounded} of \cref{assumption:H_decomposition} is $\Theta(1)$. Moreover, by leveraging a result of \cite{rouze2025efficient},
there exists a $\beta^* = \mc{O}(1)$ such that for any $\beta < \beta^*$,
\begin{align}
\Gap\lrb{\lind_{\lrb{\bra{i_A} H \ket{i_A}, \beta, \CP_B, \gamma_1}}} = \Omega(1).
\end{align}
For a proof, see \cref{cor:fast_RFA}. Hence $g_B$ of \cref{assumption:fast_mixing} is $\Omega(1)$ for $\beta < \beta^*$. 
Therefore, by \cref{thm:Gap_lower_bound}, it follows that the replica exchange Lindbladian $\lind$ in \cref{eq:total_lind_p2} has spectral gap $\Omega(1)$,  implying a mixing time that is polynomial in system size but {at most linear in $J$}. In contrast, the local barrier causes $\lind_{\lrb{H, \beta, \mathcal{P}_{[n]}, \gamma_1}}$ to have a mixing time of $\exp\!\big(\Omega(J)\big)$ by \cref{prop:slow_mixing_prop_informal}. Thus replica exchange achieves an exponential-in-$J$ acceleration, effectively removing the barrier-induced slowdown.
\begin{figure}[h]
\centering
     \centering
        \includegraphics[width=0.7\linewidth]{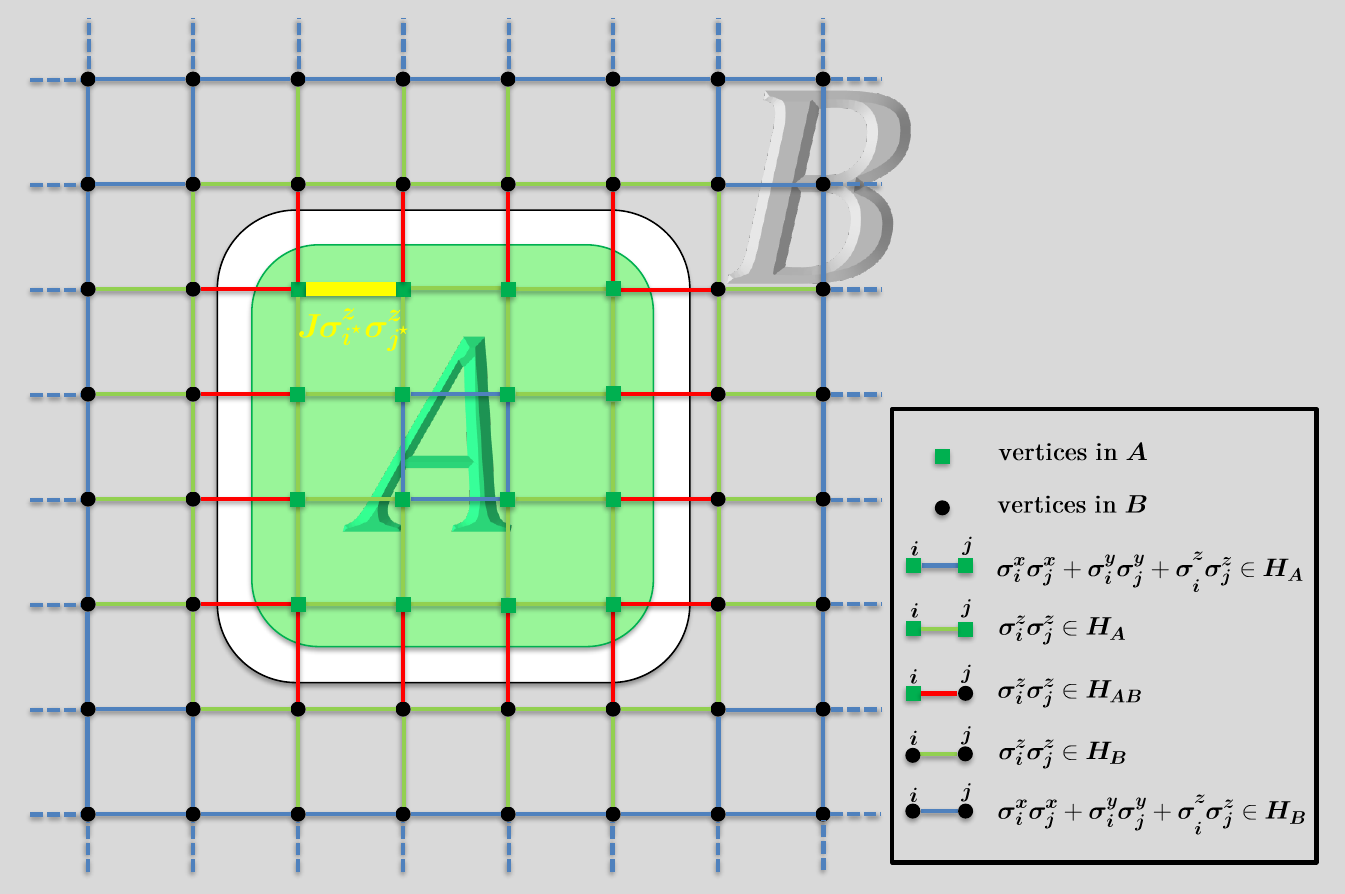}
        \caption{Heisenberg model with a commuting separating cut.
    The green vertices are in $A$, while the remaining black vertices are in $B$.
    The blue edges are $\sigma_i^x \sigma^x_j + \sigma^y_i \sigma^y_j + \sigma^z_i \sigma^z_j$,
    while the red and green edges are $\sigma^z_i \sigma^z_j$ interactions. Yellow edge is $J\sigma_{i^\star}^z\sigma_{j^\star}^z$ with a large coupling $J\gg 1$. Moreover, the red edges form $H_{AB}$.} 
        \label{fig:noncommuting_spin_example}
\end{figure}

\section*{Acknowledgements}
This work was supported in part by the Challenge Institute for Quantum Computation (CIQC) funded by National Science Foundation (NSF) through grant number OMA-2016245 (Z.C., J.B., L.L.), by the NSF under Grant No. CCF-2420130 (J.B.), and by the U.S. Department of Energy, Office of Science, Accelerated Research in Quantum Computing Centers, Quantum Utility through Advanced Computational Quantum Algorithms, grant no. DE-SC0025572 (L.L.).  L.L. is a Simons Investigator in Mathematics.  We thank Thiago Bergamaschi, Andr\'as Gily\'en and Nikhil Srivastava for insightful discussions. 

\section*{Declarations}
\subsection*{Conflict of interest} 
Authors have no conflict of interest to declare.
\subsection*{Data availability}
No datasets were generated or analyzed during the current study.

\bibliography{ref.bib}
\bibliographystyle{alpha}
\appendix
\clearpage

\phantomsection
\section*{\texorpdfstring{\rev{Notation summary}}{Notation summary}}
\label{app:notation}
\addcontentsline{toc}{section}{\protect\texorpdfstring{\protect\rev{Notation summary}}{Notation summary}}

\rev{
For quick reference, we collect below the notation used throughout the paper.
Detailed definitions remain at their first occurrence in the main text.

\medskip
\noindent\textit{Basic spaces, subsystems, and identities.}\par\nopagebreak[4]
\begin{center}
\renewcommand{\arraystretch}{1.18}
\begin{tabular}{@{}p{0.34\linewidth}p{0.61\linewidth}@{}}
$\mathcal{B}(\mathcal H)$
& the space of bounded operators on the Hilbert space $\mathcal H$ \\

$[n]$
& the set $\{1,2,\ldots,n\}$ labeling the qubits of the physical system \\

$A\sqcup B=[n]$
& a bipartition of the physical system; $A$ contains the local bottleneck and $B$ is its complement \\

$|X|$ and $d_X=2^{|X|}$
& the cardinality of a subsystem $X\subseteq[n]$ and the dimension of its Hilbert space \\

$I_X$
& the identity operator on the Hilbert space of subsystem $X$; $I_n$ denotes the identity on the full $n$-qubit system \\

$\mathbb I_X$
& the identity superoperator on operators supported on $X$, so that $\mathbb I_X(O)=O$ \\

$\mathcal P_X$
& the set $\{\sigma_i^\alpha:i\in X,\ \alpha=x,y,z\}$ of single-site Pauli coupling operators on $X$ \\

$R\succeq S$
& $R-S$ is positive semidefinite \\

$\lrs{X,Y}$
& the commutator $XY-YX$ \\

$\lrc{X,Y}$
& the anti-commutator $XY+YX$
\end{tabular}
\end{center}

\medskip
\noindent\textit{States, inner products, and convergence.}\par\nopagebreak[4]
\begin{center}
\renewcommand{\arraystretch}{1.18}
\begin{tabular}{@{}p{0.34\linewidth}p{0.61\linewidth}@{}}
$\beta$
& the inverse temperature \\

$\sigma_H$
& the Gibbs state $e^{-\beta H}/\Tr[e^{-\beta H}]$ of the Hamiltonian $H$ \\

$\sigma_2$ and $\sigma$
& the auxiliary state $\sigma_2=I_A/d_A$ and the joint state $\sigma=\sigma_H\otimes\sigma_2$ \\

$\lr{X,Y}_{\sigma_H}$
& the KMS inner product $\Tr[\sigma_H^{1/2}X^\dagger\sigma_H^{1/2}Y]$ \\

$\|X\|_{\Tr}$
& the trace norm $\Tr[\sqrt{X^\dagger X}]$ \\

$\Gap(\mathcal L)$
& the spectral gap of the Lindbladian $\mathcal L$ in the KMS inner product \\

$t_{\mathrm{mix}}(\epsilon)$
& the worst-case time required to reach trace distance at most $\epsilon$ from the fixed point
\end{tabular}
\end{center}

\medskip
\begin{samepage}
\noindent\textit{Lindbladian and frequency notation.}\par\nopagebreak[4]
\begin{center}
\renewcommand{\arraystretch}{1.18}
\begin{tabular}{@{}p{0.34\linewidth}p{0.61\linewidth}@{}}
$\mathcal L_{(H,\beta,\mathcal S,\gamma)}$
& the Heisenberg-picture CKG Lindbladian specified by the Hamiltonian $H$, inverse temperature $\beta$, coupling-operator set $\mathcal S$, and frequency-weight function $\gamma$ \\

$\mathcal L^\dagger$
& the dual Lindbladian governing Schr\"odinger-picture evolution \\

$\mathcal S=\{S^a\}$
& the set of coupling operators used to construct the Lindbladian \\

$B(H)$
& the set of Bohr frequencies $\{\lambda_i-\lambda_j:\lambda_i,\lambda_j\in\Spec(H)\}$ \\

$\gamma_G$ and $\gamma_M$
& the Gaussian- and Metropolis-type frequency-weight functions, respectively \\

$G_{\mathcal S}$
& the coherent correction term in the CKG Lindbladian, where the subscript $\mathcal S$ denotes the coupling-operator set
\end{tabular}
\end{center}
\end{samepage}

\medskip
\noindent\textit{Replica exchange notation.}\par\nopagebreak[4]
\begin{center}
\renewcommand{\arraystretch}{1.18}
\begin{tabular}{@{}p{0.34\linewidth}p{0.61\linewidth}@{}}
$H_1$ and $H_2$
& the Hamiltonians of the physical and auxiliary replicas, respectively \\

$\beta_1$ and $\beta_2$
& the inverse temperatures assigned to the two replicas \\

$\SWAP$
& the global swap unitary exchanging the two replicas \\

$\mathcal U$
& the local swap unitary exchanging the physical and auxiliary $A$ registers while leaving $B$ unchanged \\

$H_A$, $H_B$, and $H_{AB}$
& the two subsystem Hamiltonians and their interaction in $H=H_A\otimes I_B+I_A\otimes H_B+H_{AB}$ \\

$K$
& the number of product terms in the decomposition $H_{AB}=\sum_{k=1}^K V_A^{(k)}\otimes V_B^{(k)}$ \\

$V_{\max}$
& the maximum operator norm of an individual interaction term $V_A^{(k)}\otimes V_B^{(k)}$ \\

$g_B$
& the uniform spectral-gap lower bound assumed for the partial Lindbladians on subsystem $B$
\end{tabular}
\end{center}

\medskip
\noindent\textit{Asymptotic notation.}\par\nopagebreak[4]
\begin{center}
\renewcommand{\arraystretch}{1.18}
\begin{tabular}{@{}p{0.34\linewidth}p{0.61\linewidth}@{}}
$f=\mathcal O(g)$
& there exist $M>0$ and $x_0$ such that $|f(x)|\leq M|g(x)|$ for all $x\geq x_0$ \\

$f=\Omega(g)$
& $g=\mathcal O(f)$ \\

$f=\Theta(g)$
& both $f=\mathcal O(g)$ and $g=\mathcal O(f)$ \\

$\widetilde{\mathcal O}(g)$
& an $\mathcal O(g)$ bound with logarithmic factors suppressed \\

$f=o(g)$
& $f(x)/g(x)\to0$ as $x\to\infty$ whenever $g(x)\neq0$
\end{tabular}
\end{center}
}

\clearpage
\section{Auxiliary lemmas and gap estimates}\label{app:preliminaries}
This section is a collection of unrelated lemmas referred to throughout the paper.

\begin{lemma_} [Lemma 3 of \cite{ding2024polynomial}, Lemmas 1, 2 of \cite{alicki2009thermalization}]\label{lem:gap_lemma}
{
Let $A, B \in {\mathcal{B}(\mathcal{H})}$ be non-negative operators with non-trivial kernels on a Hilbert space $\mathcal{H}$, i.e., $A, B \succeq 0$. The spectral gaps of $A$ and $B$ are
their smallest strictly positive eigenvalues, denoted by ${\rm Gap} (A)$ and ${\rm Gap} (B)$, respectively.} Then, we have:

\begin{enumerate}
    \item  \label{it:gapA_B} If $\ker(A + B)$ is non-trivial such that $\ker(A + B) = \ker(B)$, then
    \begin{equation} \label{eq2}
        \Gap(A + B) \ge \Gap(B)\,.
    \end{equation}
    \item \label{it:commut_AB} If $A$ and $B$ are commuting and $\ker(A + B)$ is non-trivial, then
    \begin{equation} \label{eq3}
        \Gap(A + B) \ge \min\{\Gap(A),\Gap(B)\}\,.
    \end{equation}
    \item \label{it:inner_B} If $A$ has gap lower bound $g_A$ and $\lr{ \mathbf{p}, B \mathbf{p} } \ge g_B$ for all normalized $\mathbf{p} \in \ker(A)$, then
    \begin{equation} \label{eq4}
        A + B \succeq \frac{g_A g_B}{g_A + \norm{B}} \,,
    \end{equation}
    where $\norm{B}$ denotes the operator norm of $B$.
\end{enumerate}
\end{lemma_}
Because~\cite{ding2024polynomial} and~\cite{alicki2009thermalization} did not provide the proof for the first two points, we prove them here for completeness. The proof of the last point can be found in~\cite[Lemma 2]{alicki2009thermalization}.
\begin{proof}
\begin{enumerate}
    \item Since $A,B$ are positive \rev{operators}, $v^T (A+B) v \ge v^T B v$. The only thing that could go wrong is if $A$ introduced small nonzero eigenvalues to the kernel of $B$.
    The assumption of the kernel agreement eliminates that option. Since the gap is an eigenvalue in \rev{$\ker(B)^\bot = \ker(A+B)^\bot$} it cannot get smaller by the addition of $A$.

    \item 
    {If $A$ and $B$ commute, they are simultaneously diagonalizable:
there exists an orthonormal basis $\{e_i\}$ with
$Ae_i=a_i e_i$, $Be_i=b_i e_i$ ($a_i,b_i\ge 0$), hence
$(A+B)e_i=(a_i+b_i)e_i$.

Let
\(
\alpha\coloneqq\Gap(A)=\min\{a_i:\ a_i>0\},\text{ and  }\,
\beta \coloneqq\Gap(B)=\min\{b_i:\ b_i>0\}.
\)
For any $i$ with $a_i+b_i>0$, either $a_i>0$ or $b_i>0$.
If $a_i>0$, then $a_i+b_i\ge a_i\ge \alpha$; if $a_i=0$, then $b_i>0$ and
$a_i+b_i=b_i\ge \beta$.
Therefore $a_i+b_i\ge \min\{\alpha,\beta\}$ whenever $a_i+b_i>0$, and taking the
minimum over all such $i$ yields
\begin{align}
\Gap(A+B)=\min\{a_i+b_i:\ a_i+b_i>0\}\;\ge\;\min\{\alpha,\beta\}
=\min\{\Gap(A),\Gap(B)\}.
\end{align}
}
\end{enumerate}
\end{proof}
\begin{remark_}
    The inner product in the above lemma can be taken to be the Kubo--Martin--Schwinger (KMS) inner product $\langle X,Y\rangle_{\sigma} := \Tr[\sigma^{1/2} X^{\dagger} \sigma^{1/2} Y]$ as illustrated by \cite[Eq. (A.7)]{rouze2025efficient}.
\end{remark_}

\begin{lemma_} \label{lem:ratio_inequality}
Let \(a_1,a_2,b_1,b_2\) be positive real numbers. Then
\[
\frac{a_1+b_1}{a_2+b_2}
\;\ge\;
\min\!\left\{\frac{a_1}{a_2},\,\frac{b_1}{b_2}\right\}.
\]
\end{lemma_}
\begin{proof}
We can write the left-hand side as a convex combination of $\frac{a_1}{a_2}$ and $\frac{b_1}{b_2}$,
\begin{align}
\frac{a_1+b_1}{a_2+b_2} = \frac{a_1}{a_2}\frac{a_2}{a_2+b_2} + \frac{b_1}{b_2}\frac{b_2}{a_2+b_2},
\end{align}
which must be lower-bounded by the minimum.
\end{proof}

\begin{lemma_}[\rev{Spectral-gap bounds for the mixing time}]\label{lem:gap-mixing}
Let $\mathcal{L}$ be a primitive Lindbladian that satisfies quantum detailed balance with respect to the full-rank fixed point $\sigma_H$, and let the spectral gap be
$g=\mathrm{Gap}(\mathcal{L})$. For $\epsilon\in(0,1)$, the mixing time $t_{\mathrm{mix}}(\epsilon)$ with respect to the trace norm obeys the two-sided estimates
\begin{equation}\label{eq:tmix-upper}
t_{\mathrm{mix}}(\epsilon) \;\le\; \frac{1}{2g}\,\log\!\left(\frac{1}{\lambda_{\min}(\sigma_H)\,\epsilon^2}\right),
\end{equation}
and
\begin{equation}\label{eq:tmix-lower}
t_{\mathrm{mix}}(\epsilon) \;\ge\; \frac{1}{g}\,\log\!\left(\frac{\lambda_{\min}(\sigma_H)}{2\,\epsilon}\right).
\end{equation}
\end{lemma_}

\begin{proof}
For the upper bound \cref{eq:tmix-upper}, use the $\chi^2$-contraction for \rev{detailed-balance} semigroups (see \cite[Section IV]{kastoryano2013quantum}):
\begin{align}
    \chi^2\!\big(e^{t\lind^\dagger}(\rho),\sigma_H\big)\;\le\;e^{-2g t}\,\chi^2(\rho,\sigma_H).
\end{align}
Together with $\|\rho-\sigma_H\|_\Tr\le\sqrt{\chi^2(\rho,\sigma_H)}$ and the bound $\sup_{\rho}\chi^2(\rho,\sigma_H)=\lambda_{\min}(\sigma_H)^{-1}-1\le \lambda_{\min}(\sigma_H)^{-1}$, we get
\begin{align}
\sup_{\rho}\,\left\|e^{t\lind^\dagger}(\rho)-\sigma_H\right\|_\Tr \;\le\; e^{-g t}\,\lambda_{\min}(\sigma_H)^{-1/2}.
\end{align}
Solving for $t$ so that the right-hand side is at most $\epsilon$ gives \cref{eq:tmix-upper}.

For the lower bound \cref{eq:tmix-lower}, use that $\mathcal{L}$ is self-adjoint in the KMS inner product. Let $X$ be a Hermitian eigenoperator with $\langle I,X\rangle_{\sigma_H}=0$ and $\mathcal{L}(X)=-g X$. Normalize $Y:=\sigma_H^{1/2}X\sigma_H^{1/2}$ so that {$\|Y\|_\infty=1$}. For $0<\alpha\le \lambda_{\min}(\sigma_H)/2$, the state $\rho_0:=\sigma_H+\alpha Y$ is positive and has unit trace, since $\operatorname{Tr}(Y)=\langle I,X\rangle_{\sigma_H}=0$. By detailed balance, $e^{t\mathcal{L}^\dagger}(Y)=\sigma_H^{1/2}e^{t\mathcal{L}}(X)\sigma_H^{1/2}=e^{-g t}Y$. Hence
\begin{align}
\left\|e^{t\mathcal{L}^\dagger}(\rho_0)-\sigma_H\right\|_\Tr=\alpha e^{-g t}\,\|Y\|_\Tr\;\ge\;\alpha e^{-g t},
\end{align}
since $\|Y\|_\Tr\ge 1$ when {$\|Y\|_\infty=1$}. If this trace distance is to be at most $\epsilon$, then $t\ge g^{-1}\log(\alpha/\epsilon)$. Taking $\alpha=\lambda_{\min}(\sigma_H)/2$ yields \cref{eq:tmix-lower}.
\end{proof}

\section{Proof of \cref{prop:slow_mixing_prop_informal}} \label{sec:slow_defect}
In this section, we provide a proof of \cref{prop:slow_mixing_prop_informal}. 
Our argument builds on the framework developed in \cite{gamarnik2024slow}. {To this end, we first recall that  the induced trace norm for a superoperator  $\lind^\dagger $ is defined as 
\begin{align}
    \lrnorm{\lind^\dagger } _\Tr = \sup_{\rho } \frac{\lrnorm{\lind ^\dagger(\rho)}_\Tr}{\lrnorm{\rho}_\Tr}\,,
\end{align}
and several key lemmas and theorems from \cite{gamarnik2024slow}, which will serve as the foundation of our proof.  
}

\begin{lemma_}[Variant of Lemma 2.1 of \cite{gamarnik2024slow}]
    Given a Lindbladian \( \mathcal{L}^\dagger \) where \( \|\mathcal{L}^\dagger \| _{\Tr}  = \mc{O}(\mathrm{poly}(J)) \) and a time \( t = \exp(aJ^\alpha + o(J^\alpha)) \) for constants \( a > 0 \) and \( \alpha > 0 \), set \( \delta = \exp(-bJ^\alpha) \) for any constant \( b > a \), it holds that
\begin{align}
\exp(t \mathcal{L}^\dagger ) = \left(I + \delta \mathcal{L}^\dagger  \right)^{t/\delta} + R, \quad \|R\|_\Tr = o(1).
\end{align}
\end{lemma_}
\begin{corollary_}[Variant of Corollary 2.2 of \cite{gamarnik2024slow}]
    Given a Lindbladian \( \mathcal{L}^\dagger \) where \( \|\mathcal{L}^\dagger\| _{\Tr}  = \mc{O}(\mathrm{poly}(J)) \) and time \( t = \exp(aJ^\alpha + o(J^\alpha)) \) for constants \( a > 0 \) and \( \alpha > 0 \), set \( \delta = \exp(-bJ^\alpha) \) for any constant \( b > a \). If the discrete Gibbs sampler \( \mathcal{T} := I + \delta \mathcal{L}^\dagger \) satisfies the mixing time \( t_{\mathrm{mix}}(\mathcal{T}) = \exp\lrb{\Omega(J^\alpha)} \), then \( \mathcal{L}^\dagger \) also satisfies the mixing time \( t_{\mathrm{mix}}(\mathcal{L}) = \exp\lrb{\Omega(J^\alpha)} \).
\end{corollary_}

The proofs of the above lemma and corollary follow the arguments of \cite[Lemma 2.1]{gamarnik2024slow} and \cite[Corollary 2.2]{gamarnik2024slow}, with the only modification being that we replace the qubit number $n$ by the defect strength $J$, since in our setting the relevant scaling is with respect to $J$ rather than $n$.  
By combining these results with the proof of \cite[Corollary 3.5]{gamarnik2024slow}, we obtain the following proposition.

\begin{proposition_}[Variant of Corollary 3.5 of \cite{gamarnik2024slow}]\label{thm:slow_mixing_gamarnik}
Denote $H$ as a Hamiltonian acting on $n$ qubits with Gibbs state $\sigma _H$ at inverse temperature $\beta$.  
Let $A,B,C \subseteq \mathbb{C}^{2^n}$ be orthogonal linear subspaces of $\mathbb{C}^{2^n}$ with corresponding projectors $\Pi_A, \Pi_B, \Pi_C \in \mathcal{B}\lrb{\mathbb{C}^{2^n}}$.  
Assume that for
\begin{align}
\sigma_0 = \frac{e^{-\beta H/2} \Pi_A e^{-\beta H/2}}{\Tr\!\left[\Pi_A e^{-\beta H}\right]},
\end{align}
we have
\begin{align}
\Tr[\Pi_C \sigma_0] = o(1) 
\quad \text{and} \quad
\Tr\lrs{\Pi_C \sigma_H} = \Omega(1). 
\label{eq:assumption_slow_mixing_p3489}
\end{align}
Let $\mathcal{L}^\dagger$ be a Lindbladian Gibbs sampler with fixed point $\sigma_H$, such that $\mathcal{L}^\dagger \lrb{\sigma_H} = 0$, and $\lrnorm{\mathcal{L}^\dagger}_{\Tr} = \mc{O}({\rm{poly}}(J))$.  
Additionally, assume \rev{that} ${\mathcal{L}}$ \rev{satisfies}
\begin{align}
{\mathcal{L}}(\Pi_C) = (\Pi_C + \Pi_B) {\mathcal{L}}(\Pi_C)(\Pi_C + \Pi_B).
\label{eq:assumption_slow_mixing_L_Pi_C}
\end{align}
Also assume there exists a constant $\eta > 0$ such that
\begin{align}
\exp(J^\eta) \leq 
  \Tr\lrs{\Pi_A \sigma _H} \, \Tr\lrs{\Pi_C \sigma _H} \, \Tr\lrs{\Pi_B \sigma _H}^{-1}\,.
  \label{eq:assumption_slow_mixing_J_gamma}
\end{align}

Then, the mixing time of $\mathcal{L}$ is
$
t_{\mathrm{mix}} (\mathcal{L})= \exp\!\left( \Omega(J^\eta) \right).
$ 
\end{proposition_}

With the help of above proposition, we can now prove that a single strong ferromagnetic bond can create a slow mixing bottleneck under the dynamics generated by $\lind_{\lrb{H, \beta, \mc{S}, \gamma}}$ (\cref{eq:KMS_def}), when the coupling operators $\mc{S}$ are single-site.
\begin{proposition_} [Formal version of \cref{prop:slow_mixing_prop_informal}]\label{prop:slow_defect}
Let $H$ be an $n$-qubit Hamiltonian containing a ferromagnetic bond $-J\sigma^z_i\sigma^z_j$. Assume the rest part of the Hamiltonian, 
    \begin{align}
        H_\rest \coloneqq H - \lrb{-J\sigma^z_i\sigma^z_j},
    \end{align}
    satisfies $\lrs{H_\rest, \sigma^z_i} = \lrs{H_\rest, \sigma^z_j} = 0$, and $\|H_{\mathrm{rest}}\|\le C_0$ for a constant $C_0$ independent of $J$. 
    Then for sufficiently large $J$ and  at a fixed inverse temperature  $\beta$, we have 
\begin{align}
    t_{\mathrm{mix}} \lrb{\lind_{\lrb{H, \beta, \mc{S}, \gamma}}}= \exp\!\left( \Omega(J) \right),
\end{align} 
where the coupling operators in $\mc{S}$ are single-site.
\end{proposition_}
\begin{proof}
The commutation relations $[H_{\rest}, \sigma^z_i] = [H_{\rest}, \sigma^z_j] = 0$ imply that we can find a basis of simultaneous eigenstates for $H$, $\sigma^z_i$, and $\sigma^z_j$. Let $\Omega$ be the set of these eigenstates. 
Any such eigenstate $|\psi\rangle \in \Omega$ takes the form of a tensor product:
\begin{align}
|\psi\rangle = |s_i\rangle_i \otimes |s_j\rangle_j \otimes |\phi\rangle_{\text{rest}}
\end{align}
where $|s_i\rangle_i$ and $|s_j\rangle_j$ are eigenstates of $\sigma^z_i$ and $\sigma^z_j$ with eigenvalues $s_i, s_j \in \{\pm 1\}$ respectively, and $|\phi\rangle_{\text{rest}}\in \Omega_{\rest}$ is a state in the Hilbert space of the remaining $n-2$ qubits. The full state $|\psi\rangle$ is an eigenstate of $H_{\text{rest}}$ with eigenvalue $E_{\text{rest}}(\psi)$, i.e., $H_{\text{rest}}|\psi\rangle = E_{\text{rest}}(\psi)|\psi\rangle$.

The total energy of the state $|\psi\rangle$, denoted $E(\psi)$, is the sum of the eigenvalues from each part. Since $|\psi\rangle$ is a simultaneous eigenstate of both $H_{\text{rest}}$ and $-J\sigma^z_i\sigma^z_j$, the total energy is:
\begin{align}
    E(\psi) = -J s_i s_j + E_{\text{rest}}(\psi)
    \label{eq:eigenvalue_decomposition_est}
\end{align}
It is important to note that the eigenvalue $E_{\text{rest}}(\psi)$ can, in general, depend on the specific spin values $s_i$ and $s_j$ of qubits $i$ and $j$. A more precise notation would be $E_{\text{rest}}(s_i, s_j, \phi)$. However, since $E_{\text{rest}}(\psi)$ is an eigenvalue of the operator $H_{\text{rest}}$ and we are given $\|H_{\mathrm{rest}}\|\le C_0$, it follows that $|E_{\text{rest}}(\psi)| \le C_0$ for all $|\psi\rangle \in \Omega$.

Inspired by the above eigenvector decomposition of $H$, we choose the subspaces $A,B,C$ in \cref{thm:slow_mixing_gamarnik} as follows:
\begin{align}
A &:= \Span \left\{\,\ket{+1}_i \otimes \ket{+1}_j \otimes |\phi\rangle_{\text{rest}}
:\,\forall \phi \in \Omega_\rest\,\right\}, \label{eq:region_A}\\
C &:=  \Span \left\{\,\ket{-1}_i \otimes \ket{-1}_j \otimes |\phi\rangle_{\text{rest}}
:\,\forall \phi \in \Omega_\rest\,\right\},  \label{eq:region_C} \\ 
B &:= (A \cup C)^c = \Span  \left\{\,\ket{s_i}_i \otimes \ket{s_j}_j \otimes |\phi\rangle_{\text{rest}}
:\,   s_i\neq s_j \in \lrc{\pm 1}\,,\forall \phi \in \Omega_\rest\,\right\}.\label{eq:region_B}
\end{align}
Here, we also let $\Lambda_A, \Lambda_B,$ and $\Lambda_C$ be the sets of eigenvalues of the Hamiltonian $H$ 
whose corresponding eigenvectors lie within the subspaces $A, B,$ and $C$, respectively.

Next, we verify that the conditions in \cref{thm:slow_mixing_gamarnik} are satisfied.

{First, the condition $\lrnorm{\mathcal{L}^\dagger}_{\Tr} = \mc{O}({\rm{poly}}(J))$ is satisfied because the operator norm of each jump operator and coherent term can be bounded by $\mc{O}(1)$ according to~\cite[Page 5]{chen2023efficient} or~\cite[Page 45]{ding2025efficient}.} 

\cref{eq:assumption_slow_mixing_L_Pi_C}  is naturally satisfied, as each term in $\mathcal{L}$ flips at most one of $(i,j)$ due to the presence of single-site coupling operators $\mc{S}$. Therefore, one application of $\mathcal{L}$ on $\Pi_C$ cannot flip both $\ket{s_i}_i$ and $\ket{s_j}_j$ at the same time, making it impossible to reach the state $\ket{+1,+1}_{ij} \otimes \ket{\phi}_\rest$ in region $A$. Thus, $ {\mathcal{L}}(\Pi_C)$  still remains in region $C$ and $B$. 

Then, we check \cref{eq:assumption_slow_mixing_p3489}:
\begin{align}
\Tr\left[ \Pi_C \sigma_0\right] = \frac{1}{\Tr[\Pi_A e^{-\beta H}]} \Tr\left[ \Pi_C e^{-\beta H/2} \Pi_A e^{-\beta H/2}\right] = \frac{1}{\Tr[\Pi_A e^{-\beta H}]} \Tr\left[ \Pi_C  \Pi_A e^{-\beta H} \right] = 0,
\end{align}
where we have used the fact that $A$ and $C$ are orthogonal subspaces, and $H$ is diagonal in the basis of $A,C$. 

Next, we estimate $\Tr\lrs{\Pi_C \sigma_H}$:
\begin{align}
\Tr\left[ \Pi_C \sigma_H\right] &= \frac{1}{\Tr[e^{-\beta H}]} \Tr\left[ \Pi_C e^{-\beta H}\right] \\
&= \frac{\sum _{\lambda \in \Lambda_C } e^{-\beta \lambda}}{  \sum _{\lambda \in {\Lambda_A }} e^{-\beta \lambda} +\sum _{\lambda \in {\Lambda_B }} e^{-\beta \lambda} +\sum _{\lambda \in {\Lambda_C }} e^{-\beta \lambda}} \\
&\geq \frac{\sum _{\lambda \in \Lambda_C } e^{-\beta \max\lrb{ {\Lambda_C}}}}{  \sum _{\lambda \in {\Lambda_A }} e^{-\beta \min\lrb{\Lambda_A}} +\sum _{\lambda \in {\Lambda_B }} e^{-\beta \min\lrb{\Lambda_B}} +\sum _{\lambda \in {\Lambda_C }} e^{-\beta \min\lrb{ {\Lambda_C}}}} \\
& \geq  \frac{|C| \cdot e^{-\beta (-J+C_0)}}
{ |A|\cdot e^{-\beta (-J-C_0)} +|B|\cdot e^{-\beta \lrb{J-C_0}} +|C| \cdot e^{-\beta (-J-C_0)}} \\
&=  \frac{2^{n-2} e^{\beta J} e^{-2\beta C_0}}
{ 2^{n-2} e^{\beta J} + (2^n-2\cdot 2^{n-2})e^{-\beta J}+ 2^{n-2} e^{\beta J}  }\\
&= \Omega(1),
\label{eq:lower_bound_Tr_Pi_C_rho_beta}
\end{align}
for sufficiently large $J$. In the second inequality, we also use \cref{eq:eigenvalue_decomposition_est}. Similarly, we can also lower bound $\Tr\lrs{\Pi_A \sigma_H} = \Omega(1)$.

Then we estimate $\Tr\lrs{\Pi_B \sigma_H}$:
\begin{align}
\Tr[\Pi_B \sigma_H] &= \frac{1}{\Tr[e^{-\beta H}]} \Tr\left[ \Pi_B e^{-\beta H}\right] \\
&= \frac{\sum _{\lambda \in \Lambda_B } e^{-\beta \lambda}}{  \sum _{\lambda \in {\Lambda_A }} e^{-\beta \lambda} +\sum _{\lambda \in {\Lambda_B }} e^{-\beta \lambda} +\sum _{\lambda \in {\Lambda_C }} e^{-\beta \lambda}} \\
& \leq \frac{\sum _{\lambda \in \Lambda_B } e^{-\beta \min\lrb{ {\Lambda_B}}}}{  \sum _{\lambda \in {\Lambda_A }} e^{-\beta \max\lrb{\Lambda_A}} +\sum _{\lambda \in {\Lambda_B }} e^{-\beta \max\lrb{\Lambda_B}} +\sum _{\lambda \in {\Lambda_C }} e^{-\beta \max\lrb{ {\Lambda_C}}}} \\
&\leq \frac{(2^n-2\cdot 2^{n-2})  e^{-\beta (J-C_0)}}{2\cdot 2^{n-2} e^{-\beta (-J+C_0)} +(2^n-2\cdot 2^{n-2}) e^{-\beta (J+C_0) } } \\
&= \mc{O}(\exp(-2\beta J)),
\label{eq:lower_bound_Tr_Pi_B_rho_beta}
\end{align}
for sufficiently large $J$.

Finally, inserting \cref{eq:lower_bound_Tr_Pi_C_rho_beta,eq:lower_bound_Tr_Pi_B_rho_beta} into \cref{eq:assumption_slow_mixing_J_gamma}, we can get $\eta=1$ in \cref{eq:assumption_slow_mixing_J_gamma}. Therefore, we have $
    t_{\mathrm{mix}} \lrb{\lind_{\lrb{H, \beta, \mathcal{S}, \gamma}}}= \exp\!\left( \Omega(J) \right). 
$
\end{proof}

\section{Explicit expression of the swapping Lindbladian}
In this section, we calculate the explicit expression of the swapping Lindbladian
$\lind_{\lrb{H\otimes I_A+ I_n\otimes I_A, \beta, \{\CU\}, \gamma_M}}$, 
whose construction can be found in \cref{sec:quantum_gibbs_recall} following that of \cite{chen2023efficient}.
\begin{proposition_}
Recall $H$ from \cref{assumption:H_decomposition}, together with the definitions of $\mc{U}$ and the set $\lrc{\ket{i_A j_B}\otimes \ket{m_A}}$ from \cref{eqn:swap_operator_local}. Then, 
    by the construction of \cref{eq:KMS_def} in \cref{sec:quantum_gibbs_recall},   $\lind_{\lrb{H\otimes I_A+ I_n\otimes I_A, \beta, \{\CU\}, \gamma_M}}$ can be explicitly written as 
    \begin{align}
    \lind_{\lrb{H\otimes I_A+ I_n\otimes I_A, \beta, \{\CU\}, \gamma_M}}  =   \int_{-\infty}^\infty \d\omega \gamma_M(\omega) \left( \hat{\mc{U}}^\dagger(\omega) X \hat{\mc{U}}(\omega) -\frac{1}{2} \acomm{\hat{\mc{U}}^\dagger(\omega) \hat{\mc{U}}(\omega)}{X} \right)\,,
    \label{eq:local_swap_Lindbladian}
\end{align}
where
\begin{align}
    \hat{\mc{U}}(\omega ) = 
     \sum_{i_A,j_B,m_A} 
     \hat{f}\lrb{\omega - \lambda_{i_A,j_B,m_A} + \lambda_{m_A,j_B,i_A}} \ketbra{i_A j_B m_A}{m_A j_Bi_A}\,,
    \label{eqn:replica_exchange_davies_local_swap_general_1}
\end{align}
and 
\begin{align}
    \hat{\mathcal{U}} ^\dagger\lrb{\omega} \hat{\mathcal{U}}(\omega )
     = \sum_{i_A,j_B,m_A} 
     \abs{\hat{f}\lrb{\omega - \lambda_{i_A,j_B,m_A} + \lambda_{m_A,j_B,i_A}}}^2 
    \ketbra{m_A j_B i_A }{m_A j_B i_A} \,.
    \label{eq:U_hat_dagger_U_hat}
\end{align}
Here, $\lambda_{i_A,j_B,m_A}$ is the eigenvalue corresponding to the eigenstate $\ket{i_A j_B}\otimes \ket{m_A}$ of 
the Hamiltonian $H\otimes I_A +I_n\otimes I_A$, i.e.,
\begin{align}
    \lrb{H\otimes I_A +I_n\otimes I_A} \ket{i_A j_B} \otimes \ket{m_A} = \lambda_{i_A,j_B,m_A} \ket{i_A j_B} \otimes \ket{m_A}\,.
\end{align}
\end{proposition_}
\begin{proof}
By \cref{eq:S_hat_omega} and the definition of $\mc{U}$ in \cref{eqn:swap_operator_local},
\begin{align}
    \hat{\mc{U}}(\omega ) &= \frac{1}{\sqrt{2\pi}} \int_{-\infty}^\infty \d t~f(t) e^{\im Ht} \mc{U} e^{-\im Ht} e^{-\im \omega t}\, \\
    &= \frac{1}{\sqrt{2\pi}}\int_{-\infty}^\infty \d t \sum_{\substack{i_A,j_B,m_A \\ i_A',j_B',m_A'}}
    f(t)e^{-\im \omega t}  e^{\im t\lrb{\lambda_{i_A,j_B,m_A} - \lambda_{i_A',j_B',m_A'}}} \ketbra{i_A j_B m_A} {i_A j_B m_A} \mathcal{U} \ketbra{i_A' j_B' m_A'}{i_A' j_B' m_A'} \nonumber \\
    &=\frac{1}{\sqrt{2\pi}} \int_{-\infty}^\infty \d t \sum_{i_A,j_B,m_A} f(t)e^{-\im \omega t}  e^{\im t\lrb{\lambda_{i_A,j_B,m_A} - \lambda_{m_A,j_B,i_A}}} \ketbra{i_A j_B m_A} {m_A j_B i_A} \nonumber \\
    &= \sum_{i_A,j_B,m_A} 
    \hat{f}\lrb{\omega - \lambda_{i_A,j_B,m_A} + \lambda_{m_A,j_B,i_A}} \ketbra{i_A j_B m_A}{m_A j_Bi_A}\,.
\end{align}
Then direct calculation gives
\begin{align}
    \hat{\mathcal{U}} ^\dagger\lrb{\omega} \hat{\mathcal{U}}(\omega ) = \sum_{i_A,j_B,m_A} 
    \abs{\hat{f}\lrb{\omega - \lambda_{i_A,j_B,m_A} + \lambda_{m_A,j_B,i_A}}}^2 \ketbra{m_A j_B i_A}{m_A j_B i_A}\,.
\end{align}

Next, we will show the coherent part $G_{\lrc{\mc{U}}}$ in \cref{eq:KMS_def} is zero.
Write down $G_{\lrc{\mc{U}}}$ according to \cref{eq:G_def_CKG}:
\begin{align}
    G_{\lrc{\mc{U}}} =  \sum_{\nu_1,\nu_2\in B\lrb{H\otimes I_A +I_n\otimes I_A}} \frac{\tanh(-\beta(\nu_1 - \nu_2)/4)}{2\im } \alpha _{\nu_1,\nu_2} \lrb{\mc{U}_{\nu_2}}^{\dagger} \mc{U}_{\nu_1},
    \label{eq:G_U_def}
\end{align}
with
\begin{align}
    \mathcal{U}_{\nu_1}  
    &= \sum_{\lambda _{i_A,j_B,m_A} - \lambda _{m_A,j_B,i_A} = \nu_1} 
       \ketbra{i_A j_B\, m_A}{m_A j_B\, i_A} \,, \\
    \mathcal{U}_{\nu_2}  
    &= \sum_{\lambda _{p_A,q_B,\ell_A} - \lambda _{\ell_A,q_B,p_A} = \nu_2}  
       \ketbra{p_A q_B\, \ell_A}{\ell_A q_B\, p_A} \,.
\end{align}
Insert the above two equations into \cref{eq:G_U_def}:
\begin{align}
    G_{\lrc{\mc{U}}} =&  \sum_{\nu_1,\nu_2\in B(H)}  \frac{\tanh(-\beta(\nu_1 - \nu_2)/4)}{2\im } \alpha _{\nu_1,\nu_2} 
    \,\, \sum_{\lambda_{p_A,q_B,\ell_A} - \lambda _{\ell_A,q_B,p_A} = \nu_2}  \,\, 
    \sum_{\lambda _{i_A,j_B,m_A} - \lambda _{m_A,j_B,i_A} = \nu_1} \nonumber \\
    &
    \ketbra{\ell_A q_B p_A}{p_A q_B \ell_A}\,\ketbra{i_A j_B m_A}{m_A j_B i_A}\,,
\end{align}
where the summand is not zero only if $i_A = p_A$, $j_B = q_B$, and $m_A = \ell_A$, leading to
\begin{align}
    \nu_2 = \lambda_{p_A,q_B,\ell_A} - \lambda_{\ell_A,q_B,p_A} = \lambda_{i_A,j_B,m_A} - \lambda_{m_A,j_B,i_A} =\nu_1\,.
\end{align}
Thus, $\tanh\lrb{-\beta \lrb{\nu_1-\nu_2} /4} =\tanh(0) = 0$, deriving that 
\begin{align}
    G_{\lrc{\mc{U}}} = 0\,.
    \label{eq:G_U_zero}
\end{align}
This concludes the proof.
\end{proof}
In the later proof of the main theorem, the integral
\begin{align}
    \int_{-\infty}^{+\infty} \! \d \omega \, 
    \gamma_M(\omega)\, 
    \abs{\hat{f}\!\lrb{\omega - \lambda_{i_A,j_B,m_A} + \lambda_{m_A,j_B,i_A}} }^2 \,
\end{align}
from \cref{eq:local_swap_Lindbladian,eq:U_hat_dagger_U_hat} plays a crucial role. 
We therefore analyze its behavior in the following lemmas. For simplicity, we denote 
\begin{align}
\omega_{i_A,j_B,m_A} \coloneqq\lambda_{i_A,j_B,m_A} - \lambda_{m_A,j_B,i_A} \,.
\label{eq:omega_ijm}
\end{align}
\begin{lemma_} \label{lem:theta_calculation}
    With the Metropolis weight function $\gamma_M(\omega) = \exp\left[ -\beta \max\left\{\omega +\frac{1}{2\beta}, 0\right\} \right]$ \cref{eq:metropolis_weight_gamma}, and 
    the Gaussian filter function \cref{eq:gaussian_filter_CKG}
    \begin{align}
    \hat{f}(\omega) = \sqrt{\frac{\beta}{\sqrt{2\pi}}} \exp\left(-\frac{\beta^2 \omega^2}{4}\right),
    \end{align} 
    we have
    \begin{align}
        \theta\lrb{i_A,j_B,m_A} \coloneqq &\int_{-\infty} ^{+\infty}\d \omega \gamma_M(\omega)  \left|\hat{f}(\omega - \omega_{i_A,j_B,m_A})\right|^2 \nonumber\\
        = &\frac{1}{2} \left( 1 - \operatorname{erf}\left( \frac{1 + 2 \beta  \omega_{i_A,j_B,m_A}}{2\sqrt{2}} \right) + e^{-\beta  \omega_{i_A,j_B,m_A}} \left( 1 - \operatorname{erf}\left( \frac{1 - 2 \beta  \omega_{i_A,j_B,m_A}}{2\sqrt{2}} \right) \right) \right) \, \nonumber\\
        =& \frac{1}{2}\lrb{ \mathrm{erfc}\left( \frac{1 + 2 \beta  \omega_{i_A,j_B,m_A}}{2\sqrt{2}} \right) +e^{-\beta \omega_{i_A,j_B,m_A}} \mathrm{erfc}  \left( \frac{1 - 2 \beta  \omega_{i_A,j_B,m_A}}{2\sqrt{2}} \right) }\,,
        \label{eq:theta_erfc_p08}
    \end{align}
    where the error-function is 
    \[
        \operatorname{erf}(x) = \frac{2}{\sqrt{\pi}} \int_0^x e^{-t^2} \, dt\,,
    \]
and the complementary error-function is 
\[
   {\mathrm{erfc}}(x) = 1- \erf(x) = \frac{2}{\sqrt{\pi}} \int_x^{\infty} e^{-t^2} \, dt\, .
\]

\end{lemma_}
\begin{proof}
The proof is by standard calculation using \rev{simple changes of variables} and the definition of $\erf$ function. 
\end{proof}

\begin{remark_}
From \cref{eq:omega_ijm}, we can observe that $\omega_{i_A,j_B,i_A} = \lambda_{i_A,j_B,i_A} - \lambda_{i_A,j_B,i_A} =0$ for $\forall i_A,j_B$. In this case, 
\begin{align}
    \theta(i_A,j_B,i_A)  = \erfc\lrb{\frac{1}{2\sqrt{2}}} \approx 0.617 \,, \quad \forall i_A,j_B \,.    \label{eq:theta_iji}
\end{align}
Because $0\leq \erfc(x) \leq 2$,  \cref{eq:theta_erfc_p08} indicates that
\begin{align}
    0 \leq \theta(i_A,j_B,m_A) \leq 2 \,, \quad \forall i_A,j_B ,m_A\,.
    \label{eq:theta_bounds}
\end{align}
\end{remark_}
\begin{lemma_} \label{lem:property_erfc_p13}
    Let $\theta\lrb{i_A,j_B,m_A}$ be defined as  \cref{eq:theta_erfc_p08}, then $\theta(i_A,j_B,m_A)\leq  e^{-\frac{\beta}{2} \omega_{i_A,j_B,m_A}}$.
\end{lemma_}
\begin{proof}
    According to \cref{eq:theta_erfc_p08}, we want to show
    \begin{align}
        \frac{1}{2}\lrb{ \mathrm{erfc}\left( \frac{1 + 2 \beta  \omega_{i_A,j_B,m_A}}{2\sqrt{2}} \right) +e^{-\beta \omega_{i_A,j_B,m_A}} \mathrm{erfc}  \left( \frac{1 - 2 \beta  \omega_{i_A,j_B,m_A}}{2\sqrt{2}} \right) } \leq  e^{-\frac{\beta}{2} \omega_{i_A,j_B,m_A}}\,.
    \end{align}
    Denote $\omega_{i_A,j_B,m_A} = z$ for simplicity. Then the above inequality can be rewritten as
    \begin{align}
        f(z)\coloneqq \frac{1}{2}\lrb{ e^{\frac{\beta}{2}z }\mathrm{erfc}\left( \frac{1 + 2 \beta  z}{2\sqrt{2}} \right) +e^{-\frac{\beta }{2} z} \mathrm{erfc}  \left( \frac{1 - 2 \beta  z}{2\sqrt{2}} \right) } \leq  1\,.
    \end{align}
    Note that $f(z)$ is even, i.e., $f(z) = f(-z)$\rev{. Thus,} it suffices to show that $f(z) \leq 1$ for $z\geq 0$. By calculating derivative $\d f(z)/\d z$, we can show that $f(z)$ is non-increasing for $z \geq 0$. Thus, $\max_z f(z) = f(0) = \mathrm{erfc}\left( \frac{1}{2\sqrt{2}} \right) \approx 0.617\leq 1$.
\end{proof}

\begin{lemma_}\label{lem:property_erfc_p21}
    \begin{align}
     \lrb{\int_{-\infty}^{+\infty}\d \omega {\gamma_M (\omega)} \hat{f}  \lrb{\omega -\omega_{i_A,j_B,m_A}}  
    \hat{f}(\omega - \omega_{x_A,y_B,z_A})}^2 \leq 
    \sqrt{e^{-\beta {\omega_{i_A,j_B,m_A}}} \cdot e^{-\beta {\omega_{x_A,y_B,z_A}}}}\,.
\end{align}
\end{lemma_}
\begin{proof}
    Let $g(\omega) = \sqrt{\gamma_M(\omega)} \hat{f}(\omega - \omega_{i_A,j_B,m_A}) $ and $h(\omega) = \sqrt{\gamma_M(\omega)} \hat{f}(\omega - \omega_{x_A,y_B,z_A}) $.
    Using the Cauchy-Schwarz inequality for integrals, we have
    \begin{align}
    \lrb{\int \d \omega g(\omega) h(\omega)}^2 \leq \int \d \omega g^2(\omega) \int \d \omega h^2(\omega) \,.
    \end{align}
    Thus, it is sufficient to show that
    \begin{align}
        \int \d \omega g^2(\omega) = \theta(i_A,j_B,m_A)\leq e^{-\frac{\beta}{2} {\omega_{i_A,j_B,m_A}}} \,.
    \end{align}
    By \cref{lem:property_erfc_p13}, the above inequality holds.
\end{proof}
{
\begin{lemma_} \label{lem:erfc_lowerbound}
For any
\[
0 < p_i \leq 1, \quad 0 < p_j \leq 1, \quad\text{and}\quad p_i + p_j \leq 1 \,,
\]
the following inequality holds
\begin{align}
    p_j \, {\mathrm{erfc}}\left( \frac{1 + 2\ln(p_j/p_i)}{2\sqrt{2}} \right) 
+ 
p_i \, {\mathrm{erfc}}\left( \frac{1 - 2\ln(p_j/p_i)}{2\sqrt{2}} \right) 
\geq \frac{p_i p_j}{p_i+p_j} \,.
\label{eq:theta_lowerbound_2}
\end{align}
\end{lemma_}
}
\begin{proof}
Set the ratio
$
r := \frac{p_j}{p_i} > 0.
$
Then define
\begin{align}
A(r) := \frac{1 + 2 \ln r}{2\sqrt{2}}, \quad 
B(r) := \frac{1 - 2 \ln r}{2\sqrt{2}} = \frac{1}{\sqrt{2}} - \rev{A(r)}\,.
\end{align}
A key observation that will be used later is  
\begin{align}
    A(1/r) = B(r)\,, \quad B(1/r) = A(r).
    \label{eq:A_B_symmetry}
\end{align}  
Because $p_i r = p_j$, \cref{eq:theta_lowerbound_2} is equivalent to
\begin{align}
    r p_i\, {\erfc(A)} +p_i {\erfc(B)}  - \frac{p_i \cdot rp_i}{p_i+rp_i}\geq 0, \quad r > 0 ,\quad p_i > 0.
    \label{eq:to_prove_erfc_p09}
\end{align}
Because $p_i>0$, it is sufficient to show that the following function is non-negative for $r > 0$:
\begin{align}
F(r) := r\, {\erfc(A(r))} + {\erfc(B(r))} - \frac{r}{1 + r} \geq 0\, .  
\end{align}
Using \cref{eq:A_B_symmetry}, we can observe that
\begin{align}
    F(1/r) =  \frac{1}{r} \erfc(B(r)) + \erfc(A(r)) - \frac{1/r}{(1 + 1/r)} = \frac{F(r)}{r}\,.
\end{align}
When $0 < 1/r \leq 1$, we have $F(1/r ) = \frac{F(r)}{r}$. Thus, it suffices to prove $F(r) \geq 0$ for $r \geq 1$.
Moreover, for $r \geq 1$, $\frac{r}{1+r} < 1$. We only need to prove that, for $r \geq 1$,
\begin{align}
    S(r) \coloneqq r\, {\erfc(A(r))} + {\erfc(B(r))} - 1 \geq 0\,.
\end{align}
We first check $S(1) = 2 \erfc\lrb{\frac{1}{2\sqrt{2}}} - 1 > 0.23> 0$. So, it is sufficient to show that $S(r)$ is non-decreasing for $r \geq 1$.
Define $s= \ln(r)$ and differentiate $S$ with respect to $s$:
\begin{align}
    \frac{\d S}{\d s} = e^s \erfc(A) - e^s \frac{\sqrt{2}}{\sqrt{\pi}} e^{-A^2} + \frac{\sqrt{2}}{\sqrt{\pi}} e^{-B^2} 
    = e^s \erfc(A) -\sqrt{\frac{2}{\pi}} \lrb{e^{s} e^{-A^2} - e^{-B^2}} \,,
    \label{eq:dS_ds}
\end{align}
where we use the fact that $\frac{\d }{\d x}\erfc\lrb{x} = -\frac{2}{\sqrt{\pi}} e^{-x^2}$. Noticing that
\begin{align}
    A^2 - B^2 = (A-B) (A+B) = \lrb{\sqrt{2} s} \cdot \frac{1}{\sqrt{2}} = s = \ln(r)\,,
\end{align}
we can take exponential on both sides to obtain
\begin{align}
    e^{A^2 - B^2} = e^{s} \, \Longrightarrow \,e^{s} e^{-A^2} = e^{-B^2} \,.
\end{align}
Inserting the above equation into the second term in \cref{eq:dS_ds}, it vanishes, and we have
\begin{align}
    \frac{\d S}{\d s} = e^s \erfc(A) \geq 0, \quad \forall r \geq 1\,,
\end{align}
due to the property of the complementary error function that $\erfc(x)>0$ for $x>0$.
Next, 
\begin{align}
    \frac{\d S}{\d r} = \frac{\d S}{\d s} \frac{\d s}{\d r} = \frac{1}{r} \frac{\d S}{\d s} = \erfc(A(r)) \geq 0, \quad \forall r \geq 1\,,
\end{align}
which means $S(r)$ is non-decreasing for $r \geq 1$. 
Therefore, we complete the proof.
\end{proof}

\section{Partial Lindbladians}\label{app:partial_lindbladian}
The partial Lindbladian in \cref{assumption:fast_mixing} is motivated by the following Lemma.
\begin{lemma_}\label{lem:partial_lindbladian}
Let $H$ and $\lrc{\ket{i_A}}$ be those introduced in  \cref{assumption:H_decomposition} and \cref{eq:basis_A}. 
Given a Lindbladian 
$\lind_{\lrb{H,\beta,\CP_{B},\gamma_1}}$, a set $B \subseteq [n]$ and an operator $O_B$ with $\supp(O_B) \subseteq B$, $\lind_{(\bra{i_A} H \ket{i_A}, \beta,\CP_B,\gamma)}$ is the Lindbladian such that
\begin{align}
 \lind_{\lrb{H,\beta,\CP_{B},\gamma_1}}
  \left(\ketbra{i_A}{i_A}\otimes O_B\right) = \ketbra{i_A}{i_A}\otimes \lind_{(\bra{i_A} H \ket{i_A},\beta, \CP_B,\gamma_1)} \left(O_B\right).
 \label{eqn:partial_lindbladian}
\end{align}
\end{lemma_}

\begin{proof}
Here, we perform some calculations to explicitly justify the derivation of  \cref{eqn:partial_lindbladian}.
Recall the notation in \cref{assumption:H_decomposition,eq:basis_A,eq:basis_B}: The eigenvectors of $H_A$ are $\lrc{\ket{i_A}}$, and the eigenvectors of $H_B$ are $\lrc{\ket{j_B}}$. Additionally, denote the eigenvalues of $H$ as $\lambda_{i_A,j_B}$ with associated eigenvectors $\ket{i_A j_B}$. Then, for simplicity, we first consider the Lindbladian $\lind_{\lrb{H,\beta,I_A\otimes S^b,\gamma_1}}$ constructed by a single Pauli operator $I_A\otimes S^b$ with support on $B$. The rest discussions of other Pauli operators acting on $B$ can be done similarly.

Considering $\lind_{\lrb{H,\beta,I_A\otimes S^b,\gamma_1}}$ acting on the observable $O=\ket{\psi_{A}}\bra{\psi_{A}}\otimes O_B$, where $\ket{\psi_{A}} \in \lrc{\ket{i_A}}_{i_A}$, we have
\begin{align}
    \resizebox{0.92\linewidth}{!}{$
    \lind_{\lrb{H,I_A\otimes S^b,\gamma_1}}\lrb{O}  = \im \lrs{G_{\lrc{I_A\otimes S^b}},O} + \int_{-\infty}^\infty \d\omega~ \gamma_1(\omega) \left( \hat{S}^a(\omega)^\dagger O \hat{S}^a(\omega) -\frac{1}{2} \acomm{\hat{S}^a(\omega)^\dagger \hat{S}^a(\omega)}{O} \right)\,,
    \label{eqn:psi_1_A_partial_lindbladian}
    $
    }
\end{align}
where the unique $S^a =I_A\otimes S^b$. Then \cref{eq:S_hat_omega} gives that 
\begin{align}
    \hat{S}^a(\omega) =& \frac{1}{\sqrt{2\pi}} \int_{-\infty}^\infty \d t~f(t) 
    e^{\im H t} \lrb{I_A \otimes S^b}  e^{-\im H t} e^{-\im \omega t} \\
    =&\frac{1}{\sqrt{2\pi}}\int_{-\infty}^\infty \d t \sum_{\substack{i_A,j_B \\ i_A',j_B'}}
    f(t)e^{-\im \omega t}  e^{\im t\left(\lambda_{i_A,j_B} - \lambda_{i_A',j_B'}\right)} \ketbra{i_A j_B} {i_A j_B}  \lrb{I_A \otimes S^b} \ketbra{i_A' j_B'}{i_A' j_B'} \\
    =& \frac{1}{\sqrt{2\pi}}\int_{-\infty}^\infty \d t \sum_{{i_A,j_B ,j_B'}}
    f(t)e^{-\im \omega t}  e^{\im t\left(\lambda_{i_A,j_B} - \lambda_{i_A,j_B'}\right)} \ketbra{i_A}{i_A} \otimes \ketbra{j_B}{j_B} S^b \ketbra{j_B'}{j_B'}\\
    =& \sum_{i_A,j_B,j_B'} \hat{f} \lrb{\omega - \lambda_{i_A,j_B} + \lambda_{i_A,j_B'}} \ketbra{i_A}{i_A} \otimes \ketbra{j_B}{j_B} S^b \ketbra{j_B'}{j_B'}\,.
\end{align}

We first calculate $\hat{S}^{a\dagger}(\omega)  O \hat{S}^a(\omega)$ in the Lindbladian \cref{eqn:psi_1_A_partial_lindbladian}:
\begin{align}
    &\hat{S}^{a\dagger}(\omega) O \hat{S}^a(\omega) \nonumber \\
    =& \sum_{i_A,j_B,j_B'} \hat{f} \lrb{\omega - \lambda_{i_A,j_B} + \lambda_{i_A,j_B'}} \ketbra{i_A}{i_A} \otimes \ketbra{j_B'}{j_B'} S^b \ketbra{j_B}{j_B}
    \cdot\lrb{ \ketbra{\psi_{A}}{\psi_{A}}\otimes O_B} \nonumber \\
    & \cdot \sum _{m_A,n_B,n_B'} \hat{f} \lrb{\omega - \lambda_{m_A,n_B} + \lambda_{m_A,n_B'}} \ketbra{m_A}{m_A} \otimes \ketbra{n_B}{n_B} S^b \ketbra{n_B'}{n_B'} \\
    =& \ketbra{\psi_{A}}{\psi_{A}} \otimes
     \left(\sum_{\substack{j_B,j_B' \\ n_B,n_B' } } \hat{f} \lrb{\omega - \lambda_{\psi_{A},j_B} + \lambda_{\psi_{A},j_B'}} 
    \hat{f} \lrb{\omega - \lambda_{\psi_{A},n_B} + \lambda_{\psi_{A},n_B'}} \right. \nonumber \\
    &\left. \ketbra{j_B'}{j_B'} S^b \ketbra{j_B}{j_B} O_B \ketbra{n_B}{n_B} S^b \ketbra{n_B'}{n_B'}\right)\,,
    \label{eq:paritial_L_SXS}
\end{align}
where we used the fact that the summand is not zero only if $i_A=m_A = \psi_{A}$.

Then we calculate $\hat{S}^{a\dagger}(\omega) \hat{S}^{a}(\omega) O$ in the Lindbladian \cref{eqn:psi_1_A_partial_lindbladian}:
\begin{align}
    &\hat{S}^{a\dagger}(\omega) \hat{S}^{a}(\omega) O \nonumber\\
    =&\sum_{i_A,j_B,j_B'} \hat{f} \lrb{\omega - \lambda_{i_A,j_B} + \lambda_{i_A,j_B'}} \ketbra{i_A}{i_A} \otimes \ketbra{j_B'}{j_B'} S^b \ketbra{j_B}{j_B} \nonumber\\
    & \scalebox{0.98}{$\cdot \sum _{m_A,n_B,n_B'} \hat{f} \lrb{\omega - \lambda_{m_A,n_B} + \lambda_{m_A,n_B'}} \ketbra{m_A}{m_A} \otimes \ketbra{n_B}{n_B} S^b  \ketbra{n_B'}{n_B'}  \cdot { \lrb{ \ketbra{\psi_{A}}{\psi_{A}}\otimes O_B}}\,$} \\
    =& \ketbra{\psi_{A}}{\psi_{A}}\otimes   \nonumber\\
    & \scalebox{0.88}{$\lrb{ \sum_{{j_B,j_B' ,n_B' } } \hat{f} \lrb{\omega - \lambda_{\psi_{A},j_B} + \lambda_{\psi_{A},j_B'}} 
    \hat{f} \lrb{\omega - \lambda_{\psi_{A},j_B} + \lambda_{\psi_{A},n_B'}}  \ketbra{j_B'}{j_B'} S^b \ketbra{j_B}{j_B} S^b  \ketbra{n_B'}{n_B'} O_B}$
    }\,.
    \label{eq:paritial_L_SSX}
\end{align}
A similar calculation can be performed for $O \hat{S}^{a\dagger}(\omega) \hat{S}^{a}(\omega)$. 

Next, we examine the coherent term $\im  \lrs{G_{\lrc{I_A\otimes S^b}} ,O}$ of \cref{eqn:psi_1_A_partial_lindbladian}. Following the definition of \cref{eq:G_def_CKG}, we first write down the form of $S^a_{\nu_1}$ and $S^a_{\nu_2}$ constructed by $S^a = I_A\otimes S^b$:
\begin{align}
    S^a_{\nu_1} =& \sum_{\lambda_{i_A,j_B} - \lambda_{i_A',j_B'} = \nu_1} \ketbra{i_Aj_B}{i_Aj_B} \cdot I_A \otimes S^b \cdot \ketbra{i_A'j_B'}{i_A'j_B'}\,\\
    =& \sum_{\lambda_{i_A,j_B} - \lambda_{i_A,j_B'} = \nu_1} \ketbra{i_A}{i_A} \otimes \ketbra{j_B}{j_B} S^b \ketbra{j_B'}{j_B'}\,,\\
    S^a_{\nu_2} =& \sum_{\lambda_{m_A,n_B} - \lambda_{m_A',n_B'} = \nu_2} \ketbra{m_An_B}{m_An_B} \cdot I_A \otimes S^b \cdot \ketbra{m_A'n_B'}{m_A'n_B'}\,\\
    = &\sum_{\lambda_{m_A,n_B} - \lambda_{m_A,n_B'} = \nu_2} \ketbra{m_A}{m_A} \otimes \ketbra{n_B}{n_B} S^b \ketbra{n_B'}{n_B'}\,,
\end{align}
deriving that 
\begin{align}
    \lrb{S^a_{\nu_2}} ^\dagger S^a_{\nu_1} = \sum_{\lambda_{i_A,j_B} - \lambda_{i_A,n_B'} = \nu_2}\,\,
    \sum_{\lambda_{i_A,j_B} - \lambda_{i_A,j_B'} = \nu_1} \ketbra{i_A}{i_A} \otimes \ketbra{n_B'}{n_B'} S^b \ketbra{j_B}{j_B}  S^b \ketbra{j_B'}{j_B'}\,.
\end{align}
This can give the expression of the coherent term $\im  \lrs{G_{\lrc{I_A\otimes S^b}} ,O}$:
\begin{align}
    &\im  \lrs{G_{\lrc{I_A\otimes S^b}} ,O} \nonumber\\
     =& \sum_{\nu_1,\nu_2 \in B(H)} \frac{\tanh\lrb{-\beta\lrb{\nu_1-\nu_2}/4}}{2}
    \alpha_{\nu_1,\nu_2} \lrs{\lrb{S^a_{\nu_2}}^\dagger  S^a_{\nu_1}  ,\ketbra{\psi_A}{\psi_A} \otimes O_B} \\
    =& \sum_{\nu_1,\nu_2 \in B(H)} \frac{\tanh\lrb{-\beta\lrb{\nu_1-\nu_2}/4}}{2}
    \alpha_{\nu_1,\nu_2}  \ketbra{\psi_A}{\psi_A} \otimes \nonumber \\
    & \sum_{\lambda_{\psi_A,j_B} - \lambda_{\psi_A,n_B'} = \nu_2}  \, \,
    \sum_{\lambda_{\psi_A,j_B} - \lambda_{\psi_A,j_B'} = \nu_1} 
    \lrs{\ketbra{n_B'}{n_B'} S^b \ketbra{j_B}{j_B}  S^b \ketbra{j_B'}{j_B'} ,O_B}\,.
    \label{eq:partial_L_GX}
\end{align}

Inserting \cref{eq:paritial_L_SSX,eq:paritial_L_SXS,eq:partial_L_GX} into \cref{eqn:psi_1_A_partial_lindbladian} yields
\begin{align}
    \lind_{\lrb{H,\beta,I_A\otimes S^b,\gamma_1}} \left(\ketbra{\psi_A}{\psi_A}\otimes O_B\right) = \ketbra{\psi_A}{\psi_A}
    \otimes \hat{\mathcal{L}} \left(O_B\right),
    \label{eq:comp_or}
\end{align}
where $\hat{\mathcal{L}}$ is a superoperator acting on $O_B$ that also depends on $\ket{\psi_A}$ and $H$. In the rest part of this proof, we will show 
\begin{align}
    \hat{\mathcal{L}} \left(O_B\right) =
    \lind_{(\bra{\psi_A} H \ket{\psi_A},\beta,  S^b,\gamma_1)} \left(O_B\right).
\end{align}

We directly calculate $\lind_{(\bra{\psi_A} H \ket{\psi_A},\beta,  S^b,\gamma_1)}\!\left(O_B\right)$ by the definition \cref{eq:KMS_def} and compare it with \cref{eq:comp_or} to prove it.
For notational convenience, we denote $\tilde{H}_{B} = \bra{\psi_A} H \ket{\psi_A} $ as the effective Hamiltonian acting on $B$.
Then, by the definition of Lindbladian in \cref{eq:KMS_def}, we have
\begin{align}
   &\lind_{\lrb{\tilde{H}_{B},\beta, S^b,\gamma_1}}\!\left(O_B\right)\nonumber\\
     = &\im \lrs{G_{\lrb{\tilde{H}_{B}, S^b}},O_B} +\int_{-\infty}^\infty \d\omega~ \gamma_1(\omega) \left( \hat{S}^{\tilde{H}_{B} \dagger}(\omega) O_B \hat{S}^{\tilde{H}_{B} }(\omega) -\frac{1}{2} \acomm{\hat{S}^{\tilde{H}_{B} \dagger}(\omega) \hat{S}^{\tilde{H}_{B}}(\omega)}{O_B} \right)\,,
\end{align}
where $\hat S^{\tilde{H}_{B}} (\omega)$ is constructed by $S^b$, and \cref{eq:S_hat_omega} gives 
\begin{align}
    \hat S^{\tilde{H}_{B}} (\omega)  =& \frac{1}{\sqrt{2\pi}} \int_{-\infty}^\infty \d t~f(t) 
    e^{\im \tilde{H}_{B} t} \lrb{S^b}  e^{-\im \tilde{H}_{B} t} e^{-\im \omega t} \\
    =&\frac{1}{\sqrt{2\pi}}\int_{-\infty}^\infty \d t 
    \sum_{\substack{j_B, j_B'}}
    f(t)e^{-\im \omega t}  e^{\im t\left(\lambda^{\tilde{H}_{B}}_{j_B} - \lambda^{\tilde{H}_{B}}_{j_B'}\right)} \ketbra{j_B} {j_B}  {S^b} \ketbra{j_B'}{j_B'} \\
    =& \sum_{j_B,j_B'} 
    \hat{f} \lrb{\omega -\lambda^{\tilde{H}_{B}}_{j_B} +\lambda^{\tilde{H}_{B}}_{j_B'}} 
    \ketbra{j_B}{j_B} S^b \ketbra{j_B'}{j_B'}\,.
\end{align}
Here, $\lambda^{\tilde{H}_{B}}_{j_B}$ is the eigenvalue of the effective Hamiltonian $\tilde{H}_{B}$ corresponding to the states $\ket{j_B}$. We can check that 
\begin{align*}
    \lambda^{\tilde{H}_{B}}_{j_B} \ket{j_B} = &\tilde{H}_{B}\ket{j_B} = \lrb{\bra{\psi_A} \otimes I_B} H \lrb{\ket{\psi_A} \otimes I_B}\cdot  \ket{j_B} = \lrb{\bra{\psi_A} \otimes I_B}  H \lrb{ \ket{\psi_A} \otimes \ket{j_B}}  =\lambda_{\psi_A,j_B}\ket{j_B} \,,
\end{align*}
deducing that $\lambda^{\tilde{H}_{B}}_{j_B} = \lambda_{\psi_A,j_B}$. Here, we recall that $\lambda_{\psi_A,j_B}$ is the eigenvalue of $H$. With the help of $\lambda^{\tilde{H}_{B}}_{j_B} = \lambda_{\psi_A,j_B}$, we can check that \cref{eq:paritial_L_SXS,eq:paritial_L_SSX,eq:partial_L_GX} can match the corresponding terms in the Lindbladian $\ket{\psi_A } \bra{\psi_A} \otimes\lind_{\lrb{\tilde{H}_{B},\beta, S^b,\gamma_1}}$, i.e.,
\begin{align}
    &\hat{S}^{a\dagger }(\omega) \lrb{\ket{\psi_A}\bra{\psi_A} \otimes O_B} \hat{S}^a(\omega) 
    =\ket{\psi_A}\bra{\psi_A} \otimes \hat{S}^{\tilde{H}_{B} \dagger}(\omega) O_B \hat{S}^{\tilde{H}_{B}}(\omega)\,,\\
    &\hat{S}^{a\dagger}(\omega)\hat{S}^a(\omega)  \lrb{\ket{\psi_A}\bra{\psi_A} \otimes O_B} 
    =\ket{\psi_A}\bra{\psi_A} \otimes \hat{S}^{\tilde{H}_{B} \dagger}(\omega)  \hat{S}^{\tilde{H}_{B}}(\omega)O_B \,,\\
    &\lrb{\ket{\psi_A}\bra{\psi_A} \otimes O_B}\hat{S}^{a\dagger}(\omega)\hat{S}^a(\omega)   
    =\ket{\psi_A}\bra{\psi_A} \otimes O_B \hat{S}^{\tilde{H}_{B} \dagger}(\omega)  \hat{S}^{\tilde{H}_{B}}(\omega)\,,\\    
    & \lrs{G_{\lrc{I_A\otimes S^b}} , \lrb{\ket{\psi_A}\bra{\psi_A} \otimes O_B} }  
    = \ket{\psi_A}\bra{\psi_A} \otimes \lrs{G_{\lrb{\tilde{H}_{B}, \lrc{S^b}}},  O_B}\,.
\end{align}

Therefore, we have proven that 
\begin{align}
    \hat{\mathcal{L}} \left(O_B\right) =
    \lind_{(\bra{\psi_A} H \ket{\psi_A},\beta,  S^b,\gamma_1)}\!\left(O_B\right).
\end{align}
\end{proof}

It will be useful to know the fixed point of \rev{a} given partial Lindbladian in the proof of \cref{lem:gap_lemma_tilde_L_H1}. 

\begin{lemma_}
Instantiate the setting of \cref{lem:partial_lindbladian}. For each $\ket{i_A}$, $\lind_{(\bra{i_A} H \ket{i_A}, \beta,\CP_B,\gamma_1)}$ admits a unique fixed point $\sigma_{\ket{i_A}}$ given by
\begin{align}
    \sigma_{\ket{i_A}} \propto (\bra{i_A}\otimes I_B)\,e^{-\beta H}\,(\ket{i_A}\otimes I_{B}).
    \label{eqn:fixed_point_L_B_iA}
\end{align}
\end{lemma_}
\begin{proof}
    As discussed in the paragraph before \cref{eq:tmix},
    the unique fixed point of $\lind_{\lrb{\bra{i_A} H \ket{i_A}, \beta, \CP_B, \gamma_1}}$ is $\sigma_{\ket{i_A}} \propto e^{-\beta \bra{i_A} H \ket{i_A}}$. Then applying the fact that $H$ is block-diagonal in the $\ket{i_A}$ basis, we can move $\bra{i_A} \lrs{\cdot }\ket{i_A}$ out of the exponential, which becomes \cref{eqn:fixed_point_L_B_iA}.
\end{proof}

Now, we show that, if $\beta$ is not too large, the partial Lindbladian of a $(k,l)$-local Hamiltonian with maximum local interaction-strength $h$ has a good gap, and thus \cref{assumption:fast_mixing} is naturally satisfied. Let us first introduce the following extended result on the fast mixing property of the Lindbladian $\lind_{\lrb{H, \beta, \mathcal{P}_{[n]}, \gamma_1}}$ in \cite[Theorem II.1]{rouze2025efficient}:
\begin{proposition_}[Theorem II.1 of \cite{rouze2025efficient}] \label{thm:rouze_fast_mixing}
For any $(k, l)$-local and $n$-qubit Hamiltonian $H$ with the maximum local interaction-strength $h$, 
there exists a constant $\beta^* = \mc{O}\lrb{1/(hkl)}$ independent of $\dim(H)$ such that for any $\beta\leq \beta^* $, the spectral gap of the Lindbladian $\lind_{\lrb{H, \beta, \mathcal{P}_{[n]}, \gamma_1}}$ is $\Omega(1)$ for $\gamma_1\in\lrc{\gamma_G,\gamma_M}$.
\end{proposition_}
\begin{proof}
\rev{While \cite[Theorem II.1]{rouze2025efficient} establishes the
$\Omega(1)$ lower bound for $\gamma_1=\gamma_G$, we now show that the same
bound holds for $\gamma_1=\gamma_M$. Writing $x=\beta\omega$,
\cref{eq:metropolis_weight_gamma} gives
\begin{align}
\log\frac{\gamma_M(\omega)}{\gamma_G(\omega)}
=
\begin{cases}
\dfrac{(x+1)^2}{2}, & x\leq-\dfrac{1}{2},\\[2mm]
\dfrac{x^2}{2}, & x>-\dfrac{1}{2},
\end{cases}
\end{align}
and hence $\gamma_M(\omega)-\gamma_G(\omega)\geq0$.

The dissipative term in \cref{eq:KMS_def} is linear in $\gamma$.
Moreover, \cref{eq:alpha_omega1_omega2_CKG} shows that each coefficient
$\alpha_{\nu_1,\nu_2}$ is linear in $\gamma$, and therefore the coherent
term in \cref{eq:G_def_CKG} is linear in $\gamma$ as well. Consequently,
the full generators, including their coherent terms, satisfy
\begin{align}
\lind_{\lrb{H,\beta,\mathcal P_{[n]},\gamma_M}}
=\lind_{\lrb{H,\beta,\mathcal P_{[n]},\gamma_G}}
+\lind_{\lrb{H,\beta,\mathcal P_{[n]},\gamma_M-\gamma_G}}.
\end{align}
Because $\gamma_M-\gamma_G\geq0$, the dissipative part of the last term
is in GKLS form. Its coherent term is the difference of the Hermitian
coherent terms associated with $\gamma_M$ and $\gamma_G$, and is therefore
also Hermitian. Thus the last term is a valid Lindbladian.

Moreover, the generators associated with $\gamma_M$ and $\gamma_G$ are
self-adjoint with respect to the same KMS inner product. Their difference,
which is precisely the generator associated with $\gamma_M-\gamma_G$, also
satisfies KMS detailed balance. Consequently, its Dirichlet form is
nonnegative:
\begin{align}
\lr{X,-\lind_{\lrb{H,\beta,\mathcal P_{[n]},\gamma_M-\gamma_G}}(X)}_{\sigma_H}
\geq0
\qquad\text{for all }X.
\end{align}
It follows that
\begin{align}
\lr{X,-\lind_{\lrb{H,\beta,\mathcal P_{[n]},\gamma_M}}(X)}_{\sigma_H}
\geq
\lr{X,-\lind_{\lrb{H,\beta,\mathcal P_{[n]},\gamma_G}}(X)}_{\sigma_H}.
\end{align}
Taking the infimum over nonzero $X$ satisfying
$\lr{I,X}_{\sigma_H}=0$ gives
\begin{align}
\Gap\lrb{\lind_{\lrb{H,\beta,\mathcal P_{[n]},\gamma_M}}}
\geq
\Gap\lrb{\lind_{\lrb{H,\beta,\mathcal P_{[n]},\gamma_G}}}
=\Omega\lrb{1}.
\end{align}}
\end{proof}

To apply this result to the partial Lindbladian of a $(k,l)$-local Hamiltonian with maximum local interaction-strength $h$, we must prove that the map $\bra{i_A}[\cdot]\ket{i_A}$ preserves the $(k,l)$-locality of Hamiltonians:
\begin{lemma_}[Locality of partial Hamiltonian]\label{lem:locality_strength_partial_trace}
Let $H_A,H_B$, and $H_{AB}$ denote the Hamiltonians introduced in \cref{assumption:H_decomposition}.  Suppose $H_B+H_{AB}$ is a $(k,l)$-local Hamiltonian with the maximum local interaction-strength $h$. Then, for any $\ket{i_A}$ in \cref{eq:basis_A},
 $$(\bra{i_A} \otimes I_B) (H_B +H_{AB} ) (\ket{i_A} \otimes I_B)$$ 
 is also a $(k,l)$-local Hamiltonian with the maximum local interaction-strength $h$.
\end{lemma_}

\begin{proof}
Write the Hamiltonian
\begin{align}
    H_B+H_{AB}
   =\sum_{X\subseteq B} h_X
   +\sum_{\substack{X\subseteq [n]\\ X\cap A\neq\varnothing,\;X\cap B\neq\varnothing}} h_X,
\end{align}
where each local term $h_X$ acts non-trivially only on the qubits in $X$.  
Because $H_B+H_{AB}$ is $(k,l)$-local, every $h_X$ acts on at most $k$ qubits, every qubit is contained in at most $\ell$ such terms, and $\|h_X\|\le h$.

Fix an eigenvector $\ket{i_A}$ of $H_A$ and define the compression
\begin{align}
    \mathcal{E}_A(O)\;:=\;(\bra{i_A}\otimes I_B)\,O\,(\ket{i_A}\otimes I_B),
\qquad O\in\mathcal B\lrb{\mathbb{C}^{2^n}}.
\end{align}
Then 
\begin{align}
      \mathcal{E}_A \bigl(H_B+H_{AB}\bigr)
      =\sum_{X\subseteq B} h_X
      +\sum_{\substack{X\subseteq [n]\\ X\cap A\neq\varnothing,\;X\cap B\neq\varnothing}}
            \mathcal{E}_A(h_X).
\end{align}
\rev{If $X\subseteq B$, then $h_X$ is unchanged and remains supported on
at most $|X|\leq k$ qubits. If $X$ intersects both $A$ and $B$, then for
every operator $O$ supported on $B\setminus(X\cap B)$,
\begin{align}
\lrs{\mathcal{E}_A(h_X),O}
&=\mathcal{E}_A(h_X)O-O\mathcal{E}_A(h_X)\nonumber\\
&=(\bra{i_A}\otimes I_B)h_X(\ket{i_A}\otimes I_B)O
-O(\bra{i_A}\otimes I_B)h_X(\ket{i_A}\otimes I_B)\nonumber\\
&=(\bra{i_A}\otimes I_B)h_X(I_A\otimes O)
(\ket{i_A}\otimes I_B)-(\bra{i_A}\otimes I_B)(I_A\otimes O)h_X
(\ket{i_A}\otimes I_B)\nonumber\\
&=(\bra{i_A}\otimes I_B)
\lrs{h_X,I_A\otimes O}
(\ket{i_A}\otimes I_B)
=0,
\end{align}
because $h_X$ is supported on $X$. Therefore,
$\mathcal{E}_A(h_X)$ is supported on $X\cap B$ and acts on at most
$|X\cap B|\leq |X|\leq k$ qubits.

We keep each $\mathcal{E}_A(h_X)$ as a separately labelled local term
associated with the original term $h_X$, without combining terms that
acquire the same support after compression. For any qubit $u\in B$, a
compressed term can act on $u$ only if the corresponding original term
acted on $u$. Hence every qubit is still contained in at most $\ell$ local
terms.}

Because $\mathcal{E}_A$ is a compression by a unit vector, it is contractive in the operator norm:
\begin{align}
    \bigl\|\,\mathcal{E}_A(h_X)\bigr\|
  =\bigl\|(\bra{i_A}\otimes I_B)h_X(\ket{i_A}\otimes I_B)\bigr\|
  \le \|h_X\|\le h.
\end{align}

Combining these three observations, we complete the proof.
\end{proof}

With this, it follows that the partial Lindbladian has good gap, and thus \cref{assumption:fast_mixing} is satisfied:
\begin{corollary_}[Partial Lindbladian has good gap]\label{cor:fast_RFA}
Let $H_A,H_B,H_{AB}$, and $H$ denote the Hamiltonians introduced in \cref{assumption:H_decomposition}. Additionally,  assume that $H_B+H_{AB}$ is a $(k,l)$-local Hamiltonian with 
the maximum local interaction-strength upper bounded by $h$.  Then there exists a constant $\beta^* =\mc{O} \lrb{1/(hkl)}$ independent of $\dim(H)$ such that for any $\beta\leq \beta^* $,
\begin{align}
\Gap\lrb{\lind_{\lrb{\bra{i_A}H \ket{i_A}, \beta, \CP_B, \gamma_1}}} =\Omega(1)\,,
\end{align}
for all $\ket{i_A}$ in \cref{eq:basis_A}.
\end{corollary_}
\begin{proof}
\cref{lem:locality_strength_partial_trace} shows  that $\bra{i_A}\lrb{H_B+H_{AB}}\ket{i_A} $ is still a $(k,l)$-local Hamiltonian with the maximum local interaction-strength $h$ for any $\ket{i_A}$. What's more, the support of $\bra{i_A}\lrb{H_B+H_{AB}}\ket{i_A} $ is on $B$. 
\rev{Importantly, these locality bounds are uniform in $\ket{i_A}$:
the parameters $k$, $l$, and $h$ are the same for every compressed
Hamiltonian. Since the threshold $\beta^*$ and the implied gap constant in
\cref{thm:rouze_fast_mixing} depend only on these common locality parameters,
the same $\beta^*$ and the same constant gap lower bound hold simultaneously
for all $\ket{i_A}$.}
Therefore, \cref{thm:rouze_fast_mixing} gives that there exists a constant $\beta^* = O\lrb{1/(hkl)}$ such that for $\beta\leq \beta^* $, 
\begin{align}
\Gap\lrb{\lind_{\lrb{\bra{i_A}\lrb{H_B+H_{AB}}\ket{i_A}, \beta, \CP_B, \gamma_1}}} = \Omega(1),
\end{align}
Since $\bra{i_A}\lrb{H_B+H_{AB}}\ket{i_A} +\lambda_{i_A} = \bra{i_A}\lrb{H_A+H_B+H_{AB}}\ket{i_A}$, \rev{it} follows that
\begin{align}
   \lind_{\lrb{\bra{i_A}\lrb{H_B+H_{AB}}\ket{i_A}, \beta, \CP_B, \gamma_1}} (O_B) 
    =\lind_{\lrb{\bra{i_A}\lrb{H_A+H_B+H_{AB}}\ket{i_A}, \beta, \CP_B, \gamma_1}}(O_B),
\end{align}
and thus
\begin{align}
\Gap\lrb{\lind_{\lrb{\bra{i_A}\lrb{H_A+H_B+H_{AB}}\ket{i_A}, \beta, \CP_B, \gamma_1}}} =\Omega(1)\,,
\end{align}
as advertised.
\end{proof}

\section{Proof of \cref{thm:Gap_lower_bound}} \label{app:Gap_lower_bound}
In this section, we prove \cref{thm:Gap_lower_bound}.
{
\begin{theorem_}[\cref{thm:Gap_lower_bound}, restated]
Let $H$ be a Hamiltonian satisfying \cref{assumption:H_decomposition} with constants $K$ and $V_{\max}$ as therein, and suppose that the {partial Lindbladians} $\lind_{\lrb{\bra{i_A} H \ket{i_A}, \beta, \CP_B,\gamma_1}}$ have a gap lower bound of $g_B$ for all $\ket{i_A}$ in \cref{eq:basis_A}:
\begin{align}
\Gap\lrb{\lind_{\lrb{\bra{i_A} H \ket{i_A}, \beta, \CP_B,\gamma_1}}} \geq g_B.
\end{align}
Consider the quantum replica exchange Lindbladian $\mathcal{L}$ given by
\begin{align}
\lind = \lind_{\lrb{H, \beta, \mathcal{P}_{\lrs{n}},\gamma_1}} \otimes \BI_A + \BI_n \otimes \lind_{\lrb{I_A, \beta, \mathcal{P}_A,\gamma_2}} + \lind_{\lrb{H\otimes I_A+ I_n\otimes I_A, \beta, \{\CU\},\gamma_M}},
\label{eq:total_lind_p2_759157}
\end{align}
for any $\gamma_1,\gamma_2 \in \lrc{\gamma_G,\gamma_M}$, where $\mathcal{U}$ is the local swap operator on subsystem $A$ defined in \cref{eqn:swap_operator_local}.
Then the following gap lower bound holds:
\begin{align}
    \Gap(\lind) =\Omega\lrb{\frac{\min\left\{ g_B, 1 \right\}}{2^{|A|}  \exp(4 \beta KV_{\max})}} .
\end{align}

\end{theorem_}
}
\begin{proof}
Denote $d_A = 2^{|A|}$ and $d_B = 2^{|B|}$.
Because \cref{lem:equal_ker} shows that 
\begin{align}
    \ker\lrb{\mathcal{L}} = \ker\lrb{\lind_{\lrb{H, \beta, \mathcal{P}_B, \gamma_1}} \otimes \BI_A \;+\; \BI_n \otimes \lind_{\lrb{I_A, \beta, \mathcal{P}_A, \gamma_2}} \;+\; \lind_{\lrb{H\otimes I_A+ I_n\otimes I_A, \beta, \{\CU\}, \gamma_M}}}\,,
    \label{eqn:ker_L_reduce_equal}
\end{align}
according to \cref{it:gapA_B} of \cref{lem:gap_lemma}, it is sufficient to drop the coupling operators $\mathcal{P}_A$ on $A$ from the first term of $\mathcal{L}$ and compute the spectral gap of the remaining Lindbladian
\begin{align}
    \lind_{\lrb{H, \beta, \mathcal{P}_B, \gamma_1}} \otimes \BI_A \;+\; \BI_n \otimes \lind_{\lrb{I_A, \beta, \mathcal{P}_A, \gamma_2}} \;+\; \lind_{\lrb{H\otimes I_A+ I_n\otimes I_A, \beta, \{\CU\}, \gamma_M}}.
\end{align}

For any observable $ O_{A,B} $ in $\mathcal{B} \lrb{\mathbb{C}^{2^{n}}}$, we first rewrite it as
\begin{align}
 O_{A,B} =\sum_{{ i_A }}\ketbra{ i_A }{ i_A }\otimes \bra{i_A}  O_{A,B}  \ket{i_A}+O_{\text{off} A,B}\,,
\label{eqn:O_AB_decomp_i378}
\end{align}
where $\lrc{\ket{i_A}}$ is the eigenvector set of $H_A$ as introduced in \cref{eq:basis_A}, and $\bra{i_A} O_{A,B} \ket{i_A}=(\bra{ i_A }\otimes I_B) O_{A,B} (\ket{ i_A }\otimes I_{B})$. Here, the first summation contains terms that correspond to diagonal part in the $A$ system while the second term contains the  off-diagonal  part in the $A$ system. Thus, based on \cref{lem:partial_lindbladian}, we can rewrite $ \lind_{\lrb{H,\beta,\CP_{B},\gamma_1}}
 \!\left( O_{A,B} \right)$ as
\begin{align}
    \scalebox{0.96}{$
 \lind_{\lrb{H,\beta,\CP_{B},\gamma_1}}
 \!\left( O_{A,B} \right) =
 \sum_{i_A} \ketbra{i_A}{i_A}\otimes 
 \lind_{(\bra{i_A} H \ket{i_A},\beta, \CP_B,\gamma_1)}\!\left(\bra{i_A}  O_{A,B}  \ket{i_A}\right)
 +  \lind_{\lrb{H,\beta,\CP_{B},\gamma_1}} \lrb{O_{\offdiag A,B}}\,.
    $}
 \label{eqn:diagonal_L_1}
\end{align}

Inspired by the above equation, we introduce another operator $\tilde{\mc{L}}_{\lrb{H, \beta, \mathcal{P}_B, \gamma_1}}\colon\mathcal{B} \lrb{\mathbb{C}^{2^{n}}} \to \mathcal{B} \lrb{\mathbb{C}^{2^{n}}} $,  hereafter abbreviated as $\tilde{\mc{L}}$: 
\begin{align}
    \tilde {\mathcal{L}}( O_{A,B} )\coloneqq
     \sum_{i_A} \ketbra{i_A}{i_A}\otimes 
 \lind_{(\bra{i_A} H \ket{i_A},\beta, \CP_B,\gamma_1)}\!\left(\bra{i_A}  O_{A,B}  \ket{i_A}\right)\,,
    \label{eqn:tilde_L_H1_def}
\end{align}
which is restriction of $ \lind_{\lrb{H,\beta,\CP_{B},\gamma_1}}
 \!\left( O_{A,B} \right)$ to the $A$-diagonal sector.
By construction, $\tilde{\mc{L}}$ acts only on the components of $O_{A,B}$ that are block diagonal in $A$ and annihilates the $A$-off-diagonal components, i.e., 
\begin{align}
    \tilde{\mathcal{L}} \lrb{\ketbra{ i_A }{i'_{A}}\otimes O_B}=0\,,
\end{align}
for any two different eigenvectors $\ket{ i_A }\neq \ket{i'_{A}}$  of $H_A$, and $O_B$ is any operator in $\mc{B}\lrb{\mathbb{C} ^{{d_B}}}$.
Therefore, the kernel of $\tilde{\mathcal{L}}$ is 
\begin{align}
\mc{K}:=&\ker \left(\tilde{\mathcal{L}}\right)\nonumber\\
=&\Span \left\{\ketbra{i_{A}}{i_{A}}\otimes I_{B},\left(\ketbra{i_{A}}{i'_{A}}\otimes O_B \right)\middle|\forall \ket{i_A} \neq \ket{i_A'} \text{ in \cref{eq:basis_A}},\, \forall O_B\in\mc{B}\lrb{\mathbb{C}^{{d_B}}} \right\}\,.  
\label{eqn:ker_mathcal_K}
\end{align}

To lower bound the spectral gap of
$\lind_{\lrb{H, \beta, \mathcal{P}_B, \gamma_1}} \otimes \BI_A \;+\; \BI_n \otimes \lind_{\lrb{I_A, \beta, \mathcal{P}_A, \gamma_2}} \;+\; \lind_{\lrb{H\otimes I_A+ I_n\otimes I_A, \beta, \{\CU\}, \gamma_M}}$, it suffices to lower bound the spectral gap of
\begin{align}
    \tilde{\mc{L}} \otimes \BI_A \;+\; \BI_n \otimes \lind_{\lrb{I_A, \beta, \mathcal{P}_A, \gamma_2}} \;+\; \lind_{\lrb{H\otimes I_A+ I_n\otimes I_A, \beta, \{\CU\}, \gamma_M}}\,.
\end{align}
This is because  $\lind_{\lrb{H, \beta, \mathcal{P}_B, \gamma_1}} \otimes \BI_A$ is invariant and {non-positive} in the following two subspaces: 
\begin{align}
\Span \left\{\ketbra{i_{A}}{i_{A}}\otimes O_{B}\middle|\forall O_B\in\mc{B}\lrb{\mathbb{C}^{{d_B}}} \right\},\,\,\Span \left\{\left(\ketbra{i_{A}}{i'_{A}}\otimes O_B \right)\middle|\forall \ket{i_A} \neq \ket{i_A'},\, \forall O_B\in\mc{B}\lrb{\mathbb{C}^{{d_B}}} \right\}\,.\
\nonumber
\end{align}

Using \cref{it:inner_B} of \cref{lem:gap_lemma}, one can get
\begin{align}
     -\tilde{\mc{L}} \otimes \BI_A
     - \BI_n \otimes \lind_{\lrb{I_A, \beta, \mathcal{P}_A, \gamma_2}} 
     - \lind_{\lrb{\{\CU\}}}
      \succeq \frac{\Gap\lrb{\tilde{\mc{L}} \otimes \BI_A
     +\BI_n \otimes \lind_{\lrb{I_A, \beta, \mathcal{P}_A, \gamma_2}}}  
    g_{\mc{U}} }
    {\Gap\lrb{\tilde{\mc{L}} \otimes \BI_A
     + \BI_n \otimes \lind_{\lrb{I_A, \beta, \mathcal{P}_A, \gamma_2}}}  
     + \lrnorm{\lind_{\lrb{\{\CU\}}}}_{\sigma} }\,,
    \label{eqn:all_L_gap}
\end{align}
where 
\begin{align}
    g_{\mc{U}} = \inf_{\substack{X \in 
    \ker\lrb{\tilde{\mc{L}} \otimes \BI_A
     + \BI_n \otimes \lind_{\lrb{I_A, \beta, \mathcal{P}_A, \gamma_2}}}\,, \\
     \lr{I,X}_\sigma =0}} 
    \frac{\lr{ X, - \lind_{\lrb{\{\CU\}}} \lrb{X} }_{\sigma}}
    {\lr{X,X}_\sigma}\,.
    \label{eqn:def_g_U}
\end{align}
Here, we emphasize that the orthogonality condition $\lr{I,X}_\sigma =0$ in the above equation originates from the first step, where we calculate the gap of $\lind$ (\cref{eq:total_lind_p2_759157}) by \cref{eq:gap_def}. 

To finish the proof, we employ bounds proven in \cref{sec:auxiliary_bounds}. By \cref{it:commut_AB} of \cref{lem:gap_lemma}, we get 
\begin{align}
\Gap\lrb{\tilde{\mc{L}} \otimes \BI_A
     +\BI_n \otimes \lind_{\lrb{I_A, \beta, \mathcal{P}_A, \gamma_2}}} 
     &\geq \min \lrc{\Gap\lrb{\tilde{\mc{L}}}, 
\Gap\lrb{\lind_{\lrb{I_A, \beta, \mathcal{P}_A, \gamma_2}}}}\, \\
&\ge \min\left\{ g_B, \Theta(1) \right\}. \qquad \text{(\cref{lem:gap_lemma_tilde_L_H1} and \cref{lem:gap_lemma_H2})}
\end{align}
Additionally, $\lrnorm{\lind_{\lrb{\{\CU\}}}}_{\sigma}\leq 3$ by \cref{lem:norm_L_swap} and
$g_{\mathcal{U}} = \Omega\lrb{\frac{1}{{2^{|A|} \exp(4 \beta KV_{\max})}}}$ by \cref{lem:gap_lemma_L_swap}. Putting these together, \cref{eqn:all_L_gap} becomes 
\begin{align}
      -\tilde{\mc{L}} \otimes \BI_A
     - \BI_n \otimes \lind_{\lrb{I_A, \beta, \mathcal{P}_A, \gamma_2}} 
     - \lind_{\lrb{\{\CU\}}}
     \succeq& \frac{\min\left\{ g_B, \Theta(1) \right\}\cdot \Omega\lrb{\frac{1}{{d_A \exp(4 \beta KV_{\max})}}}}{\min\left\{ g_B, \Theta(1) \right\}+3} \\
     =& \Omega\lrb{\frac{\min\left\{ g_B, 1 \right\}}{2^{|A|}  \exp(4 \beta KV_{\max})}} \,.
\end{align}
This concludes the proof.
\end{proof}

\section{Kernel characterization and important gap/norm bounds}\label{sec:auxiliary_bounds}

This section develops the kernel structure of \cref{eqn:ker_L_reduce_equal} and collects several important gap and norm bounds, which are applied in \cref{app:Gap_lower_bound}, while the detailed calculations are deferred to \cref{app:detailthreecases}.

\begin{lemma_} \label{lem:equal_ker}
    With the notation in \cref{eqn:ker_L_reduce_equal}, we have
    \begin{align}
    \ker\lrb{\mathcal{L}} &= \ker\lrb{\lind_{\lrb{H, \beta, \mathcal{P}_B, \gamma_1}} \otimes \BI_A \;+\; \BI_n \otimes \lind_{\lrb{I_A, \beta, \mathcal{P}_A, \gamma_2}} \;+\; \lind_{\lrb{H\otimes I_A+ I_n\otimes I_A, \beta, \{\CU\}, \gamma_M}}} 
    \label{eqn:ker_L_reduce_equal_proof}\\
    &=I_n\otimes I_A\,.
\end{align}
\end{lemma_}
\begin{proof}
By the construction of \cref{eqn:diagonal_L_1,eqn:tilde_L_H1_def}, we know that 
\begin{align}
     &\lind_{\lrb{H, \beta, \mathcal{P}_B, \gamma_1}} \otimes \BI_A \;+\; \BI_n \otimes \lind_{\lrb{I_A, \beta, \mathcal{P}_A, \gamma_2}} \;+\; \lind_{\lrb{H\otimes I_A+ I_n\otimes I_A, \beta, \{\CU\}, \gamma_M}} \nonumber\\
     \preceq\,\,&
      \tilde{\mc{L}} \otimes \BI_A \;+\; \BI_n \otimes \lind_{\lrb{I_A, \beta, \mathcal{P}_A, \gamma_2}} \;+\; \lind_{\lrb{H\otimes I_A+ I_n\otimes I_A, \beta, \{\CU\}, \gamma_M}}.
\end{align}
Because each term on the second line  of the above equation is non-positive, the kernel of the sum is the intersection of the kernels of each term. By \cref{prop:kernel_local_A_swap} we know that their intersection is identity, so the second line  of the above equation is primitive, deducing that  the first line  of the above equation is primitive. 

Comparing $\lind_{\lrb{H, \beta, \mathcal{P}_B, \gamma_1}} \otimes \BI_A \;+\; \BI_n \otimes \lind_{\lrb{I_A, \beta, \mathcal{P}_A, \gamma_2}} \;+\; \lind_{\lrb{H\otimes I_A+ I_n\otimes I_A, \beta, \{\CU\}, \gamma_M}}$ with $\mathcal{L}$ defined in \cref{eq:total_lind_p2_759157}, we can see their difference is a non-positive Lindbladian 
$\lind_{\lrb{H, \beta, \mathcal{P}_A, \gamma_1}} $, which means the kernel of $\mathcal{L}$ must be a subset of the kernel of $\lind_{\lrb{H, \beta, \mathcal{P}_B, \gamma_1}} \otimes \BI_A \;+\; \BI_n \otimes \lind_{\lrb{I_A, \beta, \mathcal{P}_A, \gamma_2}} \;+\; \lind_{\lrb{H\otimes I_A+ I_n\otimes I_A, \beta, \{\CU\}, \gamma_M}}$. Thus, the only element in the kernel of $\mathcal{L}$ is $I_n\otimes I_A$, which concludes the proof.
\end{proof}

\begin{lemma_} \label{lem:gap_lemma_H2}
For any operator $X\in \mc{B} \lrb{\mathbb{C} ^{{d_A}}}$, and any  $\gamma_2\in\lrc{\gamma_G,\gamma_M}$,
    \begin{align}
        \Gap \lrb{\lind_{\lrb{I_A, \beta, \mathcal{P}_A, \gamma_2}}} 
        = \inf_{X,\, \lr{I,X}_{\sigma_2}= 0} \frac{\lr{X, -\lind_{\lrb{I_A, \beta, \mathcal{P}_A, \gamma_2}}(X)}_{\sigma_2}}{\lr{X, X}_{\sigma_2}} =\Theta\lrb{1}\,,
    \end{align}
    where $\sigma_2 = \frac{I_A}{d_A}$.
\end{lemma_}
\begin{proof}
\rev{Since $I_A$ is proportional to the identity, $B(I_A)=\lrc{0}$ and
$e^{\im I_A t}S^a e^{-\im I_A t}=S^a$. Hence
$\hat S^a(\omega)=\hat f(\omega)S^a$. Moreover, the only pair of Bohr
frequencies in \cref{eq:G_def_CKG} is $\nu_1=\nu_2=0$, so the coherent
term $G_{\mathcal P_A}$ vanishes. Since every $S^a\in\mathcal P_A$ is a
single-site Pauli operator,
\begin{align}
\lind_{\lrb{I_A,\beta,\mathcal P_A,\gamma_2}}(X)
&=c_{\gamma_2}\sum_{i\in A}\sum_{\alpha\in\lrc{x,y,z}}
\lrb{\sigma_i^\alpha X\sigma_i^\alpha-X},
\label{eqn:identity_auxiliary_lindbladian}
\end{align}
where
\begin{align}
c_{\gamma_2}
&\coloneqq\int_{-\infty}^{\infty}\d\omega\,
\gamma_2(\omega)\abs{\hat f(\omega)}^2
=
\begin{cases}
\dfrac{e^{-1/4}}{\sqrt{2}}, & \gamma_2=\gamma_G,\\[2mm]
\operatorname{erfc}\lrb{\dfrac{1}{2\sqrt{2}}}, & \gamma_2=\gamma_M.
\end{cases}
\label{eqn:identity_auxiliary_constant}
\end{align}
Here the two values follow directly from
\cref{eq:gaussian_filter_CKG,eq:metropolis_weight_gamma} after the change
of variables $x=\beta\omega$. In particular, $c_{\gamma_2}$ is a positive
constant independent of $\beta$ and $\abs{A}$.

Let $Q=\bigotimes_{i\in A}Q_i$ be a Pauli string, so each $Q_i$ is one of
$I_i,\sigma_i^x,\sigma_i^y,\sigma_i^z$. Let $w(Q)$ be the number of
nonidentity factors of $Q$. For each $i\in A$,
\begin{align}
\sum_{\alpha\in\lrc{x,y,z}}\sigma_i^\alpha Q\sigma_i^\alpha
=
\begin{cases}
3Q, & Q_i=I_i,\\
-Q, & Q_i\neq I_i.
\end{cases}
\end{align}
It follows from \cref{eqn:identity_auxiliary_lindbladian} that
\begin{align}
\lind_{\lrb{I_A,\beta,\mathcal P_A,\gamma_2}}(Q)
=-4c_{\gamma_2}w(Q)Q.
\end{align}

Because $\sigma_2=I_A/d_A$, the KMS inner product is the normalized
Hilbert--Schmidt inner product, under which the Pauli strings form an
orthonormal basis. Thus, for any
$X=\sum_Q x_QQ$ satisfying $\lr{I_A,X}_{\sigma_2}=0$, we have
$x_{I_A}=0$ and
\begin{align}
\frac{\lr{X,-\lind_{\lrb{I_A,\beta,\mathcal P_A,\gamma_2}}(X)}_{\sigma_2}}
{\lr{X,X}_{\sigma_2}}
=\frac{\sum_{Q\neq I_A}4c_{\gamma_2}w(Q)\abs{x_Q}^2}
{\sum_{Q\neq I_A}\abs{x_Q}^2}
\geq 4c_{\gamma_2}.
\end{align}
Equality is attained by any weight-one Pauli string. Therefore,
\begin{align}
\Gap\lrb{\lind_{\lrb{I_A,\beta,\mathcal P_A,\gamma_2}}}
&=4c_{\gamma_2}\nonumber\\
&=\Theta\lrb{1}.
\end{align}}

\end{proof}

{
\begin{lemma_} \label{lem:gap_lemma_tilde_L_H1}
    Under \cref{assumption:fast_mixing}, we have 
    \begin{align}
        \Gap\left(\tilde{\mathcal{L}}\right)
        &\coloneqq \inf_{\substack{ O_{A,B} \neq 0,\\
         O_{A,B}  \in \ker\lrb{\tilde{\lind}}^\perp}}
         \frac{\left\langle  O_{A,B} ,-\tilde{\mathcal{L}}( O_{A,B} )\right\rangle_{ \sigma_H }}{\left\langle  O_{A,B} , O_{A,B} \right\rangle_{ \sigma_H }} \geq g_B \,,
         \label{eq:gap_tilde_L_def_47}
    \end{align}
    where $\tilde{\mathcal{L}}$ is defined in \cref{eqn:tilde_L_H1_def}, and $\sigma_{H}= \frac{\exp(-\beta H)}{\Tr\lrs{\exp(-\beta H)}}$.
\end{lemma_}
}
\begin{proof}
    By \cref{assumption:fast_mixing}, we know the spectral gap of ${\lind_{\lrb{\bra{i_A} H \ket{i_A}, \beta, \CP_B, \gamma_1}}}$ is  at least $g_B$, i.e.,
    \begin{align}
        \inf_{\substack{O_{B}\neq 0\,,\\
        \lr{I_B,O_B}_{\sigma _{\ket{i_A}}}} = 0}
        \frac{\lr{O_B ,-{\lind_{\lrb{\bra{i_A} H \ket{i_A}, \beta, \CP_B, \gamma_1}}} (O_B)}_{\sigma_{\ket{i_{A}}}}}{\lr{O_B,O_B}_{\sigma_{\ket{i_{A}}}}}
        \geq g_B \,, \quad \forall \ket{i_A} \text{ in \cref{eq:basis_A}},
    \end{align}
    with $\sigma_{\ket{i_A}} \propto (\bra{i_A}\otimes I_B)\,e^{-\beta H}\,(\ket{i_A}\otimes I_{B}) $ given by \cref{eqn:fixed_point_L_B_iA}.
    Expand the above equation by KMS inner product:
    \begin{align}
        \inf_{\substack{O_{B}\neq 0\,,\\
        \lr{I_B,O_B}_{\sigma _{\ket{i_A}}}} = 0}
        \frac{-\Tr \lrs{  \sigma_{\ket{i_A}} ^{1/2} \cdot  O_B ^\dagger \cdot  \sigma_{\ket{i_A}} ^{1/2} 
        \cdot {\lind_{\lrb{\bra{i_A} H \ket{i_A}, \beta, \CP_B, \gamma_1}}}(O_B) }}
        {\Tr \lrs{\sigma_{\ket{i_A}} ^{1/2}  O_B ^\dagger  \sigma_{\ket{i_A}} ^{1/2} O_B }}
        \geq g_B\,.
        \label{eqn:L_gap_B_lower_bound}
    \end{align}
    Next, we write down the definition of $\Gap\lrb{\tilde{\mathcal{L}}}$ by the inner product with respect to $ \sigma_H  = \frac{\exp(-\beta H)}{\Tr\lrs{\exp(-\beta H)}}$ as in \cref{eq:gap_tilde_L_def_47}, and 
    then expand the right-hand side of it according to \cref{eqn:tilde_L_H1_def}, trying to connect it with the assumption in \cref{eqn:L_gap_B_lower_bound} to obtain a lower bound for $\Gap\lrb{\tilde{\mc{L}}}$: 
    \begin{align}
        &\frac{-\Tr\lrs{\exp(-\beta H) ^{1/2}  O_{A,B} ^\dagger \exp(-\beta H)^{1/2} \tilde{\mathcal{L}}( O_{A,B} ) }}
        {\Tr\lrs{\exp(-\beta H) ^{1/2}  O_{A,B} ^\dagger \exp(-\beta H) ^{1/2}  O_{A,B}  }} \nonumber\\
        =& \resizebox{0.88\linewidth}{!}{$
        \frac{-\Tr\lrs{\exp(-\beta H) ^{1/2}  O_{A,B} ^\dagger \exp(-\beta H) ^{1/2} 
        \cdot \sum_{\ket{i_A}} \ketbra{i_A}{i_A}
        \otimes  \lind_{(\bra{i_A} H \ket{i_A},\beta, \CP_B,\gamma_1)}\!\left(\bra{i_A}  O_{A,B}  \ket{i_A}\right)}}
        {\Tr\lrs{\exp(-\beta H) ^{1/2}  O_{A,B} ^\dagger \exp(-\beta H) ^{1/2}  O_{A,B} }}
        $}\\
        =& \frac{- \Tr _B \left[ 
            \begin{aligned}
            &\sum_{m_A} \bra{m_A} \otimes I_B \exp(-\beta H) ^{1/2}  O_{A,B}  ^\dagger \exp(-\beta H) ^{1/2} \\
        &\cdot \sum_{\ket{i_A}}  \ketbra{i_A}{i_A}
        \otimes \ \lind_{(\bra{i_A} H \ket{i_A},\beta, \CP_B,\gamma_1)}\!\left(\bra{i_A}  O_{A,B}  \ket{i_A}\right)
        \ket{m_A} \otimes I_B 
        \end{aligned}
        \right]}
        {\Tr_B 
        \lrs{ \sum_{i_A} \bra{i_A} \otimes I_B \exp(-\beta H) ^{1/2}  O_{A,B} ^\dagger \exp(-\beta H) ^{1/2}  O_{A,B}  \ket{i_A} \otimes I_B}}\\
        =& \frac{-\sum _{i_A} \Tr_B\lrs{ \sigma_{\ket{i_A}} ^{1/2}\bra{i_A}  O_{A,B} ^\dagger \ket{i_A}\sigma_{\ket{i_A}} ^{1/2} \lind_{(\bra{i_A} H \ket{i_A},\beta, \CP_B,\gamma_1)}\left(\bra{i_A}  O_{A,B}  \ket{i_A}\right)  }}
        {\Tr_B 
        \lrs{ \sum_{i_A}\lrb{ \bra{i_A} \otimes I_B} \exp(-\beta H) ^{1/2}  O_{A,B} ^\dagger \exp(-\beta H) ^{1/2}  O_{A,B} \lrb{ \ket{i_A} \otimes I_B}}}\,,
        \label{eqn:gap_tilde_L_123}
    \end{align}
    where $\Tr_B[\cdot ]$ represents the partial trace over the $B$ subsystem.
    In the last equality, the numerator is simplified because $\exp(-\beta H)^{1/2}$ is diagonal in the $\ket{i_Aj_B}$ basis\rev{; consequently, the sum over $m_A$ reduces to the single nonzero term with $m_A=i_A$}. 

    According to \cref{eq:gap_tilde_L_def_47}, we only consider $ O_{A,B}  \in \ker\lrb{\tilde{\lind}}^\perp$, where the orthogonal complement $^\perp$ is defined with respect to the $\lr{\cdot,\cdot}_{\sigma_H}$ inner product. 
Thus, \cref{eqn:ker_mathcal_K} implies  that 
\begin{align}
\mathcal K^{\perp}
  &= \operatorname{span}\Bigl\{
        \ketbra{i_A}{i_A}\otimes O_B
        \;\Big|\;\forall \ket{i_A}\, , 
        \lr{I_B,O_B} _{\sigma_{\ket{i_A}}}=0
     \Bigr\} \,.
    \label{eqn:ker_mathcal_K_perp}
\end{align}
We prove it \rev{as follows}: Let \(
\sigma_H
    =\sum_{i_A,m_B} s_{i_A,m_B}\,
      \ketbra{i_A,m_B}{i_A,m_B},
\)
and $s_{i_A,m_B}>0$.  Any operator in \(\mathcal B\bigl(\mathbb C^{d_A}\!\otimes\!\mathbb C^{d_B}\bigr)\) can be expanded by
\begin{align}
X
   \;=\;
   \sum_{i,j=1}^{d_A} \ketbra{i_A}{j_A}\otimes X_{ij},
   \qquad
   X_{ij}\in\mathcal B\bigl(\mathbb C^{d_B}\bigr).
\label{eq:generic_X}
\end{align}
Then, for any \(i_A\neq j_A\) and any \(O_B\in\mathcal B\bigl(\mathbb C^{d_B}\bigr)\),
\begin{align}
0
   &=\bigl\langle
        X,\;
        \ketbra{i_A}{j_A}\otimes O_B
     \bigr\rangle_{\sigma_H}
  \\
&=\!\!\sum_{m,n=1}^{d_B}
     \sqrt{s_{j_A,n}\,s_{i_A,m}}\;
     (X_{ij})_{mn}^\ast\,
     (O_B)_{mn}.
\end{align}
Because \(O_B\) is arbitrary, every coefficient must vanish, so
$
X_{ij}=0
\,, \forall\,i\neq j.
$
Next, consider
\(
X=\ketbra{i_A}{i_A}\otimes O_B,
\)
and for each \(i_A\)
\begin{align}
0
  &=\bigl\langle
       X,\;
       \ketbra{i_A}{i_A}\otimes I_B
    \bigr\rangle_{\sigma_H}
  \\
  &=\Tr [\sigma_{\ket{i_A} } O_B^\dagger] = \lr{O_B,I_B}_{\sigma_{\ket{i_A} }}\,.
\end{align}
Therefore, we prove \cref{eqn:ker_mathcal_K_perp}.

    Thus, for $ O_{A,B}  \in \mc {K}^{\perp}$, one can write the denominator of \cref{eqn:gap_tilde_L_123} as 
    \begin{align}
        &\lrb{\bra{i_A} \otimes I_B }\exp(-\beta H) ^{1/2}  O_{A,B} ^\dagger \exp(-\beta H) ^{1/2}  O_{A,B}  \lrb{\ket{i_A} \otimes I_B}\nonumber\\
        =& \lrb{\bra{i_A} \otimes I_B }\exp(-\beta H) ^{1/2} \lrb{\ket{i_A} \otimes I_B} \cdot
        \lrb{\bra{i_A} \otimes I_B}  O_{A,B}  ^\dagger \lrb{\ket{i_A} \otimes I_B}\nonumber\\
        &\cdot
        \lrb{\bra{i_A} \otimes I_B} \exp(-\beta H) ^{1/2} \lrb{\ket{i_A} \otimes I_B }
        \cdot\lrb{ \bra{i_A} \otimes I_B} O_{A,B} \lrb{\ket{i_A} \otimes I_B} \\
        =&\sigma_{\ket{i_A}} ^{1/2} \bra{i_A}  O_{A,B} ^\dagger \ket{i_A} \sigma_{\ket{i_A}} ^{1/2} \bra{i_A}  O_{A,B} \ket{i_A} \,,
    \end{align} 
    where we also use $ O_{A,B} $ is block diagonal in the $\ket{i_A}$ basis as illustrated by \cref{eqn:ker_mathcal_K_perp}, and $\exp(-\beta H)^{1/2}$ is also block diagonal in the same basis.
    Inserting the above equation into the denominator of \cref{eqn:gap_tilde_L_123} and combining with \cref{eqn:L_gap_B_lower_bound}, \cref{lem:ratio_inequality} implies that  $\Gap\lrb{\tilde{\mathcal{L}}}  \geq g_B$. 
\end{proof}

\begin{lemma_}\label{lem:norm_L_swap}
    Abbreviate $\lind_{\lrb{H\otimes I_A+ I_n\otimes I_A, \beta, \{\CU\}, \gamma_M}}$ as $\mathcal{L}_{\lrb{\{\CU\}}}$, then 
\begin{align}
    \lrnorm{\mathcal{L}_{\lrb{\{\CU\}}}}_{\sigma}  = \sup_{
    X\neq 0}
    \frac{\lr{X, -\mathcal{L}_{\lrb{\{\CU\}}}(X)}_{\sigma} }
    { \lr{X, X}_{\sigma} } \leq 3\,,
\end{align}
where $\sigma = \frac{\exp(-\beta H)}{\Tr\lrs{\exp(-\beta H)}} \otimes \frac{I_A}{d_A}$. 
\end{lemma_}
\begin{proof}
    We begin by bounding the second term of \cref{eq:local_swap_Lindbladian} after the KMS inner product, i.e., $ \lr{ X, \lrc{\int \d\omega \gamma_M(\omega) \hat{\mc{U}} ^\dagger \hat{\mc{U}}, X }}_{\sigma}$:
\begin{align}
&\int \d \omega {\gamma_M (\omega)} \mathrm{Tr}\left[\sigma^{1/2} 
X^\dagger
\sigma^{1/2}\left\{\hat{\mc{U}}^\dagger(\omega)\hat{\mc{U}}(\omega),X\right\}\right] \nonumber\\
= &\int \d \omega {\gamma_M (\omega)} \sum_{i_A,j_B,m_A} \abs{\hat{f}(\omega - \lambda_{i_A,j_B,m_A} + \lambda_{m_A,j_B,i_A})}^2
\cdot \mathrm{Tr}\left[\sigma^{1/2} 
X^\dagger
\sigma^{1/2}
\left\{ \ketbra{m_A j_B i_A} {m_A j_Bi_A},X\right\}\right] \nonumber\\
\leq& 2 \sum_{i_A,j_B,m_A} 
\mathrm{Tr}\left[\sigma^{1/2} 
X^\dagger
\sigma^{1/2}
\left\{ \ketbra{m_A j_B i_A}{m_A j_B i_A},X\right\}\right]\\
= & 2\Tr\lrs{  \sigma^{1/2} X^\dagger \sigma^{1/2} \lrc{I_n \otimes I_A,X}} \\
=& 4 \lr{X,X}_{\sigma}\,.
\label{eqn:UdaggerU_X}
\end{align}
Here, the first line uses \cref{eq:U_hat_dagger_U_hat}. The second line follows from \cref{eq:theta_bounds}, and the third line comes from 
\begin{align}
    \sum_{i_A,j_B,m_A} \ketbra{m_A j_B i_A}{m_A j_B i_A} = I_n \otimes I_A\,.
\end{align}

Next, we try to bound the first term of Lindbladian, i.e.,  $\lr{X, \int \d\omega \gamma_M(\omega) \hat{\mc{U}} X \hat{\mc{U}}^\dagger}_{\sigma}$ in \cref{eq:local_swap_Lindbladian}\rev{. Recall H\"older's inequality for the KMS inner product.}
\begin{align}
\left|\lr{A, B}_{\sigma}\right|=\left|\mathrm{Tr}(\sigma^{1/2}A^\dagger\sigma^{1/2} B) \right|\leq \|A\|_{\sigma}\cdot \|B\|_{\sigma}\,.
\end{align}
By setting $A=X$ and $B=\int \d \omega \gamma_M(\omega) \hat{\mathcal{U}}^\dagger(\omega)X\hat{\mathcal{U}}(\omega)$, from the above equation we obtain
\begin{align}
    &\abs{\lr{X, \int \d\omega \gamma_M(\omega) \hat{\mc{U}}^\dagger  X \hat{\mc{U}}}_{\sigma}}\nonumber \\
    =&\left|\mathrm{Tr}\left( \sigma^{1/2}  X^\dagger \sigma^{1/2} 
     \int \d\omega \gamma_M(\omega) \hat{\mathcal{U}}^\dagger  X\hat{\mathcal{U}} \right)\right| \\
    \leq &\resizebox{0.88\linewidth}{!}{$
    \sqrt{\mathrm{Tr}\left[ \sigma^{1/2} X^\dagger \sigma^{1/2} X\right]}\cdot 
    \sqrt {\mathrm{Tr}\left[\sigma^{1/2} \lrb{\int \d \nu \gamma_M(\nu)\hat{\mathcal{U}}^\dagger(\nu)  X\hat{\mathcal{U}}(\nu) }^\dagger \sigma^{1/2} \int \d \omega \gamma_M(\omega) \hat{\mathcal{U}}^\dagger(\omega ) X\hat{\mathcal{U}}(\omega )\right]}\,.
     $}
\end{align}
Then apply the fact that $2ab\leq a^2+b^2$ to the above equation, deriving that
\begin{align}
&\abs{\lr{X, \int \d\omega \gamma_M(\omega) \hat{\mc{U}}^\dagger  X \hat{\mc{U}}}_{\sigma}} \nonumber\\
\leq &
\resizebox{0.88\linewidth}{!}{$
 \frac{1}{2} 
\mathrm{Tr}\left( \sigma^{1/2} X^\dagger \sigma^{1/2} X\right)
+\frac{1}{2}
   \mathrm{Tr}\left[\sigma^{1/2} \lrb{\int \d \nu \gamma_M(\nu)\hat{\mathcal{U}}^\dagger(\nu)  X \hat{\mathcal{U}} (\nu)}^\dagger \sigma^{1/2} \int \d \omega \gamma_M(\omega) \hat{\mathcal{U}}^\dagger(\omega ) X\hat{\mathcal{U}}(\omega )\right]
$}.
   \label{eq:norm_LU_p113}
\end{align}
Note that the first term of the right-hand side is $\frac{1}{2}\lr{X,X}_{\sigma}$, so the remaining task is to bound the second term:
\begin{align}
    &\mathrm{Tr}\left[\sigma^{1/2} \lrb{\int \d \nu \gamma_M(\nu)\hat{\mathcal{U}}^\dagger(\nu)  X\hat{\mathcal{U}}(\nu) }^\dagger \sigma^{1/2} \int \d \omega \gamma_M(\omega) \hat{\mathcal{U}}^\dagger(\omega ) X\hat{\mathcal{U}}(\omega )\right]\nonumber \\
     =& \int \d \nu \gamma_M(\nu) \int \d \omega \gamma_M(\omega)
    \Tr \left[\sigma^{1/2} {\hat{\mathcal{U}}^\dagger(\nu)  X^\dagger \hat{\mathcal{U}}(\nu) }\sigma^{1/2} \hat{\mathcal{U}}^\dagger(\omega ) X\hat{\mathcal{U}}(\omega )\right]\\
    =&\int \d \nu \gamma_M(\nu) \int \d \omega \gamma_M(\omega)
    \Tr \left[\lrb{\hat{\mathcal{U}}(\omega ) \sigma^{1/2} \hat{\mathcal{U}}^\dagger(\nu)}  
    X^\dagger 
    \lrb{\hat{\mathcal{U}}(\nu) \sigma^{1/2} \hat{\mathcal{U}}^\dagger(\omega )} X\right]\,.
    \label{eq:norm_LU_p14}
\end{align}

Denote the Gibbs state $\sigma$ of $H\otimes I_A+ I_n\otimes I_A$ as 
\begin{align}
    \sigma =\sum_{i_A,j_B,m_A} s_{i_A, j_B, m_A}\, \ketbra{i_A j_B m_A}{i_A j_B m_A}\,,
\end{align}
where 
\begin{align}
    s_{i_A,j_B,m_A} = \frac{e^{-\beta \lambda_{i_A,j_B,m_A}}}{\Tr\lrs{e^{-\beta \lrb{H\otimes I_A+ I_n\otimes I_A}}}}\,.
    \label{eqn:Gibbs_state_H}
\end{align}

Expand $\hat{\mathcal{U}}(\omega ) \sigma^{1/2} \hat{\mathcal{U}}^\dagger(\nu)$ by \cref{eqn:replica_exchange_davies_local_swap_general_1}:
\begin{align}
    &\hat{\mathcal{U}}(\omega ) \sigma^{1/2} \hat{\mathcal{U}}^\dagger(\nu) \nonumber\\
    =& \sum_{i_A,j_B,m_A} \hat{f}\lrb{\omega - \lambda_{i_A,j_B,m_A} + \lambda_{m_A,j_B,i_A}} \ketbra{i_A j_B m_A}{m_A j_B i_A} \nonumber\\
    &\cdot \sum_{x_A,y_B,z_A} \sqrt{s_{x_A, y_B,z_A}} \ketbra{x_Ay_Bz_A}{x_Ay_Bz_A} \nonumber\\
    &\cdot \sum_{k_A,q_B,l_A} \hat{f}\lrb{\nu - \lambda_{k_A,q_B,l_A} + \lambda_{l_A,q_B,k_A}} \ketbra{l_A q_B k_A}{k_A q_B l_A} \,.
\end{align}
Noticing that the summands are non-zero only if $m_A = x_A = l_A$, $j_B = y_B = q_B$, and 
$i_A =z_A =  k_A$, the above equation becomes
\begin{align}
    &\hat{\mathcal{U}}(\omega ) \sigma^{1/2} \hat{\mathcal{U}}^\dagger(\nu) \nonumber\\
    =& \resizebox{0.88\linewidth}{!}{$
    \sum_{i_A,j_B,m_A} \sqrt{s_{m_A,j_B,i_A}} 
    \hat{f}(\omega - \lambda_{i_A,j_B,m_A} + \lambda_{m_A,j_B,i_A}) 
    \hat{f}(\nu - \lambda_{i_A,j_B,m_A} + \lambda_{m_A,j_B,i_A}) 
    \ketbra{i_A j_B m_A}{i_A j_B m_A} \,.
    $}
\end{align}
Similarly, one can get
\begin{align}
    &\hat{\mathcal{U}}(\nu ) \sigma^{1/2} \hat{\mathcal{U}}^\dagger(\omega) \nonumber\\
    =& \resizebox{0.88\linewidth}{!}{$
    \sum_{i_A,j_B,m_A} \sqrt{s_{m_A,j_B,i_A}} 
    \hat{f}(\nu - \lambda_{i_A,j_B,m_A} + \lambda_{m_A,j_B,i_A}) 
    \hat{f}(\omega - \lambda_{i_A,j_B,m_A} + \lambda_{m_A,j_B,i_A}) 
     \ketbra{i_A j_Bm_A}{i_A j_Bm_A}  \,.
     $}
\end{align}
Inserting the above two equations into \cref{eq:norm_LU_p14}, we have
\begin{align}
    &\mathrm{Tr}\left[\sigma^{1/2} \lrb{\int \d \nu \gamma_M\lrb{\nu}\hat{\mathcal{U}}^\dagger\lrb{\nu}  X\hat{\mathcal{U}}\lrb{\nu} }^\dagger \sigma^{1/2} \int \d \omega \gamma_M\lrb{\omega} \hat{\mathcal{U}}^\dagger\lrb{\omega} X\hat{\mathcal{U}}\lrb{\omega}\right] \nonumber\\
    =& \int \d \nu \gamma_M\lrb{\nu} \int \d \omega \gamma_M\lrb{\omega}\nonumber\\
    &\sum_{i_A,j_B,m_A} \sqrt{s_{m_A,j_B,i_A}} 
    \hat{f}\lrb{\omega - \lambda_{i_A,j_B,m_A} + \lambda_{m_A,j_B,i_A}} 
    \hat{f}\lrb{\nu - \lambda_{i_A,j_B,m_A} + \lambda_{m_A,j_B,i_A}} \nonumber\\
    & \cdot \sum_{x_A,y_B,z_A}  \sqrt{s_{z_A,y_B,x_A}} 
    \hat{f}\lrb{\nu - \lambda_{x_A,y_B,z_A} + \lambda_{z_A,y_B,x_A}} 
    \hat{f}\lrb{\omega - \lambda_{x_A,y_B,z_A} + \lambda_{z_A,y_B,x_A}} \nonumber\\
    &\cdot \mathrm{Tr} \left[ \ketbra{i_A j_B m_A}{i_A j_B m_A} X^\dagger \ketbra{x_A y_B z_A}{x_A y_B z_A} X \right] \\
    =& \sum_{\substack{i_A,j_B,m_A\\ x_A,y_B,z_A}} 
    \int \d \nu \gamma_M\lrb{\nu} 
    \hat{f}\lrb{\nu - \omega_{i_A,j_B,m_A}} 
    \hat{f}\lrb{\nu - \omega_{x_A,y_B,z_A}} \nonumber\\
    &\cdot \int \d \omega \gamma_M\lrb{\omega}
    \hat{f}\lrb{\omega - \omega_{i_A,j_B,m_A}} 
    \hat{f}\lrb{\omega - \omega_{x_A,y_B,z_A}} \nonumber\\
    &\cdot \sqrt{s_{m_A,j_B,i_A} s_{z_A,y_B,x_A}}
    \bra{i_A j_B m_A} X^\dagger \ket{x_A y_B z_A}\bra{x_A y_B z_A} X \ket{i_A j_B m_A} \\
    =& \sum_{\substack{i_A,j_B,m_A\\ x_A,y_B,z_A}} 
    \lrb{\int \d \omega \gamma_M\lrb{\omega} \hat{f}\lrb{\omega - \omega_{i_A,j_B,m_A}} \hat{f}\lrb{\omega - \omega_{x_A,y_B,z_A}}}^2 \nonumber\\
    &\cdot \sqrt{s_{m_A,j_B,i_A} s_{z_A,y_B,x_A}}
    \bra{i_A j_B m_A} X^\dagger \ket{x_A y_B z_A}\bra{x_A y_B z_A} X \ket{i_A j_B m_A}\,,
    \label{eq:norm_LU_p31}
\end{align}
Here, $\omega_{i_A,j_B,m_A} \coloneqq \lambda_{i_A,j_B,m_A} - \lambda_{m_A,j_B,i_A}$. 

Then we will compare the above expression with $\lr{X,X}_{\sigma}$:
\begin{align}
    &\lr{X,X}_{\sigma} \nonumber\\
    =& \Tr\lrs{\sigma^{1/2} X^\dagger \sigma^{1/2} X}\\
    =&\resizebox{0.88\linewidth}{!}{$
    \Tr\lrs{\sum_{i_A,j_B,m_A} \sqrt{s_{i_A, j_B, m_A}} \ket{i_Aj_Bm_A}\bra{i_Aj_Bm_A} X^\dagger 
    \sum_{x_A,y_B,z_A} \sqrt{s_{x_A, y_B, z_A}} \ket{x_Ay_Bz_A}\bra{x_Ay_Bz_A} X}
    $}
    \\
    =&\sum_{\substack{i_A,j_B,m_A\\ x_A,y_B,z_A}} 
    \sqrt{s_{i_A, j_B, m_A}s_{x_A, y_B, z_A}}
    \bra{i_Aj_Bm_A} X^\dagger \ket{x_Ay_Bz_A}\bra{x_Ay_Bz_A} X \ket{i_Aj_Bm_A}
    \label{eq:norm_LU_p32}
\end{align}

Comparing \cref{eq:norm_LU_p31} and \cref{eq:norm_LU_p32}, we find that they have the same structure $$\bra{i_Aj_Bm_A} X^\dagger \ket{x_Ay_Bz_A}\bra{x_Ay_Bz_A} X \ket{i_Aj_Bm_A} \geq 0$$ in each summand, and the only difference is the coefficient before each term. With the help of \cref{eqn:Gibbs_state_H}, we can notice that 
\begin{align}
    \frac{\sqrt{s_{i_A, j_B, m_A}s_{x_A, y_B, z_A}}}
    {\sqrt{s_{m_A,j_B,i_A} s_{z_A,y_B,x_A}} }  
    =
    \sqrt{e^{-\beta {\omega_{i_A,j_B,m_A}}} \cdot e^{-\beta {\omega_{x_A,y_B,z_A}}}}
\end{align}
The above equation, together with \cref{lem:property_erfc_p21}, demonstrate that, for any fixed $i_A,j_B,m_A,x_A,y_B,z_A$, 
\begin{align}
\resizebox{0.91\linewidth}{!}{$
     \lrb{\int \d \omega {\gamma_M  (\omega)} \hat{f}  \lrb{\omega -\omega_{i_A,j_B,m_A}}  
    \hat{f}(\omega - \omega_{x_A,y_B,z_A})}^2  
    \sqrt{s_{m_A,j_B,i_A}  s_{z_A,y_B,x_A}}  \leq \sqrt{s_{i_A, j_B, m_A}s_{x_A, y_B, z_A}}\,.
    $}
\end{align}
Using the above inequality in \cref{eq:norm_LU_p31,eq:norm_LU_p32}, and combining with 
$$\bra{i_Aj_Bm_A} X^\dagger \ket{x_Ay_Bz_A}\bra{x_Ay_Bz_A} X \ket{i_Aj_Bm_A} \geq 0\,,$$ we know that each summand in \cref{eq:norm_LU_p31} is not larger than the corresponding summand in \cref{eq:norm_LU_p32}. Thus, we conclude that \cref{eq:norm_LU_p31} is not larger than \cref{eq:norm_LU_p32}. Substituting this result back into \cref{eq:norm_LU_p113}, we have
\begin{align}
    \abs{\lr{X, \int \d\omega \gamma_M(\omega) \hat{\mc{U}}^\dagger  X \hat{\mc{U}}}_{\sigma}} 
\leq \Tr\lrs{ \sigma^{1/2}  X^\dagger \sigma^{1/2}  X  } =\lr{X,X}_{\sigma}\,.
\label{eq:bngqoguy}
\end{align}
Finally, combining \cref{eqn:UdaggerU_X} with the above results, we have 
\begin{align}
    \lr{X, - \mathcal{L}_{\lrb{\lrc{\mc{U}}} }(X)}_{\sigma} \leq 3 \lr{X,X}_{\sigma}\,.
\end{align}

\end{proof}

We now prove the key bound on the gap $g_{\mc{U}}$. 
Using the structure of the local swap $\mc{U}$ defined in \cref{eqn:swap_operator_local}, which exchanges states on subsystem $A$ while leaving $B$ invariant, we observe that if $X$ is block-diagonal (resp.\ off-diagonal) with respect to $A$ or $B$, then $\mathcal{L}_{\lrb{\lrc{\mc{U}}}}(X)$ preserves this structure. Since the inner product $\lr{\cdot, \cdot}_{\sigma}$ between a diagonal and an off-diagonal matrix is zero, the cross terms vanish. This motivates a case analysis: we estimate the gap separately on the block-diagonal and off-diagonal sectors for $A$ and for $B$. The lemma below follows this plan and rigorously justifies that it suffices to bound each sector independently.
\begin{lemma_}\label{lem:gap_lemma_L_swap}
    For $g_{\mc{U}}$ defined in \cref{eqn:def_g_U}, we have 
\begin{align}
    g_{\mc{U}} = \inf_{\substack{X \in 
    \ker\lrb{\tilde{\mc{L}} \otimes \BI_A
     + \BI_n \otimes \lind_{\lrb{I_A, \beta, \mathcal{P}_A, \gamma_2}}}\,, \\
     \lr{I,X}_\sigma =0}} 
    \frac{\lr{ X, - \lind_{\lrb{\{\CU\}}} \lrb{X} }_{\sigma}}
    {\lr{X,X}_\sigma}
     = \Omega\lrb{\frac{1}{{d_A \exp(4 \beta KV_{\max})}}}\,,
     \label{eqn:gap_lemma_L_swap}
\end{align}
where $K$ and $V_{\max}$ are those introduced in \cref{assumption:H_decomposition}, and $\beta$ is the inverse temperature.
\end{lemma_}
\begin{proof}
Because 
$\tilde{\mathcal{L}}$ in \cref{eqn:tilde_L_H1_def} and $\lind_{\lrb{I_A, \beta, \mathcal{P}_A, \gamma_2}}$ are non-positive operators, 
\begin{align}
X\in \ker\lrb{\tilde{\mathcal{L}}\otimes \BI_A + \BI_n \otimes \lind_{\lrb{I_A, \beta, \mathcal{P}_A, \gamma_2}}} 
&= \ker\lrb{ \tilde{\mathcal{L}}\otimes \BI_A  }    
\cap
\ker\lrb{ \BI_n\otimes \lind_{\lrb{I_A, \beta, \mathcal{P}_A, \gamma_2}} } \\
& = \ker\lrb{\tilde{\mathcal{L}}} \otimes \ker\lrb{\lind_{\lrb{I_A, \beta, \mathcal{P}_A, \gamma_2}}}  \\
& =  \ker\lrb{\tilde{\mathcal{L}}} \otimes I_A \,,
\end{align}
where the last equality uses $\ker\lrb{\lind_{\lrb{I_A, \beta, \mathcal{P}_A, \gamma_2}} } = I_A$ stated in \cite[Section 5]{gilyen2024quantum} \cite[Section 3]{ding2024polynomial}. Here, 
\begin{align}
\mc{K}\coloneqq &\ker \left(\tilde{\mathcal{L}}\right)\ \
=&\Span \left\{\ketbra{i_{A}}{i_{A}}\otimes I_{B},\left(\ketbra{i_{A}}{i'_{A}}\otimes O_B \right)\middle|\forall \ket{i_A} \neq \ket{i_A'},\, \forall O_B\in\mc{B}\lrb{\mathbb{C}^{{d_B}}} \right\}\,,  
\end{align}
is given by \cref{eqn:ker_mathcal_K}. For later reference, we decompose $\mathcal{K}$ into two disjoint subspaces $\mathcal{K} _{\diag A}$ and $\mathcal{K} _{\offdiag A}$ with 
$\mathcal{K} =  \mathcal{K} _{\diag A} \oplus \mathcal{K} _{\offdiag A}$. 
Here,
$\mathcal{K} _{\diag A}$ is the subspace of $\mathcal{K}$ consisting of operators that are diagonal in the $A$ subsystem:
\begin{align}
\mathcal{K} _{\diag A} \coloneqq \Span \lrc{\ketbra{i_{A}}{i_{A}}\otimes I_B \middle| \forall \ket{i_{A}} \text{ in  \cref{eq:basis_A}}}\,,
\label{eqn:diag_A_def38}
\end{align}
and $\mathcal{K} _{\offdiag A}$ is the subspace of $\mathcal{K}$ consisting of operators that are off-diagonal in the $A$ subsystem:
\begin{align}
\mathcal{K} _{\offdiag A} \coloneqq \Span \lrc{\ketbra{i_{A}}{i'_{A}}\otimes O_B \big | \forall \ket{i_A}\neq \ket{i_A'} \text{ in  \cref{eq:basis_A}} ,\,\forall O_B\in\mc{B}\lrb{\mathbb{C}^{{d_B}}}  }\,.
\label{eqn:offdiag_A_def39}
\end{align}

Write $X$ as $X=  O_{A,B} \otimes I_A \in \mathcal{K} \otimes I_A$. 
Like what we did in \cref{eqn:O_AB_decomp_i378}, we
    decompose $ O_{A,B} $ into diagonal and off-diagonal parts in the $A$ \rev{subsystem}, i.e.,
    \begin{align}
        X = O_{\diag A,B} \otimes I_A + O_{\offdiag A,B} \otimes I_A\,,
        \label{eq:X_decomp_diag_offdiagA_182}
    \end{align}
    where $ O_{\diag A,B} = \sum_{{ i_A }}\ketbra{ i_A }{ i_A }\otimes \bra{i_A}  O_{A,B}  \ket{i_A}$ as \cref{eqn:O_AB_decomp_i378}. 
Meanwhile, $O_{\diag A,B} \in \mathcal{K}_{\diag A}$ and $O_{\offdiag A,B} \in \mathcal{K}_{\offdiag A}$.

We then use the above expression of $X$ to expand \cref{eqn:gap_lemma_L_swap}: 
\begin{align}
&\frac{\lr{ O_{A,B} \otimes I_A, - \lind_{\lrb{\{\CU\}}}\lrb{O_{ A, B} \otimes I_A}}_{\sigma}}
{ \lr{ O_{A,B} \otimes I_A,  O_{A,B} \otimes I_A}_{\sigma}}\nonumber \\
=& \frac{\lr{O_{\diag A,B}\otimes I_A, - \lind_{\lrb{\{\CU\}}}\lrb{O_{\diag A , B} \otimes I_A}}_{\sigma} 
+
\lr{O_{\offdiag A,B}\otimes I_A, - \lind_{\lrb{\{\CU\}}}\lrb{O_{\offdiag A , B} \otimes I_A}}_{\sigma} }
{ \lr{O_{\diag A, B}\otimes I_A, O_{\diag A, B}\otimes I_A}_{\sigma}
+
{ \lr{O_{\offdiag A, B}\otimes I_A, O_{\offdiag A, B}\otimes I_A}_{\sigma}}
}\label{eqn:gap_lemma_L_swap_decomp_14782} \\
\geq & 
 \resizebox{0.88\linewidth}{!}{$
 \min \left\{ \frac{\lr{O_{\diag A,B}\otimes I_A, - \lind_{\lrb{\{\CU\}}}\lrb{O_{\diag A , B} \otimes I_A}}_{\sigma} }
{ \lr{O_{\diag A, B}\otimes I_A, O_{\diag A, B}\otimes I_A}_{\sigma}}\,,\frac{\lr{O_{\offdiag A,B}\otimes I_A, - \lind_{\lrb{\{\CU\}}}\lrb{O_{\offdiag A , B} \otimes I_A}}_{\sigma} }
{ \lr{O_{\offdiag A, B}\otimes I_A, O_{\offdiag A, B}\otimes I_A}_{\sigma}}\right\}
$}
\,,
\label{eqn:gap_lemma_L_swap_decomp_12}
\end{align}
where the first equality is due to \cref{lem:diag_off_diag_decomp_local_A_swap}, saying that the inner product crossing $O_{\offdiag A,B} \otimes I_A$ and $O_{\diag A,B} \otimes I_A$ vanishes. The last inequality is due to \cref{lem:ratio_inequality}.
In addition, the orthogonality $\lr{I,X}_{\sigma}=0$ in \cref{eqn:gap_lemma_L_swap} imposes one extra constraint on the first ratio in \cref{eqn:gap_lemma_L_swap_decomp_12}, namely $\lr{I,O_{\diag A,B}\otimes I_A}_{\sigma}=0$. This is because \(\lr{I,\,O_{\offdiag A,B}\otimes I_A}_{\sigma}\equiv 0\), as \(O_{\offdiag A,B}\) is off-diagonal on \(A\), yielding zero trace against \(I\) and diagonal \(\sigma\).

Next, we calculate the above two cases of \cref{eqn:gap_lemma_L_swap_decomp_12} separately. 

\begin{itemize}
\item For the first case, when $ O_{A,B} =O_{\diag A, B} \in \mathcal{K}_{\diag A}$ is block-diagonal in subsystem $A$, the lower bound   $\Omega(1/\lrb{d_A \exp(4 \beta KV_{\max})})$ is  given in \cref{prop:gap_lower_bound_local_Adiag_swap} of \cref{sec:condition_1}. 

\item For the second case when $ O_{A,B} =O_{\offdiag A,B}\in \mathcal{K}_{\offdiag A}$: We rewrite it as
\begin{align}
O_{\offdiag A,B}=O_{\offdiag A,\diag B}+O_{\offdiag A,\offdiag B}\,,
\label{eqn:O_AB_offdiagA_decomp_8f99}
\end{align}
where 
\begin{align}
    O_{\offdiag A,\diag B}=\sum_{\ket{j_B}} (I_A\otimes \bra{j_{B}}) O_{A,B} (I_A\otimes \ket{j_{B}})\otimes \ketbra{j_{B}}{j_{B}}\,.
    \label{eqn:O_offdiagA_diagB_def_899}
\end{align}
Here, $\lrc{\ket{j_B}}$ is \cref{eq:basis_B}. 

Using the above expression of $ O_{A,B} $ to expand the second ratio of \cref{eqn:gap_lemma_L_swap_decomp_12}, we have
\begin{align}
    &\frac{\left\langle O_{\offdiag A,B} \otimes I_A, -\mc{L}_{ \lrb{\lrc{\mc{U}}} }(O_{\offdiag A,B} \otimes I_A)\right\rangle_{\sigma }}{\left\langle O_{\offdiag A,B}\otimes I_A,O_{\offdiag A,B}\otimes I_A\right\rangle_{\sigma }} \nonumber\\
=&
\resizebox{0.88\linewidth}{!}{$
\frac{\left\langle O_{\offdiag A,\diag B}\otimes I_A,
-\mc{L}_{ \lrb{\lrc{\mc{U}}} }(O_{\offdiag A,\diag B}\otimes I_A)\right\rangle_{\sigma}+\left\langle O_{\offdiag A,\offdiag B}\otimes I_A,-\mc{L}_{ \lrb{\lrc{\mc{U}}} }(O_{\offdiag A,\offdiag B}\otimes I_A)\right\rangle_{\sigma }}
{\left\langle O_{\offdiag A,\diag B}\otimes I_A,O_{\offdiag A,\diag B}\otimes I_A\right\rangle_{\sigma }+\left\langle O_{\offdiag A,\offdiag B}\otimes I_A,O_{\offdiag A,\offdiag B}\otimes I_A\right\rangle_{\sigma }}\,,
$}\\
\geq& 
\resizebox{0.88\linewidth}{!}{$
\min \lrc{\frac{\left\langle O_{\offdiag A,\diag B}\otimes I_A,
-\mc{L}_{ \lrb{\lrc{\mc{U}}} }(O_{\offdiag A,\diag B}\otimes I_A)\right\rangle_{\sigma}}
{\left\langle O_{\offdiag A,\diag B}\otimes I_A,O_{\offdiag A,\diag B}\otimes I_A\right\rangle_{\sigma } } , 
\frac{\left\langle O_{\offdiag A,\offdiag B}\otimes I_A,
-\mc{L}_{ \lrb{\lrc{\mc{U}}} }(O_{\offdiag A,\offdiag B}\otimes I_A)\right\rangle_{\sigma}}
{\left\langle O_{\offdiag A,\offdiag B}\otimes I_A,O_{\offdiag A,\offdiag B}\otimes I_A\right\rangle_{\sigma } } }\,.
$}
\label{eqn:gap_leu254}
\end{align}
where \cref{lem:local_A_swap_part_B_938} is employed to cancel out the inner product of crossing terms between 
$O_{\offdiag A,\diag B}\otimes I_A$ and 
$O_{\offdiag A,\offdiag B}\otimes I_A$. 
Applying \cref{lem:ratio_inequality}, we can consider minimization over diagonal part and  off-diagonal  part in $B$ of \cref{eqn:gap_leu254} separately like what we did in \cref{eqn:gap_lemma_L_swap_decomp_12}:
\begin{itemize}
\item For the diagonal term of $B$, the lower bound $\Omega(1/d_A)$ is provided in \cref{prop:gap_lower_bound_local_Aoffdiag_swap} of  \cref{sec:condition_2}.

\item When both $A$ and $B$ are \rev{off-diagonal,} we still have the lower bound $\Omega(1/d_A)$ by \cref{prop:gap_lower_bound_local_AoffdiagBoff_swap} in~\cref{sec:condition_3}.

\end{itemize}
\end{itemize}
As \cref{eqn:gap_lemma_L_swap_decomp_12,eqn:gap_leu254}, taking the minimum over the above three cases gives $\Omega\lrb{\frac{1}{d_A \exp(4 \beta KV_{\max})}}$ as desired. 

\end{proof}

\section{Action of the swapping Lindbladian: Detailed derivations}  \label{app:detailthreecases}

This section details the action of the swapping Lindbladian by explicitly calculating its effect in three cases as described in \cref{lem:gap_lemma_L_swap}, distinguished by whether the operators $A$ and $B$ are diagonal or off-diagonal (see \cref{sec:condition_1,sec:condition_2,sec:condition_3}). These derivations are a {key component} in the proof of \cref{lem:gap_lemma_L_swap} and are essential for {characterizing} the kernel of the swapping Lindbladian $\lind_{\lrb{\lrc{\mc{U}}}}$ in \cref{lem:equal_ker}. The calculations also {formally confirm} the intuition that the swapping Lindbladian $\lind_{\lrb{\lrc{\mc{U}}}}$ exchanges states within subsystem $A$ while leaving subsystem $B$ invariant.

\subsection{Case 1: $X=  O_{\diag A, B}\otimes I_A$} \label{sec:condition_1}
In this part, we consider the simplest case where the observable $X$ is diagonal in the eigenbasis of $H_A$ and is an identity in $B$, i.e.,
$X = O_{\diag A} \otimes I_B \otimes I_A \in \mathcal{K} _{\diag A}\otimes I_A$, where $\mathcal{K} ^{\diag A}$ is defined in \cref{eqn:diag_A_def38}.
\rev{Let} $O_{\diag A} = \sum_{p_A} a _{p_A}\ketbra{p_A}{p_A} $\rev{, which is} diagonal in the eigenbasis of $H_A$.
For simplicity, we write $I_B$ as $\sum_{q_B} \ketbra{q_B}{q_B}$, because \rev{we fix $q_B$ below}. This gives
\begin{align}
    X = O_{\diag A} \otimes \sum_{q_B} \ketbra{q_B}{q_B} \otimes I_A\,,
\end{align}
deriving that
\begin{align}
    \mathcal{L}_{\lrb{\lrc{\mc{U}}}}(X) &= \sum_{q_B} \mathcal{L}_{\lrb{\lrc{\mc{U}}}}\lrb{O_{\diag A} \otimes \ketbra{q_B}{q_B} \otimes I_A}\,.
\end{align}
Next, we study the action of $\mathcal{L}_{\lrb{\lrc{\mc{U}}}}$ on each term $O_{\diag A} \otimes \ketbra{q_B}{q_B} \otimes I_A$ separately.
In fact, the action of $\mathcal{L}_{\lrb{\lrc{\mc{U}}}}$ on $O_{\diag A} \otimes \ketbra{q_B}{q_B} \otimes I_A$ is to swap the $A$ subsystems of two chains while leaving $B$ invariant, i.e., 
\begin{align}
    &\mathcal{L}_{\lrb{\lrc{\mc{U}}}}\lrb{O_{\diag A} \otimes \ketbra{q_B}{q_B} \otimes I_A} 
    =\mathcal{L}_{\lrb{\lrc{\mc{U}}} , \ket{q_B}}\lrb{O_{\diag A} \otimes I_A} \otimes \ketbra{q_B}{q_B} \,,
    \label{lem:action_A_swap_local}
\end{align}
where $\mathcal{L}_{\lrb{\lrc{\mc{U}}} , \ket{q_B}}$ is the local swap superoperator acting on the $A$ subsystems of two replicas conditioned on the state $\ket{q_B}$ of the $B$ subsystem. We can prove this by an explicit calculation as follows:

    Let's calculate each term of \cref{eq:local_swap_Lindbladian}:  The first term of the Lindbladian is
\begin{align}
    & \gamma_M(\omega) \hat{\mathcal{U}}^\dagger(\omega) O_{\diag A} \otimes \ketbra{q_B}{q_B} \otimes I_A \hat{\mathcal{U}}(\omega)\nonumber \\
  = & \gamma_M(\omega) \sum_{i_A,j_B,m_A} \hat{f}\lrb{\omega - \lambda_{i_A,j_B,m_A} + \lambda_{m_A,j_B,i_A}} \ketbra{m_A j_B i_A}{i_A j_B m_A} \nonumber \\
    & \cdot \sum_{p_A} a_{p_A} \ketbra{p_A}{p_A} \otimes \ketbra{q_B}{q_B} \otimes I_A \nonumber  \\
    & \cdot \sum_{x_A, y_B, z_A} \hat{f}\lrb{\omega - \lambda_{x_A,y_B,z_A} + \lambda_{z_A,y_B,x_A}} \ketbra{x_A y_B z_A}{z_A y_B x_A} \\
  =& \gamma_M(\omega) \sum_{i_A,m_A} \hat{f}^2\lrb{\omega - \lambda_{i_A,q_B,m_A} + \lambda_{m_A,q_B,i_A}}
     a_{i_A} \lrb{\ketbra{m_A}{m_A} \otimes \ketbra{q_B}{q_B}} \otimes \lrb{\ketbra{i_A}{i_A}} \,.
     \label{eqn:action_A_swap_local_19}
\end{align}

    The second term $\gamma_M(\omega)  \hat{\mathcal{U}} ^\dagger \hat{\mathcal{U}} O_{\diag A} \otimes \ketbra{q_B}{q_B}\otimes I_A$ of the dissipative part is
\begin{align}
    & \gamma_M(\omega)  \hat{\mathcal{U}}^\dagger \hat{\mathcal{U}} O_{\diag A} \otimes \ketbra{q_B}{q_B}\otimes I_A \nonumber \\
    =& \gamma_M(\omega) \sum_{i_A,j_B,m_A} \abs{\hat{f}\lrb{\omega - \lambda_{i_A,j_B,m_A} + \lambda_{m_A,j_B,i_A}}}^2 \ketbra{m_A j_B i_A}{m_A j_B i_A} \nonumber \\
    &\cdot \sum_{p_A} a_{p_A} \ketbra{p_A}{p_A} \otimes \ketbra{q_B}{q_B} \otimes I_A  \\
    =& \gamma_M(\omega) \sum_{i_A,m_A} 
    \abs{\hat{f}\lrb{\omega - \lambda_{i_A,q_B,m_A} + \lambda_{m_A,q_B,i_A}}}^2
    a_{m_A} \lrb{\ketbra{m_A}{m_A} \otimes \ketbra{q_B}{q_B}}   \otimes \lrb{\ketbra{i_A}{i_A}}\,.
    \label{eqn:action_A_swap_local_20}
\end{align}
The third term $\gamma_M(\omega) O_{\diag A} \otimes \ketbra{q_B}{q_B}\otimes I_A \hat{\mathcal{U}} ^\dagger \hat{\mathcal{U}}$ can be obtained similarly:
\begin{align}
    & \gamma_M(\omega) O_{\diag A} \otimes \ketbra{q_B}{q_B}\otimes I_A \hat{\mathcal{U}}^\dagger \hat{\mathcal{U}} \nonumber \\
    =& \gamma_M(\omega) \sum_{p_A} a_{p_A} \ketbra{p_A}{p_A} \otimes \ketbra{q_B}{q_B} \otimes I_A \nonumber  \\
    & \cdot \sum_{i_A,j_B,m_A} \abs{\hat{f}\lrb{\omega - \lambda_{i_A,j_B,m_A} + \lambda_{m_A,j_B,i_A}}}^2 \ketbra{m_A j_B i_A}{m_A j_B i_A} \\
    =& \gamma_M(\omega) \sum_{i_A,m_A} 
    \abs{\hat{f}\lrb{\omega - \lambda_{i_A,q_B,m_A} + \lambda_{m_A,q_B,i_A}}}^2
    a_{m_A} \lrb{\ketbra{m_A}{m_A} \otimes \ketbra{q_B}{q_B}} \otimes \lrb{\ketbra{i_A}{i_A}}\,.
    \label{eqn:action_A_swap_local_21}
\end{align}

Combining the above three terms and integrating by $\int \d \omega$, we can get \cref{lem:action_A_swap_local}.

Then we prove a lower bound of the spectral gap of $\mathcal{L}_{\lrb{\lrc{\mc{U}}}}$ when acting on $X = O_{\diag A} \otimes I_B \otimes I_A \in \mathcal{K} _{\diag A}\otimes I_A$ by using \cref{lem:action_A_swap_local}.

\begin{proposition_} \label{prop:gap_lower_bound_local_Adiag_swap}
For any observable $X = O_{\diag A} \otimes I_B \otimes I_A \in \mathcal{K} _{\diag A}\otimes I_A$, we have
 \begin{align}
     \inf_{\substack{X= O_{\diag A} \otimes I_B \otimes I_A,\\
     \lr{I,X}_{\sigma} = 0 }}
     \frac{\lr{X, - \mathcal{L}_{\lrb{\lrc{\mc{U}}}}(X)}_{\sigma}}
     { \lr{X, X}_{\sigma}} = \Omega\lrb{\frac{1}{d_A \exp(4 \beta KV_{\max})}}\,,
     \label{eqn:gap_lower_bound_local_Adiag_swap}
\end{align}
where $K$ and $V_{\max}$ are those introduced in \cref{assumption:H_decomposition}, and $\beta$ is the inverse temperature.
\end{proposition_}

\begin{proof}
Still denote $X$ as
 \begin{align}
     X = &O_{\diag A} \otimes \sum_{q_B} \ketbra{q_B}{q_B} \otimes I_A\,\\
     =&  \sum_{p_A} a _{p_A}\ketbra{p_A}{p_A} \otimes \sum_{q_B} \ketbra{q_B}{q_B} \otimes I_A\,.
\end{align}
Inserting the above expression of $X$ into the left-hand side of \cref{eqn:gap_lower_bound_local_Adiag_swap} and use \cref{lem:action_A_swap_local}, one can obtain
\begin{align}
    &\frac{\lr{X, - \mathcal{L}_{\lrb{\lrc{\mc{U}}}}(X)}_{\sigma}}
    { \lr{X, X}_{\sigma}} \nonumber\\
    =&\frac{\lr{O_{\diag A} \otimes \sum_{j_B} \ketbra{j_B}{j_B} \otimes I_A, -\sum_{q_B} \mathcal{L}_{\lrb{ \lrc{\mc{U}}} , \ket{q_B}}\lrb{O_{\diag A}  \otimes I_A} \otimes  \ketbra{q_B}{q_B}}_{\sigma} }
    {\lr{O_{\diag A} \otimes \sum_{j_B} \ketbra{j_B}{j_B} \otimes I_A, O_{\diag A} \otimes \sum_{q_B} \ketbra{q_B}{q_B} \otimes I_A}_{\sigma}} \\
    =&\frac{ \sum_{j_B} \lr{O_{\diag A} \otimes \ketbra{j_B}{j_B} \otimes I_A,  - \mathcal{L}_{\lrb{ \lrc{\mc{U}}} , \ket{j_B}}\lrb{O_{\diag A}  \otimes I_A} \otimes  \ketbra{j_B}{j_B}}_{\sigma} }
    { \sum_{j_B} \lr{O_{\diag A} \otimes \ketbra{j_B}{j_B} \otimes I_A, O_{\diag A} \otimes \ketbra{j_B}{j_B} \otimes I_A}_{\sigma} }\,,
    \label{eq:gap_ratio_final_A_diag_q_B}
\end{align}
where the last equality holds because only when $j_B =q_B$ it has a non-zero contribution to the inner product, and we also use $\sigma$ is block-diagonal under $\ket{j_B}$ basis.

Then, we calculate the inner product for each part of the Lindbladian in \cref{eq:local_swap_Lindbladian}.
Denote $Y = O_{\diag A} \otimes \ketbra{j_B}{j_B} \otimes I_A$.
\begin{align}
    \lr{Y, \int \d \omega \gamma_M(\omega) \hat{\mathcal{U}}^\dagger(\omega) Y \hat{\mathcal{U}}(\omega) }_{\sigma}
    =\int \d \omega \gamma_M(\omega)  \Tr \lrs{Y^\dagger  \hat{\mathcal{U}}^\dagger(\omega) Y \hat{\mathcal{U}}(\omega)  \sigma} \,,
    \label{eqn:local_Adiag_swap_p1}
\end{align}
where we use that both \cref{eqn:action_A_swap_local_19} and $\sigma$ are diagonal under $\ket{i_A,j_B}\otimes \ket{m_A}$ basis, deducing that
\begin{align}
    \lr{Y, \hat{\mathcal{U}} ^\dagger Y \hat{\mathcal{U}} }_{\sigma} &=
    \Tr \lrs{ \sigma^{1/2} Y^\dagger \sigma^{1/2} \hat{\mathcal{U}} ^\dagger Y \hat{\mathcal{U}} } \\
    & = \Tr \lrs{ Y^\dagger  \hat{\mathcal{U}} ^\dagger Y \hat{\mathcal{U}} \sigma }\,.
\end{align}

Similarly, we also have 
\begin{align}
    \lr{Y,\int  \d \omega \gamma_M(\omega) \hat{\mathcal{U}} ^\dagger \hat{\mathcal{U}} Y }_{\sigma} &=
    \int  \d \omega \gamma_M(\omega)  \Tr\lrs{ Y^\dagger  \hat{\mathcal{U}} ^\dagger \hat{\mathcal{U}} Y \sigma } \,,
    \label{eqn:local_Adiag_swap_p2}\\
    \lr{Y,  \int  \d \omega \gamma_M(\omega)  Y \hat{\mathcal{U}} ^\dagger \hat{\mathcal{U}} }_{\sigma}
    & = \int  \d \omega \gamma_M(\omega)  \Tr\lrs{  Y^\dagger  Y \hat{\mathcal{U}} ^\dagger \hat{\mathcal{U}}  \sigma }\,.
    \label{eqn:local_Adiag_swap_p3}
\end{align}

Next, we calculate $Y^\dagger  \hat{\mathcal{U}} ^\dagger Y \hat{\mathcal{U}} $ by using \cref{eqn:action_A_swap_local_19}:
\begin{align}
    &Y^\dagger  \hat{\mathcal{U}} ^\dagger Y \hat{\mathcal{U}}  \nonumber\\
    =& \lrb{\sum_{p_A} a _{p_A}\ketbra{p_A}{p_A} \otimes \ketbra{j_B}{j_B} \otimes I_A}^\dagger \nonumber \\
    &\cdot
    \sum_{i_A,m_A} \hat{f}^2(\omega - \lambda_{i_A,j_B,m_A} + \lambda_{m_A,j_B,i_A})
        a_{i_A} \lrb{\ketbra{m_A}{m_A} \otimes \ketbra{j_B}{j_B}}   \otimes \lrb{\ketbra{i_A}{i_A}} \\
    =& \sum_{i_A,m_A} \hat{f}^2(\omega - \lambda_{i_A,j_B,m_A} + \lambda_{m_A,j_B,i_A})
        a _{m_A}^* a_{i_A} \lrb{\ketbra{m_A}{m_A} \otimes \ketbra{j_B}{j_B}}   \otimes \lrb{\ketbra{i_A}{i_A}} \\
    =& \sum_{i_A,m_A} \hat{f}^2(\omega - \omega_{i_A,j_B,m_A}) a _{m_A}^* a_{i_A} \lrb{\ketbra{m_A}{m_A} \otimes \ketbra{j_B}{j_B}}   \otimes \lrb{\ketbra{i_A}{i_A}} \,,
\end{align}
where 
\begin{align}
     \omega_{i_A,j_B,m_A} \coloneqq\lambda_{i_A,j_B,m_A} - \lambda_{m_A,j_B,i_A} 
     = \lambda_{i_A,j_B} - \lambda_{m_A,j_B}\,.
     \label{eq:omega_ijm_after}
\end{align}
Here, we emphasize that $\lambda_{i_A,j_B,m_A}$ is the eigenvalue of $H\otimes I_A+ I_n \otimes I_A$ corresponding to the eigenvector $\ket{i_A j_B m_A}$, and $\lambda_{i_A,j_B}$ is the eigenvalue of $H$ corresponding to the eigenvector $\ket{i_A j_B}$.

Next step is to calculate $Y^\dagger  \hat{\mathcal{U}} ^\dagger \hat{\mathcal{U}} Y $ by \cref{eqn:action_A_swap_local_20}:
\begin{align}
    &Y^\dagger  \hat{\mathcal{U}} ^\dagger \hat{\mathcal{U}} Y  \nonumber\\
    =& \lrb{\sum_{p_A} a _{p_A}\ketbra{p_A}{p_A} \otimes \ketbra{j_B}{j_B} \otimes I_A}^\dagger   \nonumber \\
     & \cdot  \sum_{i_A,m_A}
    \abs{\hat{f}(\omega - \lambda_{i_A,j_B,m_A} + \lambda_{m_A,j_B,i_A})}^2
    a_{m_A} \lrb{\ketbra{m_A}{m_A} \otimes \ketbra{j_B}{j_B}}   \otimes \lrb{\ketbra{i_A}{i_A}}\\
    =& \sum_{i_A,m_A} \hat{f}^2(\omega -\omega_{i_A,j_B,m_A}) a_{m_A} ^* a_{m_A} \lrb{\ketbra{m_A}{m_A} \otimes \ketbra{j_B}{j_B}}   \otimes \lrb{\ketbra{i_A}{i_A}}\,.
\end{align}
Finally, $Y^\dagger Y \hat{\mathcal{U}}^\dagger \hat{\mathcal{U}}$ is similar,
\begin{align}
    Y^\dagger Y \hat{\mathcal{U}} ^\dagger \hat{\mathcal{U}} = \sum_{i_A,m_A} \hat{f}^2(\omega -\omega_{i_A,j_B,m_A}) a_{m_A} ^* a_{m_A} \lrb{\ketbra{m_A}{m_A} \otimes \ketbra{j_B}{j_B}}   \otimes \lrb{\ketbra{i_A}{i_A}} \,.
\end{align}

Denote the Gibbs state of $H$ by 
\begin{align}
     \sigma_H  = \frac{e^{-\beta H}}{\text{Tr}(e^{-\beta H})} = \sum_{i_{A}, j_{B}} 
 s_{i_{A},j_{B}}  \ketbra{i_{A}j_{B}} {i_A j_{B}}\,,
 \label{eq:sigma_H_diag_s}
\end{align}
where $\sum_{i_A,j_B} s_{i_A,j_B} =1\,.$

Then the inner product for each part of the Lindbladian in \cref{eq:local_swap_Lindbladian} can be calculated by \cref{eqn:local_Adiag_swap_p1,eqn:local_Adiag_swap_p2,eqn:local_Adiag_swap_p3},
\begin{align}
     &\lr{Y, \int \d \omega \gamma_M(\omega) \hat{\mathcal{U}}^\dagger(\omega) Y \hat{\mathcal{U}}(\omega) }_{\sigma}  \nonumber\\
     =& \int \d \omega \gamma_M(\omega) \Tr\lrs{ Y^\dagger  \hat{\mathcal{U}} ^\dagger Y \hat{\mathcal{U}}  \sigma } \\
     =& \int \d \omega \gamma_M(\omega) 
     \Tr\lrs{ \sum_{i_A,m_A} \hat{f}^2(\omega - \omega_{i_A,j_B,m_A}) a _{m_A}^* a_{i_A} \lrb{\ketbra{m_A}{m_A} \otimes \ketbra{j_B}{j_B}}   \otimes \lrb{\ketbra{i_A}{i_A}} \sigma } \\
     =&\sum_{i_A,m_A} \int \d \omega \gamma_M(\omega)  \hat{f}^2(\omega - \omega_{i_A,j_B,m_A}) a _{m_A}^* a_{i_A}  \nonumber\\
     & \cdot \Tr \lrs{ \lrb{\ketbra{m_A}{m_A} \otimes \ketbra{j_B}{j_B}}   \otimes \lrb{\ketbra{i_A}{i_A}} \lrb{\sum_{x_A,y_B} s_{x_A,y_B} \ketbra{x_A,y_B}{x_A,y_B}}\otimes \frac{I_A}{d_A}} \\
     =& \sum_{i_A,m_A} \int \d \omega \gamma_M(\omega)  \hat{f}^2(\omega - \omega_{i_A,j_B,m_A}) a _{m_A}^* a_{i_A} \frac{1}{d_A} s_{m_A,j_B} \,.
\end{align}
As \cref{eq:theta_erfc_p08}, we denote
\begin{align}
    \theta\lrb{i_A,j_B,m_A} \coloneqq \int \d \omega \gamma_M(\omega)  \hat{f}^2\lrb{\omega - \omega_{i_A,j_B,m_A}} \,.
    \label{eq:theta_ijm_new_Def}
\end{align}
Calculate the inner product of the second part of the Lindbladian by \cref{eqn:local_Adiag_swap_p2}:
\begin{align}
    &\lr{Y,\int  \d \omega \gamma_M(\omega) \hat{\mathcal{U}} ^\dagger \hat{\mathcal{U}} Y }_{\sigma} \nonumber \\
    =&  \int \d \omega \gamma_M(\omega)  
    \Tr\lrs{ \sum_{i_A,m_A} \hat{f}^2\lrb{\omega -\omega_{i_A,j_B,m_A}} a_{m_A} ^* a_{m_A} \ketbra{m_A j_B i_A}{m_A j_B i_A} \sigma } \\
    = & \int \d \omega \gamma_M(\omega)  \sum_{i_A,m_A} \hat{f}^2\lrb{\omega -\omega_{i_A,j_B,m_A}} a_{m_A} ^* a_{m_A}  \nonumber\\
    & \Tr \lrs{ \ketbra{m_A j_B i_A}{m_A j_B i_A}
    \lrb{{\sum_{x_A,y_B} s_{x_A,y_B} \ketbra{x_A y_B}{x_A y_B}}}
    \otimes \frac{I_A}{d_A}} \\
    =& \sum_{i_A,m_A} \int \d \omega \gamma_M(\omega)  \hat{f}^2\lrb{\omega - \omega_{i_A,j_B,m_A}} a _{m_A}^* a_{m_A} \frac{1}{d_A} s_{m_A,j_B} \,\\
    = & \sum_{i_A,m_A} \theta\lrb{i_A,j_B,m_A} a _{m_A}^* a_{m_A} \frac{1}{d_A} s_{m_A,j_B}\,.
\end{align}
The third part is similar. 
Combining the above three parts, we have
\begin{align}
    &\lr{O_{\diag A} \otimes \ketbra{j_B}{j_B} \otimes I_A, -\mathcal{L}_{\lrb{\lrc{\mc{U}}}} \lrb{O_{\diag A} \otimes \ketbra{j_B}{j_B} \otimes I_A} }_{\sigma}  \nonumber\\
    =& \sum_{i_A,m_A} \theta\lrb{i_A,j_B,m_A} 
   \lrb{a_{m_A}^* a_{m_A} -  a_{m_A}^* a_{i_A}  } \frac{1}{d_A} s_{m_A,j_B}\,.
   \label{eq:gap_numerator}
\end{align}
After taking the summation $\sum_{j_B}$, the above equation is the numerator of \cref{eq:gap_ratio_final_A_diag_q_B}. The denominator of \cref{eq:gap_ratio_final_A_diag_q_B} is
\begin{align}
    &\sum_{j_B}\lr{O_{\diag A} \otimes \ketbra{j_B}{j_B} \otimes I_A, O_{\diag A} \otimes \ketbra{j_B}{j_B} \otimes I_A}_{\sigma}  \nonumber\\
    =& \sum_{j_B} \Tr\lrs{ \sigma^{1/2} \lrb{O_{\diag A} \otimes \ketbra{j_B}{j_B} \otimes I_A}^\dagger \sigma^{1/2} O_{\diag A} \otimes \ketbra{j_B}{j_B} \otimes I_A}\\
    =& \sum_{j_B} \Tr\lrs{\sum_{i_A} a_{i_A}^* \ketbra{i_A}{i_A} \otimes \ketbra{j_B}{j_B} \otimes I_A \cdot \sum_{m_A} a_{m_A} \ketbra{m_A}{m_A} \otimes \ketbra{j_B}{j_B} \otimes I_A \cdot \sigma } \\
    =& \sum_{j_B} \sum_{i_A} a_{i_A}^* a_{i_A} s_{i_A,j_B} \,,
    \label{eq:gap_denominator_A_diag}
\end{align}
where we use the fact that $\sigma$ is diagonal under $\ket{i_Aj_B}\otimes \ket{m_A}$ in the second equality.

The orthogonality condition $\lr{I, O_{\diag A} \otimes \sum_{j_B}\ketbra{j_B}{j_B} \otimes I_A} _{\sigma} = 0$ gives
\begin{align}
    \lr{I, O_{\diag A} \otimes \sum_{j_B} \ketbra{j_B}{j_B} \otimes I_A} _{\sigma}  
    =& \sum_{j_B}\Tr\lrs { \sigma^{1/2} {I}^\dagger \sigma^{1/2} O_{\diag A} \otimes \ketbra{j_B}{j_B} \otimes I_A} \\
    =& \sum_{j_B}\Tr\lrs{\sum_{i_A} a_{i_A} \ketbra{i_A}{i_A} \otimes \ketbra{j_B}{j_B} \otimes I_A \cdot \sigma } \\
    =& \sum_{j_B} \sum_{i_A} a_{i_A} s_{i_A,j_B} \\
    =&0 \,.
    \label{eq:orthogonality_A_diag_89}
\end{align}

Dividing the numerator \cref{eq:gap_numerator} by
the denominator \cref{eq:gap_denominator_A_diag}, \cref{eq:gap_ratio_final_A_diag_q_B}
becomes
\begin{align}
    &\frac{\sum_{j_B}  \lr{O_{\diag A} \otimes \ketbra{j_B}{j_B} \otimes I_A, -\mathcal{L}_{\lrb{\lrc{\mc{U}}}} \lrb{O_{\diag A} \otimes \ketbra{j_B}{j_B} \otimes I_A} }_{\sigma}}
    {\sum_{j_B}  \lr{O_{\diag A} \otimes \ketbra{j_B}{j_B} \otimes I_A, O_{\diag A} \otimes \ketbra{j_B}{j_B} \otimes I_A}_{\sigma}}  \nonumber\\
    =& \frac{1}{d_A}
    \frac{ \sum_{j_B}  \sum_{i_A,m_A} \theta\lrb{i_A,j_B,m_A}
   \lrb{a_{m_A}^* a_{m_A} -  a_{m_A}^* a_{i_A}   } {s_{m_A,j_B}}}
    { \sum_{j_B}  \sum_{i_A} a_{i_A}^* a_{i_A}{s_{i_A,j_B}}}\,.
    \label{eq:gap_ratio_final_A_diag}
\end{align}

Because \cref{lem:diag_A_98}  shows that, in \cref{eq:gap_ratio_final_A_diag},
\begin{align}
  \frac{ \sum_{j_B}  \sum_{i_A,m_A} \theta\lrb{i_A,j_B,m_A}
   \lrb{a_{m_A}^* a_{m_A} -  a_{m_A}^* a_{i_A}   } {s_{m_A,j_B}}}
    { \sum_{j_B}  \sum_{i_A} a_{i_A}^* a_{i_A}{s_{i_A,j_B}}}
   \geq  \frac{1}{2e^{4 \beta KV_{\max}}} \,,
\end{align}
deducing that \cref{eq:gap_ratio_final_A_diag} is lower bounded by $\frac{1}{2d_A \exp(4 \beta KV_{\max})}$. Therefore, we complete the proof.
\end{proof}

{
\begin{lemma_} \label{lem:diag_A_98}
Under \cref{assumption:H_decomposition}, with constants $K$ and $V_{\max}$ as therein, and $\beta$ is the inverse temperature. If, in addition, the orthogonality condition
\begin{align}
\sum_{j_B}\sum_{i_A} a_{i_A} s_{i_A,j_B}=0
\label{eq:orthogonality_A_diag_89-again}
\end{align}
is satisfied, then the following bound holds:
\begin{align}
  \frac{ \sum_{j_B}  \sum_{i_A,m_A} \theta\lrb{i_A,j_B,m_A}
   \lrb{a_{m_A}^* a_{m_A} -  a_{m_A}^* a_{i_A}   } {s_{m_A,j_B}}}
    { \sum_{j_B}  \sum_{i_A} a_{i_A}^* a_{i_A}{s_{i_A,j_B}}}
   \geq  \frac{1}{2e^{4 \beta KV_{\max}}} \,,
    \label{eq:diag_A_981}
\end{align}
where $\theta(i_A,j_B,m_A)$ is defined in \cref{eq:theta_ijm_new_Def}.
\end{lemma_}
}
\begin{proof}
    From \cref{lem:theta_calculation}, 
    \begin{align}
     \theta\lrb{i_A,j_B,m_A}  = \frac{1}{2}\lrb{ \mathrm{erfc}\left( \frac{1 + 2 \beta  \omega_{i_A,j_B,m_A}}{2\sqrt{2}} \right) +e^{-\beta \omega_{i_A,j_B,m_A}} \mathrm{erfc}  \left( \frac{1 - 2 \beta  \omega_{i_A,j_B,m_A}}{2\sqrt{2}} \right) }\,,
    \end{align}
    where
    \begin{align}
        \beta \omega_{i_A,j_B,m_A}
        =\beta  \lrb  {\lambda_{i_A,j_B} - \lambda_{m_A,j_B}}
        = \ln \lrb{\frac{s_{m_A,j_B}}{ s_{i_A,j_B}}}\,
    \end{align}
    is defined in \cref{eq:omega_ijm_after,eq:sigma_H_diag_s}. 
    Thus,
    \begin{align}
        \theta(i_A,j_B,m_A) = \frac{1}{2}  
        \lrb{ \mathrm{erfc}\left( \frac{1 + 2 \ln\lrb{\frac{s_{m_A,j_B}}{  s_{i_A,j_B}}}}{2\sqrt{2}} \right) + \frac{s_{i_A,j_B}}{s_{m_A,j_B}}  \mathrm{erfc}  \left( \frac{1 - 2 \ln\lrb{\frac{s_{m_A,j_B}}{  s_{i_A,j_B}}}}{2\sqrt{2}} \right) }\,.
    \end{align}

    The numerator of \cref{eq:diag_A_981} becomes 
    \begin{align}
        &\sum_{j_B}\sum_{i_A,m_A} \theta\lrb{i_A,j_B,m_A} 
   \lrb{a_{m_A}^* a_{m_A} -  a_{m_A}^* a_{i_A}  } 
    {s_{m_A,j_B}}   \nonumber\\
        =&
        \sum_{j_B}\sum_{i_A,m_A}
        \frac{1}{2}  
        \lrb{ \mathrm{erfc}\left( \frac{1 + 2 \ln\lrb{\frac{s_{m_A,j_B}}{  s_{i_A,j_B}}}}{2\sqrt{2}} \right) + \frac{s_{i_A,j_B}}{s_{m_A,j_B}}  \mathrm{erfc}  \left( \frac{1 - 2 \ln\lrb{\frac{s_{m_A,j_B}}{  s_{i_A,j_B}}}}{2\sqrt{2}} \right) }\nonumber\\
        &\cdot \lrb{a_{m_A}^* a_{m_A} -  a_{m_A}^* a_{i_A}  } s_{m_A,j_B}
          \\
        =&
        \sum_{j_B}\sum_{i_A,m_A}
        \frac{1}{2}  
        \lrb{ {s_{m_A,j_B}} \mathrm{erfc}\left( \frac{1 + 2 \ln\lrb{\frac{s_{m_A,j_B}}{  s_{i_A,j_B}}}}{2\sqrt{2}} \right) 
        + {s_{i_A,j_B}} \mathrm{erfc}  \left( \frac{1 - 2 \ln\lrb{\frac{s_{m_A,j_B}}{  s_{i_A,j_B}}}}{2\sqrt{2}} \right) }    \nonumber\\ 
        &\cdot \lrb{a_{m_A}^* a_{m_A} -  a_{m_A}^* a_{i_A}  }
        \,.
    \end{align}
    Denote $S$ as the above equation. We can also write $S$ into another form by swapping the indices $i_A$ and $m_A$ in the summation:
    \begin{align}
        S 
        \coloneqq & 
         \sum_{j_B}\sum_{i_A,m_A}
        \frac{1}{2}  
        \lrb{ {s_{m_A,j_B}} \mathrm{erfc}\left( \frac{1 + 2 \ln\lrb{\frac{s_{m_A,j_B}}{  s_{i_A,j_B}}}}{2\sqrt{2}} \right) 
        + {s_{i_A,j_B}} \mathrm{erfc}  \left( \frac{1 - 2 \ln\lrb{\frac{s_{m_A,j_B}}{  s_{i_A,j_B}}}}{2\sqrt{2}} \right) }    \nonumber\\ 
        &\cdot \lrb{a_{m_A}^* a_{m_A} -  a_{m_A}^* a_{i_A}  }\\
        =& 
         \sum_{j_B}\sum_{m_A,i_A}
        \frac{1}{2}
        \lrb{ {s_{i_A,j_B}} \mathrm{erfc}\left( \frac{1 + 2 \ln\lrb{\frac{s_{i_A,j_B}}{  s_{m_A,j_B}}}}{2\sqrt{2}} \right)
        + {s_{m_A,j_B}} \mathrm{erfc}  \left( \frac{1 - 2 \ln\lrb{\frac{s_{i_A,j_B}}{  s_{m_A,j_B}}}}{2\sqrt{2}} \right) }    \nonumber\\
        &\cdot \lrb{a_{i_A}^* a_{i_A} -  a_{i_A}^* a_{m_A}  }\,.
    \end{align}
    By adding the above two equivalent expressions for $S$, we get:
    \begin{align}
    2S =&\frac{1}{2}  \sum_{j_B}\sum_{i_A,m_A}
        \lrb{ {s_{m_A,j_B}} \mathrm{erfc}\left( \frac{1 + 2 \ln\lrb{\frac{s_{m_A,j_B}}{  s_{i_A,j_B}}}}{2\sqrt{2}} \right) 
        + {s_{i_A,j_B}} \mathrm{erfc}  \left( \frac{1 - 2 \ln\lrb{\frac{s_{m_A,j_B}}{  s_{i_A,j_B}}}}{2\sqrt{2}} \right) }    \nonumber\\ 
    & \cdot \lrb{a_{m_A}^* a_{m_A} -  a_{m_A}^* a_{i_A} +a^*_{i_A} a_{i_A} - a_{i_A}^* a_{m_A} }\\
    =& 
    \frac{1}{2}  \sum_{j_B}\sum_{i_A,m_A}
        \lrb{ {s_{m_A,j_B}} \mathrm{erfc}\left( \frac{1 + 2 \ln\lrb{\frac{s_{m_A,j_B}}{  s_{i_A,j_B}}}}{2\sqrt{2}} \right) 
        + {s_{i_A,j_B}} \mathrm{erfc}  \left( \frac{1 - 2 \ln\lrb{\frac{s_{m_A,j_B}}{  s_{i_A,j_B}}}}{2\sqrt{2}} \right) }    \nonumber\\ 
     &\cdot \abs{a_{m_A}-a_{i_A}  }^2   
     \,.
     \label{eq:S_mid974}
    \end{align}
    
    \Cref{lem:erfc_lowerbound} implies that the above equation can be lower bounded as 
    \begin{align}
        S \geq \frac{1}{4}\sum_{j_B} \sum_{i_A,m_A} \frac{s_{m_A,j_B}
     s_{i_A,j_B}}{s_{m_A,j_B}+s_{i_A,j_B}}\cdot  \abs{a_{m_A}-a_{i_A}  }^2 \,.
     \label{eq:S_lowerbound_1mi34}
    \end{align}
    Since the term with $m_A = i_A$ vanishes (because $\abs{a_{m_A}-a_{i_A}}^2=0$), we focus on $m_A \neq i_A$ from now on.
Then we claim that (we will prove it later), for all $i_A\neq m_A$, 
\begin{align}
    \frac{(\sum_{j_B} s_{m_A,j_B})(\sum_{k_B} s_{i_A,k_B})}{\sum_{j_B} \frac{s_{m_A,j_B}s_{i_A,j_B}}{s_{m_A,j_B}+s_{i_A,j_B}}}\leq e^{4 \beta KV_{\max}}\,,
    \label{eq:claim_ratio_bound}
\end{align}
where $K$ and $V_{\max}$ are as in \cref{item:H_AB_bounded} of \cref{assumption:H_decomposition}. Granting this claim, \cref{eq:S_lowerbound_1mi34} further yields
\begin{align}
    S\geq \frac{1}{4e^{4 \beta KV_{\max}}} \sum_{i_A,m_A} {\lrb{\sum_{j_B} s_{m_A,j_B}}\lrb{\sum_{k_B} s_{i_A,k_B}}}\cdot  \abs{a_{m_A}-a_{i_A}  }^2 \,.
\end{align}
Now separate the summations:
\begin{align}
    =& \frac{1}{4e^{4 \beta KV_{\max}}}  \sum_{i_A,m_A} {\lrb{\sum_{j_B} s_{m_A,j_B}}\lrb{\sum_{k_B} s_{i_A,k_B}}}\lrb{ a_{m_A} ^* a_{m_A} - a_{m_A} ^* a_{i_A} -a_{m_A} a^*_{i_A} +a^*_{i_A} a_{i_A} } \\
    =& \frac{1}{4e^{4 \beta KV_{\max}}} \left(
    \lrb{\sum_{m_A, j_B} s_{m_A,j_B}a_{m_A} ^* a_{m_A}} \lrb{\sum_{i_A, k_B} s_{i_A,k_B}}
    + 
    \lrb{\sum_{m_A, j_B} s_{m_A,j_B} } \lrb{\sum_{i_A, k_B} s_{i_A,k_B}a_{i_A} ^* a_{i_A}}
    \right. \nonumber\\
    &\left. - \lrb{\sum_{m_A, j_B} s_{m_A,j_B}a^*_{m_A}} \lrb{\sum_{i_A, k_B} s_{i_A,k_B}a_{i_A}} 
    - 
    \lrb{\sum_{m_A, j_B} s_{m_A,j_B}a_{m_A}} \lrb{\sum_{i_A, k_B} s_{i_A,k_B}a^*_{i_A}}
    \right)\\
    =& \frac{1}{2e^{4 \beta KV_{\max}}} 
    \lrb{\sum_{i_A, j_B} s_{i_A,j_B}a_{i_A} ^* a_{i_A}} \,,
\end{align}
where, in the last step, we used the orthogonality condition \cref{eq:orthogonality_A_diag_89} together with the normalization of probability $\sum_{i_A,j_B} s_{i_A,j_B}=1$. The above expression of numerator is precisely the denominator of \cref{eq:diag_A_981}, up to the factor $\frac{1}{2e^{4 \beta KV_{\max}}}$, and thus \cref{eq:diag_A_981} follows.

It remains to establish the claim \cref{eq:claim_ratio_bound}.
To avoid ambiguity, we decorate the superscripts of eigenvalues to indicate the Hamiltonians: $\lambda_{i_A} ^{H_A}$ is the eigenvalue of $H_A$ corresponding to $\ket{i_A}$ in \cref{eq:basis_A}; $\lambda_{j_B}^{H_B}$ is the eigenvalue of $H_B$ corresponding to $\ket{j_B}$ in \cref{eq:basis_B}; and $\lambda_{i_A j_B}^{H_{AB}}$ is the eigenvalue of $H_{AB}$ corresponding to $\ket{i_A}\otimes \ket{j_B}$. 
By \cref{item:H_AB_bounded} of \cref{assumption:H_decomposition} and \cref{eq:H_AB_bounded}, we have $\lrnorm{H_{AB}} \leq KV_{\max}$, indicating that $\abs{\lambda_{i_A j_B} ^{H_{AB}}} \leq KV_{\max}$ for all $i_A,j_B$.
Recalling the definition of $s_{i_A,j_B}$ in \cref{eq:sigma_H_diag_s} and the decomposition $H = H_A\otimes I_B + I_A\otimes H_B +H_{AB}$ from \cref{eq:H_decomposition_2547}, we have
\begin{align}
    \sigma^{H} &= \frac{e^{-\beta H}}{\Tr(e^{-\beta H})} 
    = \sum_{i_{A}, j_{B}} 
 s_{i_{A},j_{B}}  \ketbra{i_{A}j_{B}} {i_A j_{B}}  \\
    &= \frac{\lrb{e^{-\beta H_A} \otimes e^{-\beta H_B} } \cdot {e^{-\beta H_AB}} }{\Tr\lrs{\lrb{e^{-\beta H_A} \otimes e^{-\beta H_B} } \cdot {e^{-\beta H_AB}} }} 
    = \frac{1}{Z}\sum_{i_A,j_B} e^{-\beta \lrb{\lambda_{i_A}^{H_A} + \lambda_{j_B}^{H_B} + \lambda_{i_A j_B}^{H_{AB}}}}  \ketbra{i_A j_B}{i_A j_B}\,,
\end{align}
where $Z = \sum_{i_A,j_B}e^{-\beta \lrb{\lambda_{i_A}^{H_A} + \lambda_{j_B}^{H_B} + \lambda_{i_A j_B}^{H_{AB}}}} $ is the partition function. In the third equality we used the commutativity of $H_A\otimes I_B$, $I_A\otimes H_B$, and $H_{AB}$, guaranteed by \cref{item:mutual_commute} of \cref{assumption:H_decomposition}. 
Consequently,
\begin{align}
    s_{i_A,j_B} = \frac{1}{Z} e^{-\beta \lrb{\lambda_{i_A}^{H_A} + \lambda_{j_B}^{H_B} + \lambda_{i_A j_B}^{H_{AB}}}} \,,
\end{align}
and hence
\begin{align}
\frac{s_{m_A,j_B}}{s_{i_A,j_B}}= \frac{e^{-\beta \lambda ^{H_A}_{m_A} } e^{-\beta \lambda ^{H_B}_{j_B} }  e^{-\beta \lambda ^{H_{AB}}_{m_Aj_B} }}
{e^{-\beta \lambda ^{H_A}_{i_A} } e^{-\beta \lambda ^{H_B}_{j_B} }  e^{-\beta \lambda ^{H_{AB}}_{i_Aj_B} }} 
= e^{-\beta \lrb{\lambda ^{H_A}_{m_A} - \lambda ^{H_A}_{i_A}}} e^{-\beta  \lrb{\lambda ^{H_{AB}}_{m_Aj_B} - \lambda ^{H_{AB}}_{i_Aj_B}}}\,.
\end{align}
Let $r_{j_B} \coloneqq \frac{s_{m_A,j_B}}{s_{i_A,j_B}}$, omitting its dependence on $i_A,m_A$ for notational convenience. Since $\abs{\lambda_{i_A j_B} ^{H_{AB}}} \leq KV_{\max}$ for all $i_A,j_B$, we obtain
\begin{align}
e^{-\beta \lrb{\lambda ^{H_A}_{m_A} - \lambda ^{H_A}_{i_A}}} e^{-2\beta KV_{\max} }\leq r_{j_B}\coloneqq \frac{s_{m_A,j_B}}{s_{i_A,j_B}} \leq e^{-\beta \lrb{\lambda ^{H_A}_{m_A} - \lambda ^{H_A}_{i_A}}} e^{2\beta KV_{\max}} 
\end{align}
Define $c_1 \coloneqq e^{-\beta \lrb{\lambda ^{H_A}_{m_A} - \lambda ^{H_A}_{i_A}}} e^{-2\beta KV_{\max} } $ and $c_2 \coloneqq e^{-\beta \lrb{\lambda ^{H_A}_{m_A} - \lambda ^{H_A}_{i_A}}} e^{2\beta K V_{\max}}$. Note that $c_1$ and $c_2$ still depend on $i_A,m_A$, but we omit them for notational convenience. Then $c_2/c_1 \equiv e^{4 \beta K V_{\max}}$. 

Next, we expand the left-hand side of \cref{eq:claim_ratio_bound} by $ s_{m_A,j_B} =  r_{j_B} s_{i_A,j_B}$:
\begin{align}
\frac{(\sum_{j_B} s_{m_A,j_B})(\sum_{k_B} s_{i_A,k_B})}
{\sum_{j_B} \frac{s_{m_A,j_B}s_{i_A,j_B}}{s_{m_A,j_B}+s_{i_A,j_B}}}  
=& 
\frac{\lrb{\sum_{j_B} r_{j_B}s_{i_A,j_B}}}
{\sum_{j_B} \frac{r_{j_B}}{r_{j_B}+1}s_{i_A,j_B}}\lrb{\sum_{k_B} s_{i_A,k_B}}
\tag{$ s_{m_A,j_B} =  r_{j_B} s_{i_A,j_B}$}\\
\leq &\frac{c_2\sum_{j_B} s_{i_A,j_B}}
{\frac{c_1}{(c_1+1)}\sum_{j_B} s_{i_A,j_B}}\lrb{\sum_{k_B} s_{i_A,k_B}}
\tag{$c_1\leq r_{j_B}\leq c_2$, $\min \frac{r_{j_B}}{r_{j_B}+1} = \frac{c_1}{c_1+1}$}\\
\leq &\frac{c_2(c_1+1)}{c_1}\lrb{\sum_{k_B} s_{i_A,k_B}}    \\
\leq &e^{4 \beta K V_{\max}} (c_1+1)\lrb{\sum_{j_B} s_{i_A,j_B}}
\tag{$c_2/c_1 \equiv e^{4 \beta K V_{\max}}$}\\
\leq& e^{4 \beta K V_{\max}}  \lrb{\sum_{j_B} s_{m_A,j_B}+\sum_{j_B} s_{i_A,j_B}}
\tag{$ s_{m_A,j_B}\geq c_1  s_{i_A,j_B}$} \nonumber\\
\leq & e^{4 \beta K V_{\max}} 
\tag{$\sum_{j_B} s_{m_A,j_B}+\sum_{j_B} s_{i_A,j_B}\leq 1$ when $m_A\neq i_A$ } \,.
\end{align}
This proves the claim \cref{eq:claim_ratio_bound} and completes the argument.

\end{proof}

\subsection{Case 2:  $X= O_{\offdiag A,\diag B}\otimes I_A$} \label{sec:condition_2}
\rev{In this part, we consider the case of \cref{eqn:O_offdiagA_diagB_def_899}, where the observable $X$ is off-diagonal in the eigenbasis of $H_A$ and diagonal in the eigenbasis of $H_B$. If let  
\begin{align}
    X& = O_{\offdiag A,\diag  B}\otimes I_A \\
    & =\sum_{j_B}  O_{\offdiag A} \otimes \ketbra{j_B}{j_B} \otimes I_A \,\\
    &= \sum_{j_B} \sum_{i_A\neq m_A} A_{i_A,m_A,j_B} \ketbra{i_A}{m_A} \otimes\ketbra{j_B}{j_B} \otimes I_A \,,
    \label{eqn:observable_offdiag_A_diag_B}
\end{align}
then, 
\begin{align}
    \mathcal{L}_{\lrb{\lrc{\mc{U}}}}(X) &= \sum_{j_B}  \mathcal{L}_{\lrb{\lrc{\mc{U}}}}\lrb{O_{\offdiag A} \otimes \ketbra{j_B}{j_B} \otimes I_A}.
\end{align}
Note that, here, the off-diagonal operator $O_{\offdiag A}$ actulally depends on $j_B$ as shown in \cref{eqn:observable_offdiag_A_diag_B}, but for simplicity of notation we omit this dependence in the following discussion.

Actually, the action of $\mathcal{L}_{\lrb{\lrc{\mc{U}}}}$ on $O_{\offdiag A} \otimes \ketbra{j_B}{j_B} \otimes I_A$ is to swap the $A$ subsystems of two chains while leaving $B$ invariant, so we have 
\begin{align}
&\mathcal{L}_{\lrb{\lrc{\mc{U}}}}\lrb{O_{\offdiag A} \otimes \ketbra{q_B}{q_B} \otimes I_A} 
=\mathcal{L}_{\lrb{\lrc{\mc{U}}}, \ket{q_B}}\lrb{O_{\offdiag A} \otimes I_A} \otimes \ketbra{q_B}{q_B} \,,
\label{lem:action_A_swap_local_offdiag}
\end{align}
where $\mathcal{L}_{\lrb{\lrc{\mc{U}}}, \ket{q_B}}$ is the local swap superoperator acting on the $A$ subsystem conditioned on the state $\ket{q_B}$ of the $B$ subsystem.
We can prove the above equation by an explicit calculation as follows:

    By \cref{eqn:replica_exchange_davies_local_swap_general_1}, we can calculate the first part of the Lindbladian \cref{eq:local_swap_Lindbladian}:
\begin{align}
    & \gamma_M\lrb{\omega} \hat{\mathcal{U}} ^\dagger\lrb{\omega} O_{\offdiag A} \otimes \ketbra{q_B}{q_B} \otimes I_A \hat{\mathcal{U}}\lrb{\omega} \nonumber \\
    =&  \gamma_M\lrb{\omega}  \sum_{i_A,j_B,m_A } 
    \hat{f}\lrb{\omega - \lambda_{i_A,j_B,m_A} + \lambda_{m_A,j_B,i_A}}
    \ketbra{m_A j_B i_A}{i_A j_B m_A} \nonumber\\
    & \cdot \sum_{r_A\neq s_A} A_{r_A,s_A,q_B} \ketbra{r_A}{s_A} \otimes \ketbra{q_B}{q_B} \otimes I_A \nonumber\\
    & \cdot \sum_{x_A,y_B,z_A} \hat{f}\lrb{\omega - \lambda_{x_A,y_B,z_A} + \lambda_{z_A,y_B,x_A}} \ketbra{x_A y_B z_A}{z_A y_B x_A} \\
    =&\gamma_M\lrb{\omega} \sum_{i_A \neq x_A,m_A} 
    \hat{f}\lrb{ \omega - \lambda_{i_A,q_B,m_A} + \lambda_{m_A,q_B,i_A}} 
    \hat{f}\lrb{ \omega - \lambda_{x_A,q_B,m_A} + \lambda_{m_A,q_B,x_A}}\nonumber \\
    & \cdot A_{i_A,x_A,q_B}
    \ketbra{m_A}{m_A} \otimes \ketbra{q_B}{q_B} \otimes \ketbra{i_A}{x_A}
    \label{eq:off_A_p1}
\end{align}

Then, calculating the second part of Lindbladian of \cref{eq:local_swap_Lindbladian}, we have
\begin{align}
    & \gamma_M\lrb{\omega}  \hat{\mathcal{U}}^\dagger  \hat{\mathcal{U}} O_{\offdiag A} \otimes \ketbra{q_B}{q_B}\otimes I_A \nonumber\\
    = & \gamma_M\lrb{\omega} \sum_{i_A,j_B,m_A} \abs{\hat{f}\lrb{\omega - \lambda_{i_A,j_B,m_A} + \lambda_{m_A,j_B,i_A}}}^2 \ketbra{m_A j_B i_A}{m_A j_B i_A}\nonumber\\
    & \cdot  \sum_{r_A\neq s_A} A_{r_A,s_A,q_B} \ketbra{r_A}{s_A} \otimes \ketbra{q_B}{q_B} \otimes I_A\\
    =& \gamma_M\lrb{\omega} \sum_{r_A\neq s_A ,i_A}  
    \abs{\hat{f}\lrb{\omega - \lambda_{i_A,q_B,r_A} + \lambda_{r_A,q_B,i_A}} }^2
    A_{r_A,s_A,q_B} \ketbra{r_A}{s_A} \otimes \ketbra{q_B}{q_B} \otimes \ketbra{i_A}{i_A} \\
    =& \gamma_M\lrb{\omega} \sum_{i_A\neq x_A,m_A} 
    \abs{\hat{f}\lrb{\omega - \lambda_{m_A,q_B,i_A} + \lambda_{i_A,q_B,m_A}} }^2  A_{i_A,x_A,q_B} \ketbra{i_A}{x_A} \otimes \ketbra{q_B}{q_B} \otimes \ketbra{m_A}{m_A}\,.
    \label{eq:off_A_p2}
\end{align}

Similarly, the third part is
\begin{align}
    &\gamma_M\lrb{\omega}  O_{\offdiag A} \otimes \ketbra{q_B}{q_B} \otimes I_A \hat{\mathcal{U}}^\dagger\hat{\mathcal{U}}\nonumber \\
    = & \gamma_M\lrb{\omega}\!\! \sum_{i_A\neq x_A,m_A} 
    \abs{\hat{f}\lrb{\omega - \lambda_{m_A,q_B,x_A} + \lambda_{x_A,q_B,m_A}} }^2 A_{i_A,x_A,q_B}
    \ketbra{i_A}{x_A} \otimes \ketbra{q_B}{q_B} \otimes \ketbra{m_A}{m_A}.
    \label{eq:off_A_p3}
\end{align}
Combining the above three terms, we can get \cref{lem:action_A_swap_local_offdiag}, \rev{which yields} the following proposition of the spectral gap lower bound for the case of \cref{eqn:O_offdiagA_diagB_def_899}.

\begin{proposition_} \label{prop:gap_lower_bound_local_Aoffdiag_swap}
For any observable $X = O_{\offdiag A,\diag B}\otimes I_A$ in \cref{eqn:O_offdiagA_diagB_def_899}, where $O_{\offdiag A,\diag B}$ is off-diagonal in the eigenbasis of $H_A$, and diagonal in the eigenbasis of $H_B$, we have
\begin{align}
    \inf_{\substack{X= O_{\offdiag A,\diag B}\otimes I_A }} 
    \frac{\lr{X, - \mathcal{L}_{\lrb{\lrc{\mc{U}}}}(X)}_{\sigma}}
    { \lr{X, X}_{\sigma}} = \Omega\lrb{\frac{1}{d_A}}\,.
    \label{eq:8491}
\end{align}
\end{proposition_}

\begin{proof}
    Like what we did in \cref{prop:gap_lower_bound_local_Adiag_swap}, we still want to fix each $j_B$ and calculate the minimization for each fixed $j_B$ separately. Expand \cref{eq:8491} by \cref{lem:action_A_swap_local_offdiag} and \cref{eqn:observable_offdiag_A_diag_B}:
    \begin{align}
    &\frac{\lr{X, -  \mathcal{L}_{\lrb{\lrc{\mc{U}}}}(X)}_{\sigma}}
    { \lr{X, X}_{\sigma}} \nonumber\\
    =&\frac{ \lr{ \sum_{j_B} O_{\offdiag A} \otimes   \ketbra{j_B}{j_B} \otimes I_A, -\sum_{q_B}  \mathcal{L}_{\lrb{\lrc{\mc{U}}}, \ket{q_B}}\lrb{O_{\offdiag A}  \otimes I_A} \otimes  \ketbra{q_B}{q_B}}_{\sigma} }
    {\lr{\sum_{j_B} O_{\offdiag A} \otimes  \ketbra{j_B}{j_B} \otimes I_A, 
    \sum_{q_B}  O_{\offdiag A} \otimes   \ketbra{q_B}{q_B} \otimes I_A}_{\sigma}} \\
    =&\frac{ \sum_{j_B}    \lr{O_{\offdiag A} \otimes \ketbra{j_B}{j_B} \otimes I_A,  -\mathcal{L}_{\lrb{\lrc{\mc{U}}}, \ket{j_B}}\lrb{O_{\offdiag A}  \otimes I_A} \otimes  \ketbra{j_B}{j_B}}_{\sigma} }
    { \sum_{j_B}  \lr{O_{\offdiag A} \otimes \ketbra{j_B}{j_B} \otimes I_A, O_{\offdiag A} \otimes \ketbra{j_B}{j_B} \otimes I_A}_{\sigma} }\,.
    \end{align}
Then we apply \cref{lem:ratio_inequality} to the above equation, which shows that it is sufficient to calculate the infimum over all fixed $j_B$ for 
\begin{align}
    &\frac{  \lr{O_{\offdiag A} \otimes \ketbra{j_B}{j_B} \otimes I_A,  -\mathcal{L}_{\lrb{\lrc{\mc{U}}}, \ket{j_B}}\lrb{O_{\offdiag A}  \otimes I_A} \otimes  \ketbra{j_B}{j_B}}_{\sigma} }
    { \lr{O_{\offdiag A} \otimes \ketbra{j_B}{j_B} \otimes I_A, O_{\offdiag A} \otimes \ketbra{j_B}{j_B} \otimes I_A}_{\sigma} }\,. \label{eq:gap_ratio_offdiagA}
\end{align}

    Next, we calculate the inner product of each part in the numerator: Using \cref{eq:off_A_p1}, the first part of Lindbladian after inner product gives
\begin{align}
     &\lr{O_{\offdiag A} \otimes \ketbra{j_B}{j_B} \otimes I_A,
     \gamma_M\lrb{\omega} \hat{\mathcal{U}}^\dagger\lrb{\omega} O_{\offdiag A} \otimes \ketbra{j_B}{j_B} \otimes I_A  \hat{\mathcal{U}}\lrb{\omega}   } _{\sigma} \nonumber\\
     = & \Tr\lrs{\sigma^{1/2} \lrb{O_{\offdiag A} \otimes \ketbra{j_B}{j_B} \otimes I_A} ^\dagger \sigma^{1/2} \cdot \gamma_M\lrb{\omega} \hat{\mathcal{U}}^\dagger\lrb{\omega} O_{\offdiag A} \otimes \ketbra{j_B}{j_B} \otimes I_A  \hat{\mathcal{U}}\lrb{\omega} } \\
     = & \gamma_M\lrb{\omega} \sum_{i_A \neq x_A,m_A} 
     \hat{f}\lrb{ \omega - \lambda_{i_A,j_B,m_A} + \lambda_{m_A,j_B,i_A}} 
     \hat{f}\lrb{ \omega - \lambda_{x_A,j_B,m_A} + \lambda_{m_A,j_B,x_A}} \nonumber\\
     & \cdot \Tr\left[\sigma^{1/2} \lrb{O_{\offdiag A} \otimes \ketbra{j_B}{j_B} \otimes I_A} ^\dagger \sigma^{1/2} 
     \cdot A_{i_A,x_A,j_B} \ketbra{m_A}{m_A} \otimes \ketbra{j_B}{j_B} \otimes \lrb{\ketbra{i_A}{x_A} } \right] \\
     =&0\,.
\end{align}
Here, the last equality holds because $\sigma = \frac{\exp(-\beta H) }{\Tr\lrs{\exp(-\beta H)}} \otimes \sigma_2$, implying that $\Tr\lrs{\sigma_2^{1/2} I_A \sigma_2^{1/2} \ketbra{i_A}{x_A}} = 0$ for ${i_A} \neq {x_A}$, given that $\sigma_2 =I_A/d_A$.

Then calculate the second part of Lindbladian after KMS inner product by \cref{eq:off_A_p2}:
\begin{align}
     &\lr{O_{\offdiag A} \otimes \ketbra{j_B}{j_B} \otimes I_A,
     \gamma_M\lrb{\omega}\hat{\mathcal{U}}^\dagger\lrb{\omega} \hat{\mathcal{U}}\lrb{\omega} O_{\offdiag A} \otimes \ketbra{j_B}{j_B} \otimes I_A  }_{\sigma} \nonumber\\
     = & \Tr\left[ \sigma^{1/2} 
     \lrb{\sum_{r_A\neq s_A} A^*_{r_A,s_A,j_B} \ketbra{s_A}{r_A}  \otimes \ketbra{j_B}{j_B}\otimes  I_A} \sigma^{1/2}   \right. \nonumber\\
     &\resizebox{0.87\linewidth}{!}{$
     \left.\gamma_M\lrb{\omega} \sum_{i_A\neq x_A,m_A} 
     \abs{\hat{f} \lrb{\omega - \lambda_{m_A,j_B,i_A} + \lambda_{i_A,j_B,m_A}} }^2  A_{i_A,x_A,j_B} \ketbra{i_A}{x_A} \otimes \ketbra{j_B}{j_B} \otimes \lrb{\ketbra{m_A}{m_A}} \right] 
     $}\,.
     \label{eq:off_A_8faf910}
\end{align}
Denote the Gibbs state of $H$ as 
\begin{align}
      \sigma_H  =\sum_{i_A,j_B} s_{i_Aj_B} \ketbra{i_Aj_B}{i_Aj_B}\,,
\end{align}
and the Gibbs state of the second system is $\sigma_2 = I_A/d_A$, then $\sigma =  \sigma_H  \otimes \sigma_2$. 
Inserting $\sigma$ into \cref{eq:off_A_8faf910}, we have
\begin{align}
     =& \frac{1}{d_A} \sum_{i_A\neq x_A} \left|A_{i_A,x_A,j_B}\right|^2  \lrb{s_{x_Aj_B} }^{1/2} \lrb{s_{i_Aj_B} }^{1/2} \sum_{m_A} \gamma_M\lrb{\omega} \abs{\hat{f} \lrb{\omega - \omega_{m_A,j_B,i_A}} }^2 \,,
\end{align}
where 
$\omega_{m_A,j_B,i_A} = \lambda_{m_A,j_B,i_A} - \lambda_{i_A,j_B,m_A}$. 

Similarly, we can calculate the third part of Lindbladian after inner product,
\begin{align}
    &\lr{O_{\offdiag A} \otimes \ketbra{j_B}{j_B} \otimes I_A,
    \gamma_M(\omega) O_{\offdiag A} \otimes \ketbra{j_B}{j_B} \otimes I_A  \hat{\mathcal{U}}^\dagger(\omega) \hat{\mathcal{U}}(\omega)}_{\sigma} \nonumber\\
    = & \frac{1}{d_A} \sum_{i_A\neq x_A} \left|A_{i_A,x_A,j_B}\right|^2  \lrb{s_{x_Aj_B} }^{1/2} \lrb{s_{i_Aj_B} }^{1/2}  \sum_{m_A} \gamma_M(\omega) \abs{\hat{f} (\omega - \omega_{m_A,j_B,x_A}) }^2 \,.
    \label{eq:off_A_8911}
\end{align}

Combining the above three parts, we have
\begin{align}
    &\lr{O_{\offdiag A} \otimes \ketbra{j_B}{j_B} \otimes I_A,
    -\mathcal{L}_{\lrb{\lrc{\mc{U}}}}\lrb{O_{\offdiag A} \otimes \ketbra{j_B}{j_B} \otimes I_A} }_{\sigma} \nonumber\\
    = & \frac{1}{2d_A} \sum_{i_A\neq x_A} \left|A_{i_A,x_A,j_B}\right|^2 \lrb{s_{x_Aj_B} }^{1/2} \lrb{s_{i_Aj_B} }^{1/2}  \nonumber\\
    & \sum_{m_A}\int \d \omega \gamma_M(\omega) \lrb{\abs{ \hat{f}(\omega - \omega_{m_A,j_B,i_A})}^2 +\abs{ \hat{f}(\omega - \omega_{m_A,j_B,x_A})}^2 } \,.
\end{align}
By \cref{eq:theta_iji}, we know that 
\begin{align}
    &\sum_{m_A}\int \d \omega \gamma_M(\omega) \lrb{\abs{ \hat{f}(\omega - \omega_{m_A,j_B,i_A})}^2 +\abs{ \hat{f}(\omega - \omega_{m_A,j_B,x_A})}^2 } \nonumber\\
    \geq& \int \d \omega \gamma_M(\omega) \lrb{\hat{f}^2 (\omega-\omega_{i_A,j_B,i_A} ) +\hat{f}^2 (\omega-\omega_{x_A,j_B,x_A} ) } \\
    =& \theta (i_A,j_B,i_A) +\theta (x_A,j_B,x_A) \\
    \geq& 1\,,
\end{align}
deriving that the numerator of \cref{eq:gap_ratio_offdiagA} can be lower bounded as
\begin{align}
    &\lr{O_{\offdiag A} \otimes \ketbra{j_B}{j_B} \otimes I_A,
    -\mathcal{L}_{\lrb{\lrc{\mc{U}}}}\lrb{O_{\offdiag A} \otimes \ketbra{j_B}{j_B} \otimes I_A} } \nonumber\\
    \geq &\frac{1}{2d_A} \sum_{i_A\neq x_A} \left|A_{i_A,x_A,j_B}\right|^2  
    \lrb{s_{x_Aj_B} }^{1/2} \lrb{s_{i_Aj_B} }^{1/2} \,.
\end{align}

The final step is to calculate the denominator of \cref{eq:gap_ratio_offdiagA}:
\begin{align}
    &\lr{O_{\offdiag A} \otimes \ketbra{j_B}{j_B} \otimes I_A,
    O_{\offdiag A} \otimes \ketbra{j_B}{j_B} \otimes I_A}_{\sigma} \nonumber\\
    =&
    \resizebox{0.86\linewidth}{!}{$
    \Tr\left[{\sigma}^{1/2} 
    \lrb{\sum_{r_A\neq s_A} A^{*}_{r_A,s_A,j_B} \ketbra{s_A}{r_A}  \otimes \ketbra{j_B}{j_B}\otimes  I_A} 
    {\sigma}^{1/2}  \sum_{i_A\neq x_A} A_{i_A,x_A,j_B} \ketbra{i_A}{x_A} \otimes \ketbra{j_B}{j_B} \otimes I_A\right] 
    $}\\
    =&\sum_{i_A\neq x_A} \left|A_{i_A,x_A,j_B}\right|^2 \lrb{s_{x_Aj_B} }^{1/2} \lrb{s_{i_Aj_B} }^{1/2} \,.
\end{align}
Dividing numerator by denominator, we have the lower bound $\Omega(1/d_A)$ as expected. 
\end{proof}

\subsection{Case 3: $X= O _{\offdiag A,\offdiag B}\otimes I_A$} \label{sec:condition_3}
In this case, both $A$ and $B$ are off-diagonal in the eigenbasis of $H_A$ and $H_B$.  Specifically, let 
\begin{align}
    X &= O_{\offdiag A,\offdiag B}\otimes I_A \\
    & =\sum_{i_A \neq x_A} \sum_{q_B\neq h_B}  
    A_{i_A,x_A,q_B,h_B} \ketbra{i_A}{x_A} \otimes \ketbra{q_B}{h_B} \otimes I_A\,. 
\end{align}
Then, we prove the following proposition:

\begin{proposition_} \label{prop:gap_lower_bound_local_AoffdiagBoff_swap}
For any observable $X = O_{\offdiag A,\offdiag  B}\otimes I_A $, where $O_{\offdiag A,\offdiag B}$ is off-diagonal in the eigenbasis of $H_A$, and off-diagonal in the eigenbasis of $H_B$, we have
\begin{align}
    \inf_{\substack{X= O_{\offdiag A,\offdiag  B}\otimes I_A  }}  \frac{\lr{X, - \mathcal{L}_{\lrb{\lrc{\mc{U}}}}(X)}_{\sigma}}
     { \lr{X, X}_{\sigma}} = \Omega\lrb{\frac{1}{d_A}}\,.
     \label{eqn:AoffdiagBoff_swap_p12}
\end{align}
\end{proposition_}

\begin{proof}
Let's calculate the action of $\mathcal{L}_{\lrb{\lrc{\mc{U}}}}$ on $X$: By \cref{eqn:replica_exchange_davies_local_swap_general_1}, the first part of Lindbladian is 
\begin{align}
    &\gamma_M\lrb{\omega} \hat{\mathcal{U}}^\dagger\lrb{\omega} X \hat{\mathcal{U}}\lrb{\omega} \nonumber\\
    =& \gamma_M\lrb{\omega}  \sum_{i_A,j_B,m_A } 
     \hat{f}\lrb{\omega - \lambda_{i_A,j_B,m_A} + \lambda_{m_A,j_B,i_A}}
     \ketbra{m_A j_B i_A}{i_A j_B m_A} \nonumber\\
    & \cdot \sum_{r_A\neq s_A}  \sum_{q_B\neq h_B}  A_{r_A,s_A,q_B,h_B} \ketbra{r_A}{s_A} \otimes  \ketbra{q_B}{h_B} \otimes I_A \nonumber\\
    & \cdot \sum_{x_A,y_B,z_A} \hat{f}\lrb{\omega - \lambda_{x_A,y_B,z_A} + \lambda_{z_A,y_B,x_A}} \ketbra{x_A y_B z_A}{z_A y_B x_A} \\
    =& \gamma_M\lrb{\omega} \sum_{i_A\neq x_A} \sum_{q_B\neq h_B} \sum_{m_A}
     \hat{f}\lrb{\omega - \lambda_{i_A,q_B,m_A} + \lambda_{m_A,q_B,i_A}}  
     \hat{f}\lrb{\omega - \lambda_{x_A,h_B,m_A} + \lambda_{m_A,h_B,x_A}} \nonumber\\
    & \cdot A_{i_A,x_A,q_B,h_B} 
     \ketbra{m_A}{m_A} \otimes \ketbra{q_B}{h_B} \otimes \ketbra{i_A}{x_A} 
     \label{eq:off_AB_p1}
\end{align}

The second part of Lindbladian is
\begin{align}
    & \gamma_M\lrb{\omega} \hat{\mathcal{U}}^\dagger \hat{\mathcal{U}} X \nonumber\\
    =& \gamma_M\lrb{\omega} \sum_{i_A,j_B,m_A} \abs{\hat{f}\lrb{\omega - \lambda_{i_A,j_B,m_A} + \lambda_{m_A,j_B,i_A}}}^2 \ketbra{m_A j_B i_A}{m_A j_B i_A} \nonumber \\
    &\cdot \sum_{r_A\neq s_A} \sum_{q_B\neq h_B} A_{r_A,s_A,q_B,h_B} \ketbra{r_A}{s_A} \otimes  \ketbra{q_B}{h_B} \otimes I_A \\
    =& \gamma_M\lrb{\omega} \sum_{m_A\neq s_A} \sum_{q_B\neq h_B} \sum_{i_A}
     \abs{\hat{f}\lrb{\omega - \lambda_{i_A,q_B,m_A} + \lambda_{m_A,q_B,i_A}}}^2 \nonumber\\
    & \cdot A_{m_A,s_A,q_B,h_B} \ketbra{m_A}{s_A} \otimes \ketbra{q_B}{h_B} \otimes \ketbra{i_A}{i_A} \\
    =& \gamma_M\lrb{\omega} \sum_{i_A\neq x_A} \sum_{q_B\neq h_B} \sum_{m_A}
     \abs{\hat{f}\lrb{\omega - \lambda_{m_A,q_B,i_A} + \lambda_{i_A,q_B,m_A}}}^2 A_{i_A,x_A,q_B,h_B} \nonumber\\
    & \ketbra{i_A}{x_A} \otimes \ketbra{q_B}{h_B} \otimes \ketbra{m_A}{m_A}\,.
     \label{eq:off_AB_p2}
\end{align}

Similarly, the third part is
\begin{align}
    & \gamma_M\lrb{\omega}  X \hat{\mathcal{U}}^\dagger\lrb{\omega}  \hat{\mathcal{U}}\lrb{\omega} \nonumber\\
    =& \gamma_M\lrb{\omega} \sum_{r_A\neq s_A}   \sum_{q_B\neq h_B} A_{r_A,s_A,q_B,h_B} \ketbra{r_A}{s_A} \otimes  \ketbra{q_B}{h_B} \otimes I_A \nonumber\\
    & \sum_{i_A,j_B,m_A} \abs{\hat{f}\lrb{\omega - \lambda_{i_A,j_B,m_A} + \lambda_{m_A,j_B,i_A}}}^2 \ketbra{m_A j_B i_A}{m_A j_B i_A} \\
    = & \gamma_M\lrb{\omega} \sum_{r_A\neq s_A} \sum_{q_B\neq h_B} \sum_{i_A} 
    \abs{\hat{f}\lrb{\omega - \lambda_{i_A,h_B,s_A} + \lambda_{s_A,h_B,i_A}}}^2 
    A_{r_A,s_A,q_B,h_B}\nonumber\\
    & \cdot \ketbra{r_A}{s_A} \otimes \ketbra{q_B}{h_B} \otimes \lrb{\ketbra{i_A}{i_A}} \\
    = & \gamma_M\lrb{\omega} \sum_{i_A\neq x_A} \sum_{q_B\neq h_B} \sum_{m_A}
    \abs{\hat{f}\lrb{\omega - \lambda_{m_A,h_B,x_A} + \lambda_{x_A,h_B,m_A}}}^2  
    A_{i_A,x_A,q_B,h_B} \nonumber\\
    &
    \ketbra{i_A}{x_A} \otimes \ketbra{q_B}{h_B} \otimes \lrb{\ketbra{m_A}{m_A}}\,.
    \label{eq:off_AB_p3}
\end{align}

Next, we calculate the inner product of each part in the numerator of \cref{eqn:AoffdiagBoff_swap_p12}: The first part of Lindbladian after inner product gives
\begin{align}
    &\int \d \omega \lr{X ,  \gamma_M\lrb{\omega} \hat{\mathcal{U}}^\dagger\lrb{\omega} X \hat{\mathcal{U}}\lrb{\omega} }_{\sigma}\nonumber \\
    =&\int \d \omega \Tr\left[\sigma^{1/2}
        \lrb{O_{\offdiag A,\offdiag B}\otimes I_A}^\dagger  
        \sigma^{1/2} \right. \nonumber\\
    &\gamma_M\lrb{\omega} \sum_{i_A\neq x_A} \sum_{q_B\neq h_B} \sum_{m_A}
        \hat{f}\lrb{\omega - \lambda_{i_A,q_B,m_A} + \lambda_{m_A,q_B,i_A}}  
        \hat{f}\lrb{\omega - \lambda_{x_A,h_B,m_A} + \lambda_{m_A,h_B,x_A}} \nonumber \\
    &\left.  \cdot A_{i_A,x_A,q_B,h_B} 
        \ketbra{m_A}{m_A} \otimes \ketbra{q_B}{h_B} \otimes \ketbra{i_A}{x_A} 
       \right]\\
    =&0\,.
\end{align}
Here, the last equality holds because $\Tr\lrs{{\sigma_2^{1/2} I_A {\sigma_2}^{1/2} \ketbra{i_A}{x_A}}} = 0$ for ${i_A} \neq {x_A}$, given that $\sigma_2 =I_A/d_A$.

Next, calculate the second part of Lindbladian after inner product:
\begin{align}
    &\int\d\omega \lr{X ,\gamma_M\lrb{\omega} \hat{\mathcal{U}}^\dagger \hat{\mathcal{U}} X }_{\sigma} \nonumber\\
    =&\int\d\omega \gamma_M\lrb{\omega}\Tr\left[\sigma^{1/2}\cdot 
    \sum_{n_A \neq \ell_A} \sum_{e_B\neq f_B}  A^{*}_{n_A,\ell_A,e_B,f_B} \ketbra{\ell_A}{n_A}\otimes   \ketbra{f_B}{e_B} \otimes I_A \cdot 
    \sigma^{1/2}\right. \nonumber\\
    &\resizebox{0.87\linewidth}{!}{$
    \left.   \sum_{i_A\neq x_A}  \sum_{q_B\neq h_B}  \sum_{m_A}
    \abs{\hat{f}\lrb{\omega - \omega_{m_A,q_B,i_A}}}^2  A_{i_A,x_A,q_B,h_B} 
    \ketbra{i_A}{x_A} \otimes \ketbra{q_B}{h_B} \otimes \lrb{\ketbra{m_A}{m_A}}\right]
    $}\,.
\end{align}
From the above equation, because $ \sigma_H  \otimes \sigma_2$ is diagonal under the basis $\ket{i_Aj_B m_A}$, we can see that only when $i_A = n_A$ and $x_A = \ell_A$, $q_B = e_B$, and $h_B = f_B$ the summands are not zero. 
Here, we still let that 
\begin{align}
     \sigma_H  =\sum_{i_A,j_B} s_{i_Aj_B} \ketbra{i_Aj_B}{i_Aj_B}\,,
\end{align}
inserting it in the previous equation gives
\begin{align}
    &\int\d\omega \lr{X ,\gamma_M\lrb{\omega} \hat{\mathcal{U}}^\dagger \hat{\mathcal{U}} X }_{\sigma} \nonumber\\
    =& \frac{1}{d_A } \sum_{i_A \neq x_A}  \sum_{q_B\neq h_B} \abs{A_{i_A,x_A,q_B,h_B}}^2 
   \lrb{s_{x_Ah_B} s_{i_Aq_B}}^{1/2}
    \sum_{m_A}  \int\d\omega \gamma_M\lrb{\omega}  \abs{\hat{f}\lrb{\omega - \omega_{m_A,q_B,i_A}}}^2   \,.
    \label{eq:off_A_off_B34}
\end{align}
Because 
\begin{align}
    \sum_{m_A}\int\d\omega \gamma_M(\omega)  \abs{\hat{f}(\omega - \omega_{m_A,q_B,i_A})}^2   \geq \int \d \omega \gamma_M(\omega)  \abs{\hat{f}(\omega - \omega_{i_A,q_B,i_A})}^2  \geq \frac{1}{2}\,
\end{align}
by \cref{eq:theta_iji}, we have
\begin{align}
    &\int\d\omega \lr{X ,\gamma_M(\omega) \hat{\mathcal{U}}^\dagger \hat{\mathcal{U}} X }_{\sigma } 
    \geq   \frac{1}{2d_A } \sum_{i_A \neq x_A}  \sum_{q_B\neq h_B} \abs{A_{i_A,x_A,q_B,h_B}}^2 
   \lrb{s_{x_A,h_B} s_{i_A,q_B}}^{1/2}\,.
\end{align}
The third part of the Lindbladian can be calculated in a similar way:
\begin{align}
    \int \d \omega \lr{X , \gamma_M\lrb{\omega}  X \hat{\mathcal{U}}^\dagger  \hat{\mathcal{U}}}_{\sigma} \geq \frac{1}{2d_A } \sum_{i_A \neq x_A}  \sum_{q_B\neq h_B} \abs{A_{i_A,x_A,q_B,h_B}}^2 
   \lrb{s_{x_A,h_B} s_{i_A,q_B}}^{1/2}\,.
\end{align}
Combining all three terms,  we have gotten the numerator of \cref{eqn:AoffdiagBoff_swap_p12} can be lower bounded by
\begin{align}
    \lr{X,-\mathcal{L}_{\lrb{\lrc{\mc{U}}}}(X)}_{\sigma} \geq \frac{1}{2d_A } \sum_{i_A \neq x_A}  \sum_{q_B\neq h_B} \abs{A_{i_A,x_A,q_B,h_B}}^2 
   \lrb{s_{x_A,h_B} s_{i_A,q_B}}^{1/2}\,.
\end{align}

Finally, we calculate the denominator of \cref{eqn:AoffdiagBoff_swap_p12}:
\begin{align}
    &\lr{X , X}_{\sigma} \nonumber\\
= &\Tr\left[ \sigma^{1/2}\cdot \sum_{n_A \neq \ell_A} \sum_{e_B\neq f_B} A^{*}_{n_A,\ell_A,{e_B,f_B}} \ketbra{\ell_A}{n_A}\otimes   \ketbra{f_B}{e_B} \otimes I_A \cdot \sigma^{1/2} \right.  \nonumber\\
&\left. \sum_{i_A \neq x_A}\sum_{q_B\neq h_B}  A_{i_A,x_A,q_B,h_B} \ketbra{i_A}{x_A}\otimes  \ketbra{q_B}{h_B} \otimes I_A \right]\, \\
=& \sum_{i_A \neq x_A} \sum_{q_B\neq h_B}  \abs{A_{i_A,x_A,q_B,h_B}}^2  
   \lrb{s_{x_A,h_B} s_{i_A,q_B}}^{1/2}   \,.
\end{align}

Dividing the numerator by the denominator, we have $\Omega(1/d_A)$ lower bound. 
\end{proof}
}

\subsection{Vanishing cross-terms between diagonal and  off-diagonal  components}
\label{sec:cross_terms_vanish}

Based on the calculations in \cref{sec:condition_1,sec:condition_2,sec:condition_3}, we find that $\mathcal{L}_{\lrb{\lrc{\mc{U}}}}$ swaps  subsystem~A while leaving subsystem~B invariant. Since the inner product between a diagonal and an off-diagonal matrix is zero, the cross terms vanish. 
The following \cref{lem:diag_off_diag_decomp_local_A_swap,lem:local_A_swap_part_B_938} formalize this behavior. For the sake of clarity,  
here we emphasize that, we say one operator $Y$ is diagonal in the subsystem $A$ means that 
\begin{align}
    \resizebox{0.88\linewidth}{!}{$
    Y \in \DIAG_A \coloneqq \Span \lrc{\ketbra{i_A}{i_A} \otimes  O_B  \otimes \ketbra{i_A'}{i_A'}\,:\,   \forall \ket{i_A},\ket{i_A'} \text{ in \cref{eq:basis_A}}, \, \forall  O_B  \in \mc{B}\lrb{\mathbb{C}^{d_B}}}\,;
    $}
\end{align}
one operator $Y$ is off-diagonal in the  subsystem $A$ means that
\begin{align}
Y \in \OFFDIAG_A \coloneqq \Span \Bigl\{\, &
\ketbra{i_A}{i_A} \otimes  O_B  \otimes \ketbra{i_A'}{m_A'},\ 
\ketbra{i_A'}{m_A'} \otimes  O_B  \otimes \ketbra{i_A}{i_A}, \ketbra{i_A}{m_A} \otimes  O_B  \otimes \ketbra{i_A'}{m_A'} \nonumber\\
& \;:\;
\forall \ket{i_A}\neq \ket{m_A},\ \ket{i_A'}\neq \ket{m_A'} \text{ in \cref{eq:basis_A}},\ 
\forall  O_B  \in \mc{B}\lrb{\mathbb{C}^{d_B}}
\Bigr\}\,;
\end{align}
one operator $Y$ is diagonal in the subsystem $B$ means that
\begin{align}
    Y \in \DIAG_B \coloneqq \Span \lrc{O_A \otimes \ketbra{j_B}{j_B} \otimes O_A'\,:\,   \forall \ket{j_B} \text{ in \cref{eq:basis_B}}, \, \forall O_A,O_A' \in \mc{B}\lrb{\mathbb{C}^{d_A}}}\,;
\end{align}
one operator $Y$ is off-diagonal in the  subsystem $B$ means that
\begin{align}
    \resizebox{0.88\linewidth}{!}{$
    Y \in \OFFDIAG_B \coloneqq \Span \lrc{O_A \otimes \ketbra{j_B}{j_B'} \otimes O_A'\,:\,   \forall \ket{j_B}\neq \ket{j_B'} \text{ in \cref{eq:basis_B}}, \, \forall O_A,O_A' \in \mc{B}\lrb{\mathbb{C}^{d_A}}}\,.
    $}
\end{align}

\begin{lemma_} \label{lem:diag_off_diag_decomp_local_A_swap}
Following the definitions of  $\mathcal{K} _{\diag A}$, $\mathcal{K} _{\offdiag A}$, and $\mathcal{K} = \mathcal{K} _{\diag A} \oplus \mathcal{K} _{\offdiag A}$ in \cref{eqn:diag_A_def38,eqn:offdiag_A_def39,eqn:ker_mathcal_K}, for any operator $ O_{A,B}  \in \mathcal{K} $, we can decompose it as
\begin{align}
 O_{A,B}  = O_{\diag A , B} + O_{\offdiag A,B}\,,
\end{align}
where $O_{\diag A , B} \in \mathcal{K} _{\diag A} $ and $O_{\offdiag A,B} \in \mathcal{K} _{\offdiag A} $.
Then one can prove
\begin{align}
&\lr{O_{\diag A,B}\otimes I_A, - \lind_{\lrb{\{\CU\}}}\lrb{O_{\offdiag A,B}\otimes I_A}}_{\sigma} = 0\,,\\
&\lr{O_{\offdiag A,B}\otimes I_A, - \lind_{\lrb{\{\CU\}}}\lrb{O_{\diag A,B}\otimes I_A}}_{\sigma} = 0\,,\\
&\lr{O_{\diag A,B}\otimes I_A, O_{\offdiag A,B}\otimes I_A}_{\sigma} = 0 \, , \label{eq:faoi}\\
&\lr{O_{\offdiag A,B}\otimes I_A, O_{\diag A,B}\otimes I_A}_{\sigma} = 0\,.
\label{eq:4y1n}
\end{align}
\end{lemma_}
\begin{proof}
    We will first recall a useful fact that the trace of a product of a diagonal matrix and an off-diagonal matrix is zero, which will be used in the following proof. From \cref{eq:off_A_p1,eq:off_A_p2,eq:off_A_p3,eq:off_AB_p1,eq:off_AB_p2,eq:off_AB_p3}, we can see that
$
  \mathcal{L}_{\lrb{\lrc{\mc{U}}}}\lrb{O_{\offdiag A,B}\otimes I_A}
$
always produces an operator that is off-diagonal in the \(A\) subsystems. Therefore, after multiplying with
$
  O_{\diag A,B}\otimes I_A
  \text{ and }
  {\sigma}^{1/2},
$
which are always diagonal in the \(A\) subsystem, the result will be zero due to the aforementioned fact. Similar reasoning applies to the second equation by observing \cref{eqn:action_A_swap_local_19,eqn:action_A_swap_local_20,eqn:action_A_swap_local_21}.

The analysis of
\cref{eq:faoi,eq:4y1n}
can be done similarly by considering the diagonal and off-diagonal behaviour in \(A\).

\end{proof}

\begin{lemma_} \label{lem:local_A_swap_part_B_938}
    Let $O_{\offdiag A, \offdiag B}$ and $O_{\offdiag A, \diag B}$ be defined as \cref{eqn:O_AB_offdiagA_decomp_8f99}, then 
\begin{align}
&\left\langle O_{\offdiag A,\offdiag B}\otimes I_A,-\mc{L}_{\lrb{\lrc{\mc{U}}}}\lrb{O_{\offdiag A,\diag B}\otimes I_A}\right\rangle_{\sigma}  = 0\,,\\
&\left\langle O_{\offdiag A,\diag B}\otimes I_A,-\mc{L}_{\lrb{\lrc{\mc{U}}}}\lrb{O_{\offdiag A,\offdiag B}\otimes I_A}\right\rangle_{\sigma}  = 0\,,\\
&\left\langle O_{\offdiag A,\offdiag B}\otimes I_A,O_{\offdiag A,\diag B}\otimes I_A\right\rangle_{\sigma}  = 0\,, \label{eq:fnaoigfh}\\
&\left\langle O_{\offdiag A,\diag B}\otimes I_A, O_{\offdiag A,\offdiag B}\otimes I_A\right\rangle_{\sigma}  = 0\, \label{eq:1n9oth}.
\end{align}
\end{lemma_}
\begin{proof}

    This proof is very similar to the proof of \cref{lem:diag_off_diag_decomp_local_A_swap}. From \cref{eq:off_A_p1,eq:off_A_p2,eq:off_A_p3}, we can see that
$
  \mathcal{L}_{\lrb{\lrc{\mc{U}}}}\lrb{O_{\offdiag A,\diag B}\otimes I_A}
$
always produces an operator that is diagonal in the \(B\) subsystem. Multiplication with diagonal
$
  {\sigma}^{1/2}
$
still produces a diagonal operator in the \(B\) subsystem of \(H\). Therefore, after multiplying with
$
  O_{\offdiag A,\offdiag B}\otimes I_A,
$
which is always off-diagonal in the \(B\) subsystem of \(H\), the result will be zero due to the fact that the trace of a product of a diagonal matrix and an off-diagonal matrix is zero. Similar reasoning applies to the second equation by observing \cref{eq:off_AB_p1,eq:off_AB_p2,eq:off_AB_p3}.

The analysis of
\cref{eq:fnaoigfh,eq:1n9oth}
can be done similarly by considering the diagonal and off-diagonal behaviour in \(B\).
\end{proof}

\subsection{Kernel of the swapping Lindbladian} \label{sec:kernel_local_A_swap}
In this section, we prove a proposition concerning the kernel of $\mathcal{L}_{\lrb{\lrc{\mc{U}}}}$, a result that is central to the analysis in \cref{app:Gap_lower_bound}. Our proof builds on the observations from \cref{sec:condition_1,sec:condition_2,sec:condition_3}.

The preceding subsections have shown that the action of $\mathcal{L}_{\lrb{\lrc{\mc{U}}}}$ on an operator of the form $X=O_{A,B}\otimes I_A$ is to swap the eigenstates of subsystem~$A$ while leaving subsystem~$B$ invariant. This implies that $\mathcal{L}_{\lrb{\lrc{\mc{U}}}}$ effectively functions as a local ``Lindbladian" on subsystem $A$. A fundamental property of any Lindbladian $\lind$ is that the identity operator lies in its kernel; that is, $I \in \ker\lrb{\lind}$. This observation motivates the conjecture that,
$\exists  O_B  \in \mc{B}\lrb{\mathbb{C}^{d_B}}, \text{ s.t. } I_A\otimes  O_B  \otimes I_A \in \ker \lrb{\mathcal{L}_{\lrb{\lrc{\mc{U}}}}}$.

We will now prove that this conjecture holds. In fact, we  established the following stronger result:
\begin{proposition_}\label{prop:kernel_local_A_swap}
Let $\lind_{\lrb{H\otimes I_A+ I_n\otimes I_A, \beta, \{\CU\}, \gamma_M}}$ be defined as in \cref{eq:local_swap_Lindbladian}, then
\begin{align}
    &\ker\left(\lind_{\lrb{H\otimes I_A+ I_n\otimes I_A, \beta, \{\CU\}, \gamma_M}}\right)\cap \mc{K}\otimes I_A
    = \rev{\operatorname{span}}\left\{I_A\otimes I_B \otimes {I}_A\right\}
\end{align}
where $\mc{K}=\Span \left\{\ketbra{i_{A}}{i_{A}}\otimes I_{B} ,\left(\ketbra{i_{A}}{i'_{A}}\otimes  O_B  \right)\middle|\forall \ket{i_A} \neq \ket{i_A'},\, \forall  O_B \in\mc{B}\lrb{\mathbb{C}^{{d_B}}} \right\}$ is defined in \cref{eqn:ker_mathcal_K}.
\end{proposition_}
\begin{proof}
Like what we did in \cref{eq:X_decomp_diag_offdiagA_182,eqn:O_AB_offdiagA_decomp_8f99}, we can write any operator $X\in \mc{K} \otimes I_A$ as
\begin{align}
    X = O_{\diag A}\otimes I_B \otimes I_A + O_{\offdiag A,\diag B}\otimes I_A + O_{\offdiag A,\offdiag B}\otimes I_A\,.
\end{align} 
The action of $\mathcal{L}_{\lrb{\lrc{\mc{U}}}}$ on each term of  $X$ can \rev{be} found in \cref{sec:condition_1,sec:condition_2,sec:condition_3}.

Similar to \cref{sec:cross_terms_vanish},
if $X$ is (off-)diagonal in the eigenbasis of subsystem $A$ (resp.\ $B$), then 
$\mathcal{L}_{\lrb{\lrc{\mc{U}}}}(X)$ is likewise (off-)diagonal in the subsystem $A$ (resp.\ $B$). 
Specifically, we have 
\begin{align}
    &\mathcal{L}_{\lrb{\lrc{\mc{U}}}}\lrb{O_{\diag A}\otimes I_B \otimes I_A } \in \DIAG_A \cap \DIAG_B\,,\\
    &\mathcal{L}_{\lrb{\lrc{\mc{U}}}}\lrb{O_{\offdiag A,\diag B}\otimes I_A} \in \OFFDIAG_A \cap \DIAG_B\,,\\
    &\mathcal{L}_{\lrb{\lrc{\mc{U}}}}\lrb{O_{\offdiag A,\offdiag B}\otimes I_A} \in \OFFDIAG_A \cap \OFFDIAG_B\,,
\end{align}
where $\DIAG_A$, $\OFFDIAG_A$, $\DIAG_B$, and $\OFFDIAG_B$ are defined in \cref{sec:cross_terms_vanish}.
Since the above three subspaces are mutually orthogonal, to get the kernel of $\mathcal{L}_{\lrb{\lrc{\mc{U}}}}$, it is sufficient to consider the kernel of $\mathcal{L}_{\lrb{\lrc{\mc{U}}}}$ restricted to each of the above three subspaces.

Let us first check the kernel of $\mathcal{L}_{\lrb{\lrc{\mc{U}}}}$ restricted to $\OFFDIAG_A \cap \DIAG_B$ as discussed in \cref{sec:condition_2}. 
{To have $\mathcal{L}_{\lrb{\lrc{\mc{U}}}}\lrb{O_{\offdiag A,\diag B}\otimes I_A}=0$, which is equivalent to 
$\text{\cref{eq:off_A_p1}}-\tfrac{1}{2}\lrb{\text{\cref{eq:off_A_p2}}+\text{\cref{eq:off_A_p3}}}=0$, 
note that \cref{eq:off_A_p1} yields terms of the form $\ketbra{m_A}{m_A}\otimes\ketbra{i_A}{x_A}$, whereas \cref{eq:off_A_p2,eq:off_A_p3} yield $\ketbra{i_A}{x_A}\otimes\ketbra{m_A}{m_A}$. Since $i_A\neq x_A$, these two families lie in orthogonal operator subspaces (off-diagonal on different tensor legs) and therefore cannot cancel.
Hence, the kernel of $\mathcal{L}_{\lrb{\lrc{\mc{U}}}}$ restricted to $\OFFDIAG_A \cap \DIAG_B$ is \rev{$\{0\}$}. 
}
Similar analysis applies to the kernel of $\mathcal{L}_{\lrb{\lrc{\mc{U}}}}$ restricted to $\OFFDIAG_A \cap \OFFDIAG_B$ as discussed in \cref{eq:off_AB_p1,eq:off_AB_p2,eq:off_AB_p3} of \cref{sec:condition_3}, which is also \rev{$\{0\}$}.
The remaining subspace \rev{is} $\DIAG_A \cap \DIAG_B$: 
From \cref{eqn:action_A_swap_local_19,eqn:action_A_swap_local_20,eqn:action_A_swap_local_21} in \cref{sec:condition_1}, we can see that only when $X= I_A \otimes \ketbra{q_B}{q_B}\otimes I_A$, 
\begin{align}
    &\mathcal{L}_{\lrb{\lrc{\mc{U}}}}\lrb{I_A \otimes \ketbra{q_B}{q_B}\otimes I_A} = 0, \, \forall \ket{q_B} \text{ in \cref{eq:basis_B}}\,.
\end{align}
Since $\mathcal{K} =  \Span \left\{\ketbra{i_{A}}{i_{A}}\otimes I_{B} ,\left(\ketbra{i_{A}}{i'_{A}}\otimes  O_B  \right)\right\}$ and we are considering $\ker\lrb{\mathcal{L}_{\lrb{\lrc{\mc{U}}}}}\cap \mathcal{K} \otimes I_A$, we have
\begin{align}
    \ker\left(\mathcal{L}_{\lrb{\lrc{\mc{U}}}}\right)\cap \mc{K}\otimes I_A
    = \Span \left\{I_A\otimes I_B \otimes {I}_A\right\}\,.
\end{align}

\end{proof}

\end{document}